\documentclass{aa}  
\usepackage{graphicx}
\usepackage{txfonts}
\begin{document}
   \title{High precision X-ray logN-logS distributions: implications for the obscured AGN population}
   \subtitle{}

   \author{
     S. Mateos  \inst{1}
     \and
     R.S. Warwick \inst{1}
     \and
     F. J. Carrera \inst{2}
     \and
     G.C. Stewart \inst{1}
     \and
     J. Ebrero \inst{1,}\inst{2}
     \and
     R. Della Ceca \inst{3}
     \and
    A. Caccianiga \inst{3}
     \and
     R. Gilli \inst{4}
     \and
     M.J. Page \inst{5}
     \and
     E. Treister \inst{6}
     \and
     J.A. Tedds \inst{1}
     \and
     M.G. Watson  \inst{1}
     \and
     G. Lamer \inst{7}
     \and
     R.D. Saxton \inst{8}
     \and
     H. Brunner \inst{9}
     \and
     C.G. Page \inst{1}
    }

           \authorrunning{S. Mateos et al.}

   \offprints{S. Mateos, \email{sm279@star.le.ac.uk}}
   \institute{
     X-ray Astronomy Group, Department of Physics and Astronomy, Leicester University, Leicester LE1 7RH, UK
     \and Instituto de F\'\i sica de Cantabria (CSIC-UC), 39005 Santander, Spain
     \and INAF-Osservatorio Astronomico di Brera, via Brera 28, I-20121 Milan, Italy
     \and Istituto Nazionale di Astrofisica (INAF) - Osservatorio Astronomico di Bologna, via Ranzani 1, 40127 Bologna, Italy
     \and Mullard Space Science Laboratory, University College London, Holmbury St. Mary, Dorking, Surrey RH5 6NT, UK
     \and European Southern Observatory, Casilla 19001, Santiago 19, Chile
     \and Astrophysikalisches Institut Potsdam, An der Sternwarte 16, 144482, Potsdam, Germany
     \and XMM SOC, ESAC, Apartado 78, 28691 Villanueva de la Cañada, Madrid, Spain
     \and Max-Planck-Institut f\"{u}r Extraterrestrische Physik, Giessenbachstrasse, Garching D-85748, Germany
     }
 \date{10 September 2008}

  \abstract
      {Our knowledge of the properties of AGN, especially those of optical type-2 objects is very incomplete. 
	Extragalactic source count distributions are dependent on the cosmological and statistical properties of AGN, 
	and therefore provide a direct method of investigating the underlying source populations.} 
   {We aim to constrain the extragalactic source count distributions over a broad range 
     of X-ray fluxes and in various energy bands to test 
     whether the predictions from X-ray background synthesis models agree with the 
    observational constraints provided by our measurements.
   }
   {We have used 1129 XMM-{\it Newton} observations at $|b|>20{^\circ}$ covering a total sky area of 
132.3 ${\rm deg^2}$ to compile the largest complete samples of X-ray selected objects to date both in the 0.5-1 keV, 1-2 keV, 2-4.5 keV, 4.5-10 keV bands employed in standard XMM-{\it Newton} data processing and in the 0.5-2 keV and 2-10 keV energy bands more usually considered in source count studies.
Our survey includes in excess of 30,000 sources and spans fluxes from $\sim$${\rm 10^{-15}}$ to ${\rm 10^{-12}\,erg\,cm^{-2}\,s^{-1}}$ below 2 keV 
and from $\sim$${\rm 10^{-14}}$ to ${\rm 10^{-12}\,erg\,cm^{-2}\,s^{-1}}$ above 2 keV where the bulk of the CXRB energy density is produced.}
   {The very large sample size we obtained means our results are not limited by cosmic variance or low counting statistics. A break in the 
     source count distributions was detected in all energy bands except the 4.5-10 keV band. We find that an analytical model comprising 2 
     power-law components cannot adequately describe the curvature seen in the source count distributions.
The shape of the logN($>$S)-logS is strongly dependent on the energy band
with a general steepening apparent as we move to higher energies. This is due to the fact 
that non-AGN populations, comprised mainly of stars and clusters of galaxies,
contribute up to 30\% of the source population at energies $<$2 keV and at fluxes 
$\ge$${\rm 10^{-13}\,erg\,cm^{-2}\,s^{-1}}$, and these populations of objects have 
significantly flatter source count distributions than AGN. 
We find a substantial increase in the relative fraction of hard X-ray sources 
at higher energies, from $\ge$55\% below 2 keV to $\ge$77\% above 2 keV. However the majority of sources 
detected above 4.5 keV still have significant flux below 2 keV. 
Comparison with predictions from the synthesis models suggest that the models 
might be overpredicting the number of faint 
absorbed AGN, which would call for fine adjustment of some model parameters such as the obscured to unobscured AGN ratio 
and/or the distribution of column densities at intermediate obscuration.
}
   {}


   \maketitle
%
\section{Introduction}
The deepest X-ray surveys carried out to date by {\it Chandra} (Chandra Deep Field North, CDF-N; Alexander et al.~\cite{Alexander03} and Chandra Deep Field South, CDF-S; Giacconi et al.~\cite{Giacconi02}, Lou et al.~\cite{Lou08}) 
and XMM-{\it Newton} (Hasinger et al.~\cite{Hasinger01}) have 
resolved up to 90\% of the Cosmic X-ray background (CXRB) at energies below $\sim$5 keV into discrete 
sources reaching limiting fluxes of ${\rm \sim2\times10^{-17}\,erg\,cm^{-2}\,s^{-1}}$ 
in the 0.5-2 keV band and ${\rm \sim2\times10^{-16}\,erg\,cm^{-2}\,s^{-1}}$ in the 2-8 keV band 
(Bauer et al.~\cite{Bauer04}). 

However, above $\sim$5 keV the fraction of CXRB resolved into sources is substantially lower (see e.g. Worsley et al.~\cite{Worsley04}, Worsley et al.~\cite{Worsley05}) although precise estimates are hampered by the remaining uncertainty in the absolute normalisation 
of the CXRB (see e.g. Cowie et al.~\cite{Cowie02}). Additional uncertainties originate due to variations of the source counts between surveys 
arising from both the impact of the large scale structure of the Universe on the source distribution
(Gilli et al.~\cite{Gilli03}) and, more mundanely, on cross calibration uncertainties 
between different missions (Barcons et al.~\cite{Barcons00}, De Luca \& Molendi~\cite{De Luca04}).

Follow-up campaigns targeted at the sources detected in deep-medium X-ray surveys  
have shown that at high Galactic latitudes Active Galactic Nuclei (AGN) dominate the X-ray sky. 
At bright X-ray fluxes (${\rm\gtrsim 10^{-14}\,erg\,cm^{-2}\,s^{-1}}$),
unabsorbed or mildly absorbed AGN, spectroscopically identified as type-1 AGN represent 
the bulk of the population (see e.g. Shanks et al.~\cite{Shanks91}, Barcons et al.~\cite{Barcons07}, Caccianiga et al.~\cite{Caccianiga07}). 
At intermediate fluxes, absorbed AGN (optical type-2 AGN) at low redshifts (z$\lesssim$1) 
become dominant, while at fluxes ${\rm\lesssim 10^{-16}\,erg\,cm^{-2}\,s^{-1}}$ a population of 
`normal' galaxies starts to emerge (Barger et al.~\cite{Barger03}, Hornschemeier et al.~\cite{Hornschemeier03}, 
Bauer et al.~\cite{Bauer04}).  

   \begin{figure}[]
   \centering
   \includegraphics[angle=-90,width=0.45\textwidth]{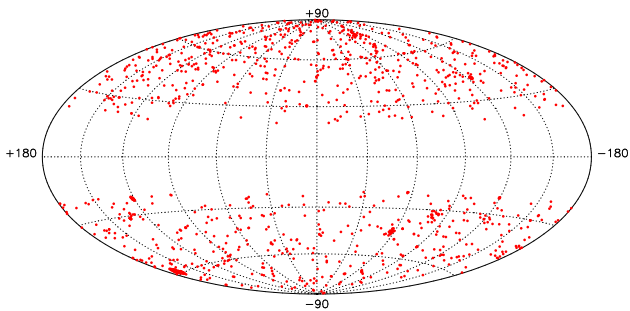}
   \caption{Sky distribution in Galactic coordinates of the selected observations.
     The high density of pointings in some areas of the sky correspond to planned surveys of relatively large sky areas (e.g. 
       the XMM-{\it Newton} Large Scale Survey, Pierre et al.~\cite{Pierre04}).}
              \label{skymap}%
    \end{figure}

Although the nature of the sources that dominate the CXRB is reasonably clear, there are still large uncertainties 
in the cosmological and statistical properties of the objects, especially for type-2 AGN for which 
the redshift and column density distributions are rather poorly determined to date.

One of the most important open issues regarding the population of absorbed AGN is 
whether the relative fraction of obscured AGN varies with redshift or X-ray luminosity.
Some results suggest that this fraction is independent of the X-ray luminosity and redshift (Dwelly \& Page~\cite{Dwelly06}), 
while others point to a decrease in the fraction of absorbed AGN 
with the X-ray luminosity (Ueda et al.~\cite{Ueda03}, Barger et al.~\cite{Barger05}, Hasinger et al.~\cite{Hasinger05}, 
La Franca et al.~\cite{La Franca05}, Akilas et al.~\cite{Akylas06}, Della Ceca et al.~\cite{Ceca08}) and/or an increase with redshift 
(La Franca et al.~\cite{La Franca05}, Ballantine et al.~\cite{Ballantine06}, Treister \& Urry~\cite{Treister06}). 
This issue has been recently 
investigated by Della Ceca et al.~(\cite{Ceca08}) via the analysis of a complete spectroscopically identified 
($\sim$97\% completeness) sample of bright X-ray sources 
(${\rm >7\times10^{-14}\,erg\,cm^{-2}\,s^{-1}}$) selected in the 4.5-7.5 keV band. The sources in this study were bright enough to obtain their absorption 
properties from a detailed analysis of their X-ray spectra. They report a dependence of the fraction of obscured AGN 
on both the luminosity and redshift, the measured evolution being consistent with that proposed by Treister \& Urry~(\cite{Treister06}).
A dependence of the fraction of absorbed AGN on the luminosity has also been suggested by observations in 
the optical and mid-infrared (Simpson et al.~\cite{Simpson05}, Maiolino et al.~\cite{Maiolino07}).

X-ray surveys are the best way to understand the properties (i.e. X-ray absorption distributions) and cosmological evolution 
(i.e. X-ray luminosity functions) of AGN and to test the predictions of the synthesis models of the CXRB 
(Treister \& Urry~\cite{Treister06}, Gilli et al.~\cite{Gilli07}). However in order to provide 
strong observational constraints these analyses require complete samples with a high fraction of sources spectroscopically 
identified. This is a difficult and very 
time consuming task that can only be achieved for relatively small samples of objects.
Source count distributions are dependent on the cosmological and statistical properties of 
AGN, and therefore provide a rather direct method of investigating the properties of the underlying source populations.
Constraining the shape of the source counts is fundamental for cosmological studies of AGN, as it provides 
strong observational constraints for the synthesis models of the CXRB. The general shape of the source counts in the 
0.5-2 keV and 2-10 keV bands is well determined from deep and medium depth X-ray surveys. The results 
of these analyses show that at fluxes $\sim$${\rm 10^{-15}-10^{-14}\,erg\,cm^{-2}\,s^{-1}}$ 
the source count distributions can be reproduced well with broken power-law shapes (i.e two power-law, hereafter 
broken power-law, Baldi et al.~\cite{Baldi02}, Cowie et al.~\cite{Cowie02}, 
Cappelluti et al.~\cite{Cappelluti07}, Carrera et al.~\cite{Carrera07}, Brunner et al.~\cite{Brunner08}). 
However, mostly due to poor statistics, a proper determination of the 
analytical form of the source count distributions is still unavailable, especially at high energies.

Deep pencil beam surveys are important to study the populations of sources 
at the faintest accessible fluxes and therefore they are best suited to constrain the faint-end slope of the source counts. However these observations only sample small sky areas and therefore suffer from significant cosmic variance.
For example, the CDF-N and CDF-S source counts deviate by  
more than 3.9$\sigma$ at the faintest flux levels (Bauer et al.~\cite{Bauer04}).
On the other hand, wide shallow surveys cover much larger areas of the sky and therefore are less affected 
by cosmic variance. However, they only sample sources at relatively bright fluxes which only contribute a small 
fraction of the CXRB emission. 
Surveys at intermediate fluxes, $\sim$${\rm 10^{-15}-10^{-13}\,erg\,cm^{-2}\,s^{-1}}$, of the type 
reported here, fill the gap 
between deep pencil beam and shallow surveys and sample the fluxes at the break in the source count
distributions, i.e. where the bulk of the CXRB energy density should be produced. 
These surveys are therefore appropriate to accurately determine the position of the break and bright-end slope of source count distributions.

In this paper we put strong constraints on the analytical shape of the 
extragalactic source count distributions in 
a number of different energy bands and over a wide range of fluxes.
For this purpose we have compiled the largest complete samples of X-ray selected 
objects to date, ensuring that our results are not limited by low 
counting statistics or cosmic variance effects. Taking advantage of our large samples,
we have investigated how the underlying population of X-ray sources changes as we move to higher energies.
Finally we have used the new observational constraints provided by our analysis to check the predictions of current 
synthesis models of the CXRB.

   \begin{figure*}[!th]
   \centering
   \hbox{
   \includegraphics[angle=90,width=0.5\textwidth]{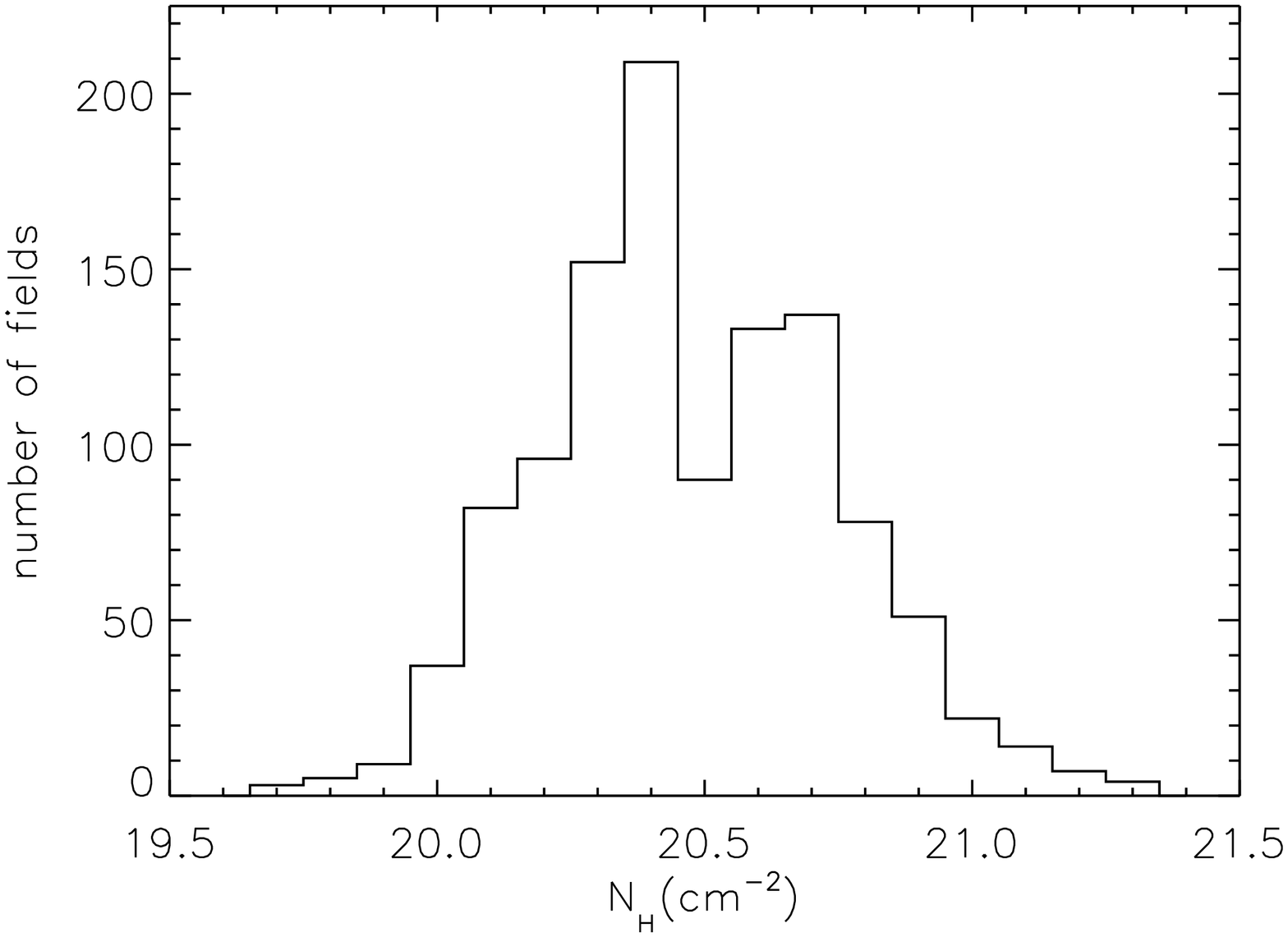}
   \includegraphics[angle=90,width=0.5\textwidth]{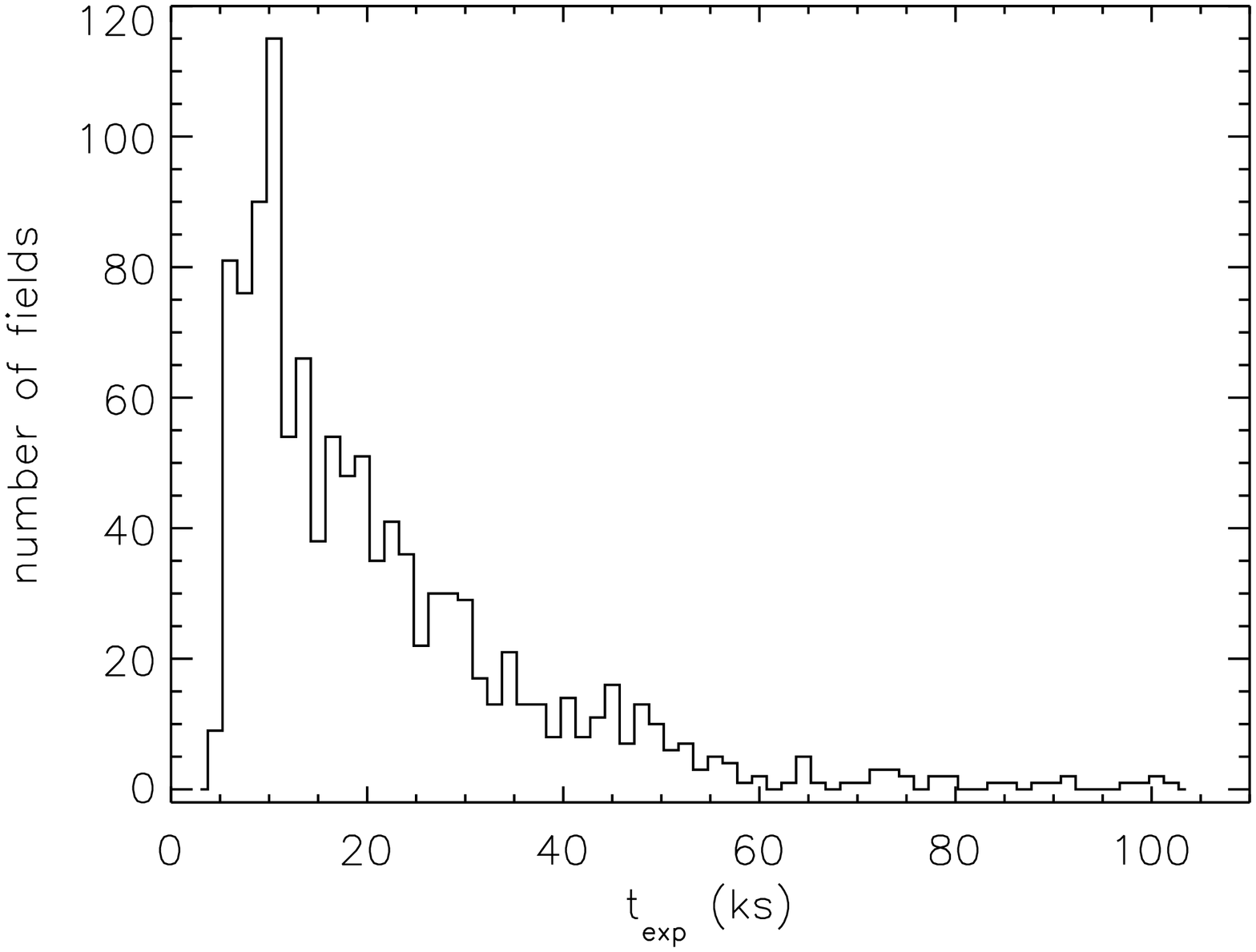}}
   \caption{Left: Distribution of Galactic hydrogen column density (in log units) along the line of sight taken from 
     the 21 cm radio measurements of Dickey \& Lockman~(\cite{Dickey90}).
     Right: Distribution of the exposure times (after filtering).}
              \label{nh_gal_dist}%
    \end{figure*}


The paper is organised as follows: In \textsection2 we describe the selection and processing of
the XMM-{\it Newton} observations (\textsection2.1), the source detection procedure and criteria 
for selection of sources (\textsection2.2 and \textsection2.3), we discuss how we calculated the 
fluxes of the sources from their count rates (\textsection2.4) and we explain how the sky coverage 
was calculated as a function of the X-ray flux (\textsection2.5). 
In \textsection3 we describe the different representations of source counts used in this work (\textsection3.1), 
present source count distributions in different energy bands (\textsection3.2 and \textsection3.3), 
discuss the impact on our source count distributions of biases inherent in our source detection procedure (\textsection3.4), 
describe the approach used to fit our distributions (\textsection3.5) and discuss the fractional X-ray background 
contributed by our sources (\textsection3.6). In \textsection4 we summarise the X-ray spectral properties of our objects.
The implications of our analysis for the cosmic X-ray background synthesis models are presented in \textsection5.
Finally, the summary and conclusions of our analysis are given in \textsection6. 
Appendix A describes the empirical approach used to obtain the sky coverage as a function of the 
X-ray flux. In Appendix B we 
compare our source count distributions with those obtained using data taken directly from the {\tt 2XMM} catalogue.

\section{Data processing and analysis}

\subsection{The XMM-{\it Newton} observations}
\label{observations}
The observations employed in this study are a subset of those utilised in producing the second XMM-{\it Newton} serendipitous source 
catalogue, {\tt 2XMM}\footnote{{\tt http://xmmssc-www.star.le.ac.uk/Catalogue/2XMM/}} (Watson et al.~\cite{Watson08}, submitted). 
{\tt 2XMM} is based on observations from the three European Photon Imaging Cameras (EPIC)
that were public by first of May 2007\footnote{XMM-{\it Newton} observations started on January 2000.}.
Here for simplicity, we have only used data from the EPIC pn camera (Turner et al.~\cite{Turner01}). 
The data have been processed using the XMM-{\it Newton} Science Analysis System (SAS, v7.1.0, Gabriel et al.~\cite{Gabriel04}). 
Because all the observations have been reprocessed using the same pipeline configuration this guarantees a uniform data set. 
Observations have been filtered for high particle background 
periods by excluding the time intervals where the 7-15 keV count rate was higher than 10 pn cts/arcmin$^2$/ks. 

The aim of this study is to constrain source count distributions for 
serendipitous AGN, hence we have selected only observations that fulfil the following criteria:
\begin{enumerate}
\item High galactic latitude fields, $|b|>20{^\circ}$ (so as to obtain samples 
with the contamination from Galactic stars minimised and with low Galactic absorbing column densities).
\item Fields with at least 5 ks of clean pn exposure time.
\item Fields free of bright and/or extended X-ray sources, i.e. 
  where most of the field of view (FOV) can be used for serendipitous source detection.
\end{enumerate}

We have not merged observations carried out at the same sky position. In these cases 
we removed the overlapping area from the observation with the shortest clean exposure time. 
The resulting sample comprises 1129 observations. The sky distribution of the pointings is shown in Fig.~\ref{skymap}.
The distribution of Galactic absorption along the line of sight and the distribution of clean exposure times,
for the set of observations are shown in Fig.~\ref{nh_gal_dist}. 

\subsection{Source detection}
\label{sources}
We have carried out source count analysis using both the 
'standard' 0.5-2 keV and 2-10 keV energy bands and also the narrower 
energy bands 0.5-1 keV, 1-2 keV, 2-4.5 keV and 4.5-10 keV\footnote{In the 4.5-10 keV band pn photons with energies 
between 7.8-8.2 keV were excluded in order to avoid the instrumental background produced 
by Cu K-lines (Lumb et al.~\cite{Lumb02}).}. The former allow comparison with previous results, whereas 
the later allow a more detailed study of the spectral characteristics of the underlying source populations. 
In order to make source 
lists we run the {\tt 2XMM} source detection algorithm on all energy bands simultaneously.
In Appendix~\ref{2xmm_vs_mydet} we compare our 0.5-2 keV and 2-10 keV source count distributions 
with those obtained combining data from the energy bands used to make the {\tt 2XMM} catalogue.

Images were created for each energy band. Only pn events with {\tt PATTERN $\le$4} (single and double events) were selected. 
The SAS task {\tt emask} was used to create a detection mask for each observation, which 
defines the area of the detector suitable for source detection. 
Energy dependent exposure maps were computed using the {\tt SAS} task {\tt eexpmap}, 
using the latest calibration information on the mirror vignetting, quantum efficiency and filter 
transmission\footnote{{\tt Eexpmap} calculates the mirror vignetting function at one single energy, 
the centroid of the energy band. Because the mirror vignetting is a strong function of energy, in the 
cases where the energy band is broad, this approach produces a less accurate determination of the 
effective exposure across the FOV. This is more important at high energies, where 
the dependence of the vignetting on the energy is much stronger. In order to reduce this effect, for 
the energy bands 0.5-2 keV, 4.5-10 keV and 2-10 keV we first computed exposure maps in narrower energy 
bands: 0.5-1 keV and 1-2 keV for band 0.5-2 keV; 2-4.5, 4.5-6 keV, 6-8 keV and 8-10 keV for 
band 2-10 keV, and 4.5-6 keV, 6-8 keV and 8-10 keV for band 4.5-10 keV. Then we 
used the weighted mean of these maps to get the exposure maps in the broader energy bands. In 
order to weight the 
maps we used the number of counts that we should have detected in each narrow band for a source with a power-law spectrum of photon index $\Gamma$=1.9 at energies below 2 keV and $\Gamma$=1.6 at energies above 2 keV (the
same spectral model we used to convert the count rates of the sources to fluxes, see Sec.~\ref{flx_cal}). 
The resulting exposure maps do not strongly depend on the assumed spectral slope. For example, 
$\Delta\Gamma$=$\pm$0.3 changes the exposure map by $\lesssim$0.003\% in the 0.5-2 keV band and 
$\lesssim$1.3\% in the 2-10 keV and 4.5-10 keV bands respectively.}.
Source lists were obtained using the {\tt SAS} task {\tt eboxdetect}, which performs source detection using 
a simple sliding box cell detection algorithm. Background maps were obtained with the {\tt SAS} 
task {\tt esplinemap}. The sources detected by {\tt eboxdetect} are masked out and then {\tt esplinemap}
 performs a spline fit on the resulting image producing a smoothed background map.
{\tt Eboxdetect} is run a second time using the background maps produced by {\tt esplinemap}, which 
increases the sensitivity of the source detection. The final list of objects is obtained from a maximum likelihood 
fit of the distribution of source counts on the images by the {\tt SAS} task {\tt emldetect}. {\tt Emldetect} provides source parameters by fitting the distribution 
of counts of the sources detected by {\tt eboxdetect} with the instrumental point spread 
function (PSF). In addition {\tt emldetect} carries out a fit with the PSF convolved with a 
beta-model profile in order to search for sources extended in X-rays. Source positions, 
count rates corrected for PSF losses and vignetting, extent and detection likelihoods are some 
of the more important source parameters provided by {\tt emldetect}. 
\begin{table*}
  \caption{Summary of the source detection results.}
\label{table:1}      
\centering                          
\begin{tabular}{c c c c c c c c}        
\hline\hline                 
Energy band &  ${\rm N_{tot}}$     &  ${\rm N_{sel}}$ & ${\rm f_{ext}}$ & ${\rm S_{min}/S_{med}}$ & ${\rm cts_{min}/cts_{med}}$ & ${\rm N(>S_{min})}$\\
(keV)   &  &  & (\%) &  ${\rm (10^{-15}\,cgs)}$ & & ${\rm deg^{-2}}$\\
(1) & (2) & (3) & (4) & (5) & (6) & (7)\\
\hline                        
0.5-1 & 21311 & 20694 & 3.6 & 1.0/5.7 & 11/42 & $417.6\pm2.9$\\
1-2   & 21848 & 21185 & 2.4 & 1.2/6.0 & 10/40 & $470.7\pm3.2$\\
2-4.5 &  9926 &  9564 & 1.0 & 3.7/15 & 12/38  & $302.4\pm3.1$\\
4.5-10&  1973 &  1895 & 0.2 & 14.0/50 & 15/46 & $92.2\pm2.1$\\
0.5-2 & 32665 & 31837 & 3.0 & 1.4/8.4 & 13/56 & $605.7\pm3.4$\\
2-10  &  9702 &  9431 & 0.7 & 9.0/35 & 17/57  & $315.6\pm3.2$\\
\hline                                
\end{tabular}
\\
(1) Energy band definition (in keV). (2) Total number of sources detected in the band with a significance of 
detection $\mathcal{L}$$\ge$15 (after excluding the targets of the observations, see Sec.~\ref{srcs_sel}). (3) Final number of sources 
selected to compute the source count 
distributions (see Sec.~\ref{srcs_sel} and Sec.~\ref{sky_coverage} for details). 
(4) Fraction of sources in the sample detected as extended in X-rays.
(5) Minimum and median flux of the sources selected to compute the source count distributions. 
(6) Minimum and median of the distribution of total pn counts in the band (background subtracted) of the sources selected 
to compute the source count distributions. Note that the minimum number of counts correspond to a 
small fraction of sources in the samples. More than 92\% of the sources have at least 20 counts in the 0.5-1 keV, 1-2 keV and 2-4.5 keV bands and 
this fraction increases to more than 98\% in the 0.5-2 keV, 2-10 keV and 4.5-10 keV energy bands.
(7) Cumulative sky density of sources in the various energy bands at the flux limits of our survey.
\end{table*}

   \begin{figure*}
   \centering
   \hbox{
   \includegraphics[angle=90,width=0.5\textwidth]{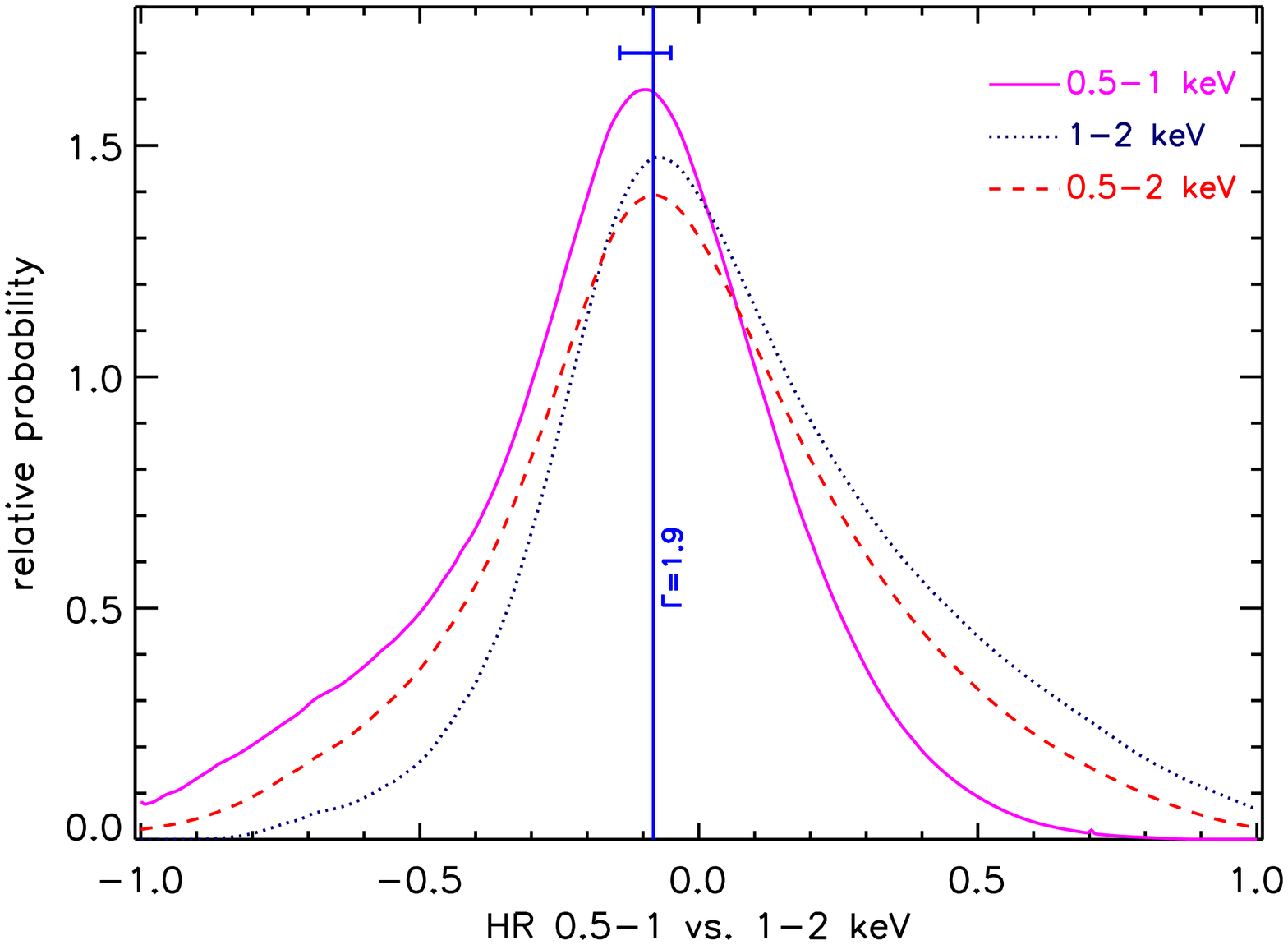}
   \includegraphics[angle=90,width=0.5\textwidth]{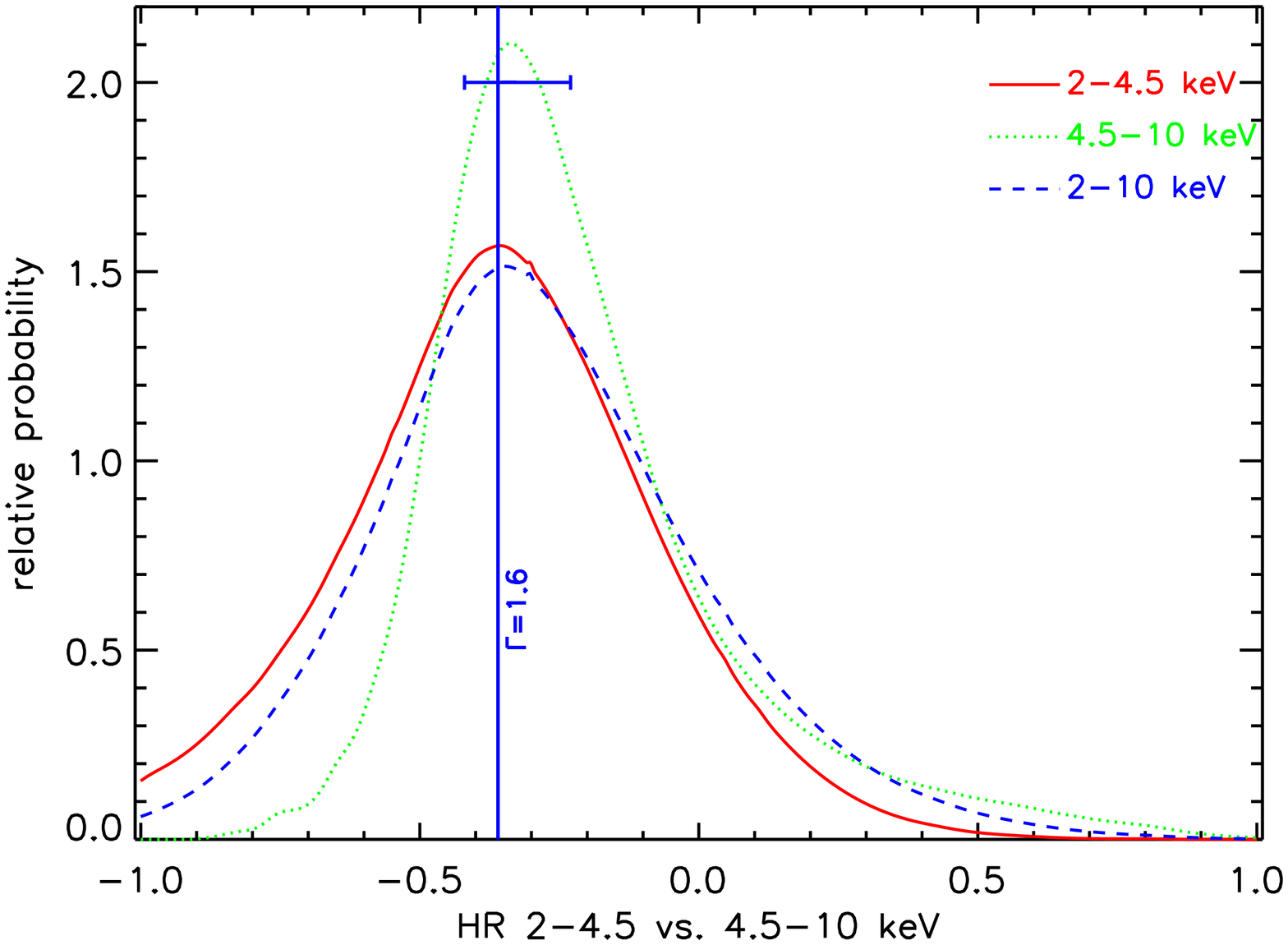}}
   \caption{Probability density distributions of the X-ray colour for sources detected in each energy band (i.e. with a detection likelihood in the band $\mathcal{L}$$\ge$15). 
     Left panel for bands $<$2 keV: the vertical line shows the hardness ratio which corresponds to a 
     source with a spectral slope $\Gamma$=1.9 
       subject to the average Galactic absorption over the sample of objects (${\rm N_H\sim2\times10^{20}\,cm^{-2}}$). 
       Right panel for bands $>$2 keV: the vertical line shows the hardness ratio which corresponds to a 
       source with a spectral slope $\Gamma$=1.6.
     The horizontal error bars show the change in X-ray colour for $\Delta\Gamma$=$\pm$0.3.}
              \label{col_dists}%
    \end{figure*}

\subsection{Selection of sources}
\label{srcs_sel}
We filtered the source lists in several ways in order to ensure the good quality 
of the data used for the analysis. First for each energy band we selected only those sources with 
a detection likelihood $\mathcal{L}$$\ge$15. 
This value is related to the probability that a Poissonian 
random fluctuation caused the observed counts, $P_{random}$, as $\mathcal{L}$=$-\log\,(P_{random})$ and 
corresponds roughly to a 5$\sigma$ significance of detection for $\mathcal{L}$=15 (Cash et al.~\cite{Cash79}). 

The uncertainties in source parameters become much larger for sources falling near CCD gaps. 
In order to remove these sources from our sample we created new detection masks for each observation with the CCD gaps 
increased by an amount equivalent to the radius encompassing 80\% of the counts of a point source at that local position.
All sources falling in the enlarged CCD gaps were masked out.
Photons registered during the readout of the pn CCDs are assigned the wrong position in the 
readout direction. The background produced by these so called out-of-time events is included in the modelling of 
the background maps. However if there is pileup for the source responsible for the out-of-time events then the 
correction is underestimated. In these cases the regions of the FOV affected by out-of-time events were masked out manually. 

Because the targets of the observations are likely to be biased towards 
certain populations of X-ray objects, the targets (and target related sources) together with the areas of the FOV 
contaminated by their emission have been excluded from the analysis. 
The number of sources at the brightest 
fluxes sampled by our survey is rather small. This together with the fact that a 
significant fraction of the X-ray brightest sources in our samples are the target of the observation, 
means that our survey is not a proper unbiased and complete statistical sample of sources 
at the brightest flux levels. Because of this 
we restricted our analysis to sources with fluxes ${\rm \le10^{-12}\,erg\,cm^{-2}\,s^{-1}}$ in each energy band.

A summary of the source detection results for each energy band 
is given in Table~\ref{table:1}. 

\subsection{Count rate to flux conversion factors}
\label{flx_cal}
One important issue in this analysis is the conversion from count rates to fluxes.
Ideally we should obtain the fluxes from the best fit model of the spectrum of each individual 
object. However, because the majority of the sources in our analysis are very faint, we cannot reproduce 
well the spectral complexity often observed in the broad band X-ray spectra of extragalactic 
objects (see e.g. Caccianiga et al.~\cite{Caccianiga04}, Mateos et al.~\cite{Mateos05b}, Mainieri et al.~\cite{Mainieri07}). 
Therefore we have made the reasonable assumption that the spectra of our objects can be well described with a simple power-law model 
absorbed by the Galactic column density along the line of sight.

We have investigated the spectral slope that best reproduces the X-ray colour\footnote{The X-ray colour or hardness ratio is
defined as the normalised ratio of the count rates in two energy bands, HR=(H-S)/(H+S), where H and S are the 
count rates in the harder and softer of the two energy bands respectively.} distribution of the objects 
detected in each energy band. 
In order to provide a better determination of the spectral shape of the sources 
we defined the X-ray colours using count rates in energy bands as close as possible to the band of interest: 
(0.5-1 keV vs. 1-2 keV) for bands 0.5-1 keV, 1-2 keV and 0.5-2 keV and 
(2-4.5 keV vs. 4.5-10 keV) for the 2-4.5 keV, 4.5-10 keV and 2-10 keV energy bands. 
Because our sources are typically faint the uncertainties on their measured X-ray colours can be 
large\footnote{The mean error of the X-ray colours was found to be $\sim$0.1-0.15 for sources detected in the 0.5-2 keV, 0.5-1 keV, 1-2 keV and 4.5-10 keV energy bands and  $\sim$0.2 for sources detected in the 2-4.5 keV and 2-10 keV energy bands.}. In 
order to account for 
this we have calculated the distribution of X-ray colours by adding the 
probability density distributions of the X-ray colour of each individual source. 
For a given source this distribution was defined as a 1-d Gaussian 
with mean and dispersion equal to the value of the X-ray colour and its respective error. 
The resulting 'integrated' probability density distributions are shown in Fig.~\ref{col_dists}.    
The spectral parameters that best characterise the distribution of X-ray colours of our sources are $\Gamma$=1.9 and 
Galactic absorption ${\rm N_H=2\times10^{20}\,cm^{-2}}$ (the average value over the sample of objects)
at energies below 2 keV, and $\Gamma$=1.6 at energies above 2 keV. 
We note that fixing $\Gamma$ at 1.9 but varying ${\rm N_H}$ by a factor of 2 results in a shift of the HR 0.5-1 vs. 1-2 keV
value of just $\sim0.025$. The values of the X-ray colour that correspond to the selected spectral model 
are shown with vertical solid lines in Fig.~\ref{col_dists}. 
Earlier spectral studies have in fact confirmed that such values of $\Gamma$ are representative of the 
average spectra of sources in the flux range of this analysis (Mateos et al.~\cite{Mateos05b}, Carrera et al.~\cite{Carrera07}).
We note that there is a hardening of the effective spectral slope of the sources at energies $\ge$2 keV. This can be easily 
explained as due to the spectral complexity of the X-ray emission of the sources at the energies sampled by our analysis.
For example, the signatures of soft excess emission 
are mostly detected at rest-frame energies below $\sim$2 keV while Compton reflection is only important at rest-frame energies 
$\gtrsim$10 keV. On the other hand X-ray absorption can affect the observed X-ray spectra of the sources over a broad range of energies depending on both the amount of X-ray absorption and the redshift of the objects.

We have investigated the effect on our derived source counts of varying the choice of mean spectral index by  
calculating the change in flux associated with the change in the power-law shape. 
We find that for $\Delta\Gamma$=$\pm$0.3 the largest effect is in the 4.5-10 keV and 2-10 keV 
energy bands where the fluxes can change by up to $\sim$9\%. In the other energy bands the effect is much 
less important, i.e. roughly 1-2\%. 
This is an expected result since the effective area of the EPIC pn detector is fairly flat from $\sim$0.5 keV 
to $\sim$5 keV.

However we see in Fig.~\ref{col_dists} that a change in the power-law continuum by $\Delta\Gamma$=$\pm$0.3 cannot 
explain the dispersion in the observed distribution of X-ray colour of the sources.
In order to account better for the large dispersion in the X-ray colour of the sources for a given object we use a 
count rate to flux conversion based on the value of $\Gamma$ derived from its X-ray colour (instead of a fixed value for all sources). 
The effect on the source counts is negligible in all energy 
bands except in the 2-10 keV and 4.5-10 keV energy bands, where a change in the 
normalisation of the distributions $\lesssim$20\% is observed. 
However, as most of the sources in our analysis are typically faint, 
in the majority of the cases they have a significance of detection well below our selection 
threshold in at least one of the energy bands used to calculate their X-ray colours. Hence
the estimation of their spectral slope on the basis of their X-ray colour could be highly uncertain.
On this basis we adopt the conservative approach of assuming the same spectral model for all sources.

   \begin{figure*}[!ht]
   \centering
   \includegraphics[angle=90,width=0.5\textwidth]{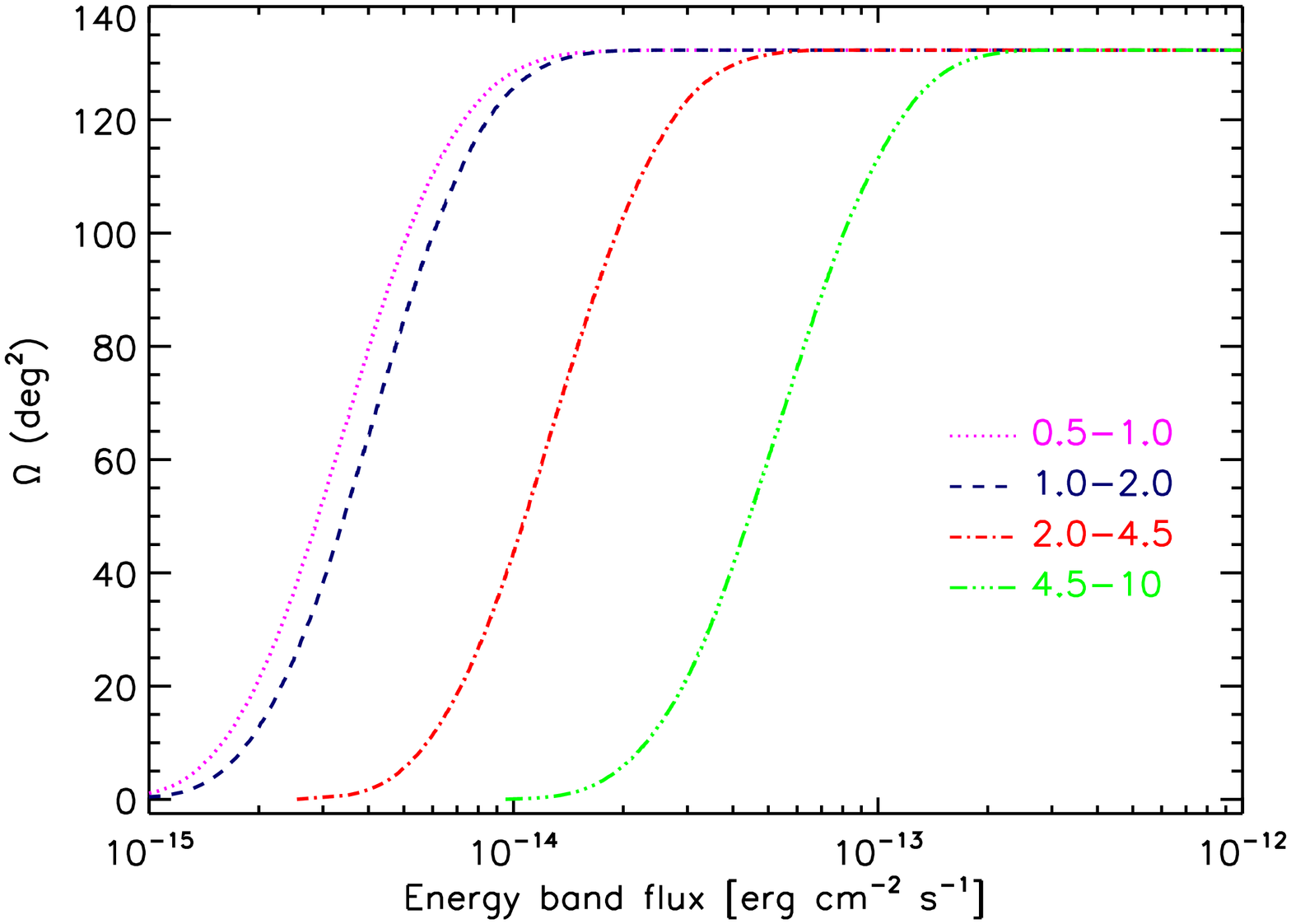}
   \includegraphics[angle=90,width=0.5\textwidth]{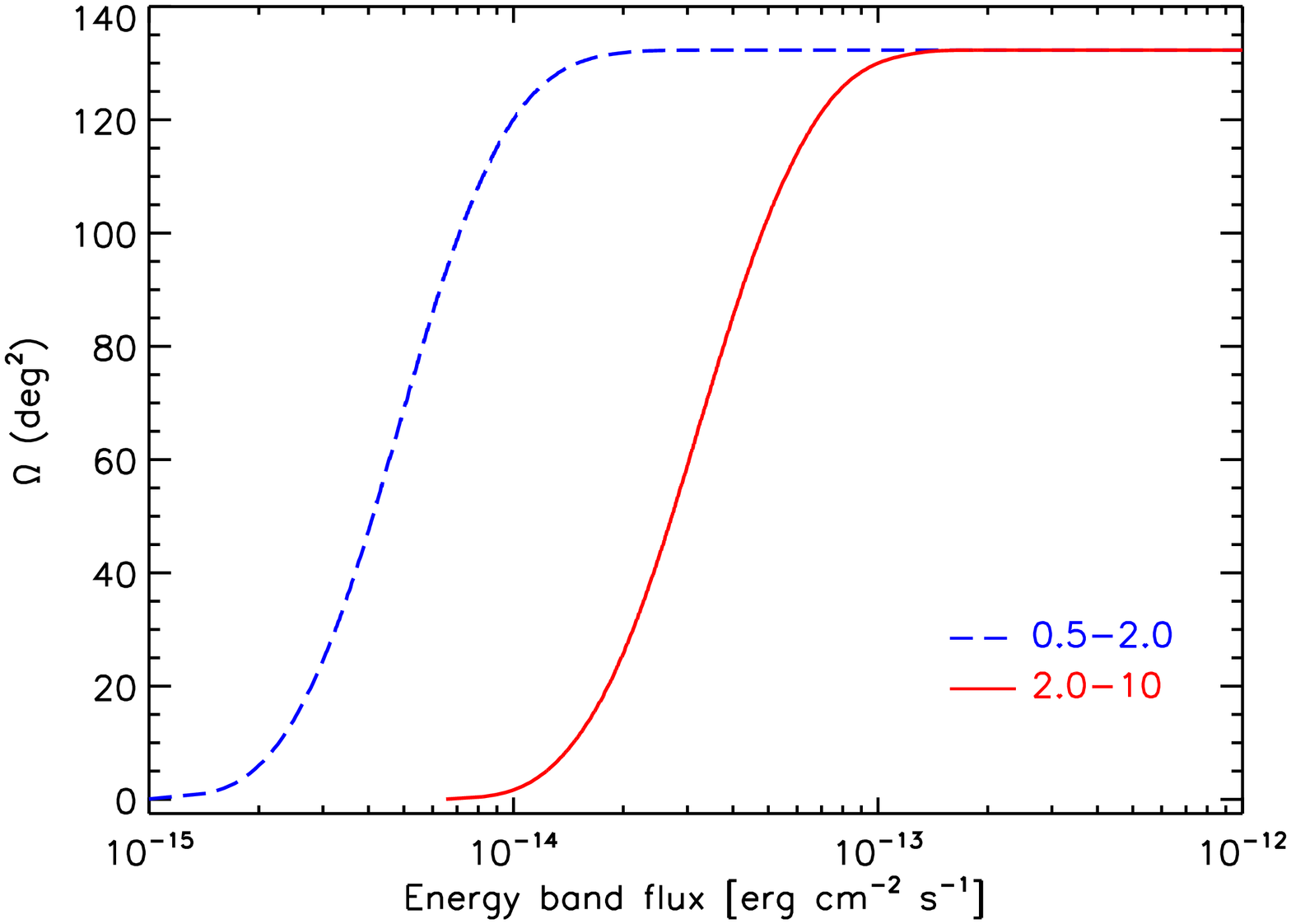}
   \caption{Distributions of the sky coverage as a function of X-ray flux for the different energy bands.}
              \label{omega_plot}%
    \end{figure*}

Energy conversion factors ($ecf$) between count rates and fluxes (corrected for the Galactic absorption 
along the line of sight) were computed for each observation. These values depend on the amount of Galactic absorption along the line of sight, the observing mode and the filter utilised in the observation.
By computing an $ecf$ for each observation we account for the changing 
sensitivity in our softer energy bands (resulting from variations in the Galactic absorption) in the sky coverage 
calculation. 

On the basis of the above, in order to calculate the $ecf$ we have assumed that the spectrum of our 
objects can be well described with a simple power-law model of photon index $\Gamma$=1.9 for energy bands 
0.5-1 keV, 1-2 keV and 0.5-2 keV and $\Gamma$=1.6 for energy bands 2-4.5 keV, 4.5-10 keV and 2-10 keV and 
the corresponding Galactic column density along the line of sight. 

The latest public pn on-axis redistribution matrices (for single and double events, v6.9) 
available at the time of this analysis were used in the computation 
of the energy conversion factors for each field together with on-axis 
effective area files produced by the {\tt SAS} task {\tt arfgen}. The count rates from {\tt emldetect} 
are corrected for the exposure map (which includes vignetting and bad pixel corrections) and the PSF enclosed energy 
fraction. Hence the effective areas were generated disabling these corrections, as indicated in the {\tt arfgen} 
documentation (see also Carrera et al.~\cite{Carrera07} for details).

\subsection{Sky coverage calculation}
\label{sky_coverage}
We have used an empirical approach to obtain the sky coverage as a function of the X-ray flux for 
the selected threshold in detection significance ($\mathcal{L}$$\ge$15). 
We compute a ``sensitivity map'' for each observation which describes 
the minimum count rate that a source must have at each position 
in the FOV to be detected with significance $\mathcal{L}$$\ge$15, taking into account both the local 
effective exposure and background level.
Full details of the method can be found in Appendix A of Carrera et al~(\cite{Carrera07}) and 
also in Appendix~\ref{esens_calc} of this paper.
The count rates of the ``sensitivity maps'' are converted to fluxes as specified in Sec.~\ref{flx_cal}.

In order to make our source lists consistent with the results of the sky coverage calculation we excluded from the 
computation of the source count distributions all sources having actual count rates below the computed minimum value for detection
at the source position. The fraction of sources removed is less than 
4\% in the energy bands 0.5-1 keV, 1-2 keV and 0.5-2 keV, $\sim$5\% in bands 2-4.5 keV and 2-10 keV 
and $\sim$7\% in band 4.5-10 keV (cf. columns 2 \& 3 in Table~\ref{table:1}).

The dependence of the sky coverage on the flux for the various energy bands is shown in Fig.~\ref{omega_plot}; our survey covers 
a total sky area of 132.3 ${\rm deg^2}$. In order to avoid uncertainties in the computation of the source count distributions associated with low count statistics
or inaccuracy in the sky coverage calculation at the very faint detection limits, we have only used sources if they 
were detectable over at least 1 deg$^2$ of sky; the result was that less than 0.5\% of sources where removed 
in all energy bands except in the 4.5-10 keV band where the fraction was $\sim$1.5\% while
the change in the flux limits of the survey was negligible.
This constraint gives rise to the energy band flux limits listed in column 5 of Table~\ref{table:1}. 

   \begin{figure*}
   \centering
   \hbox{
   \includegraphics[angle=90,width=0.5\textwidth]{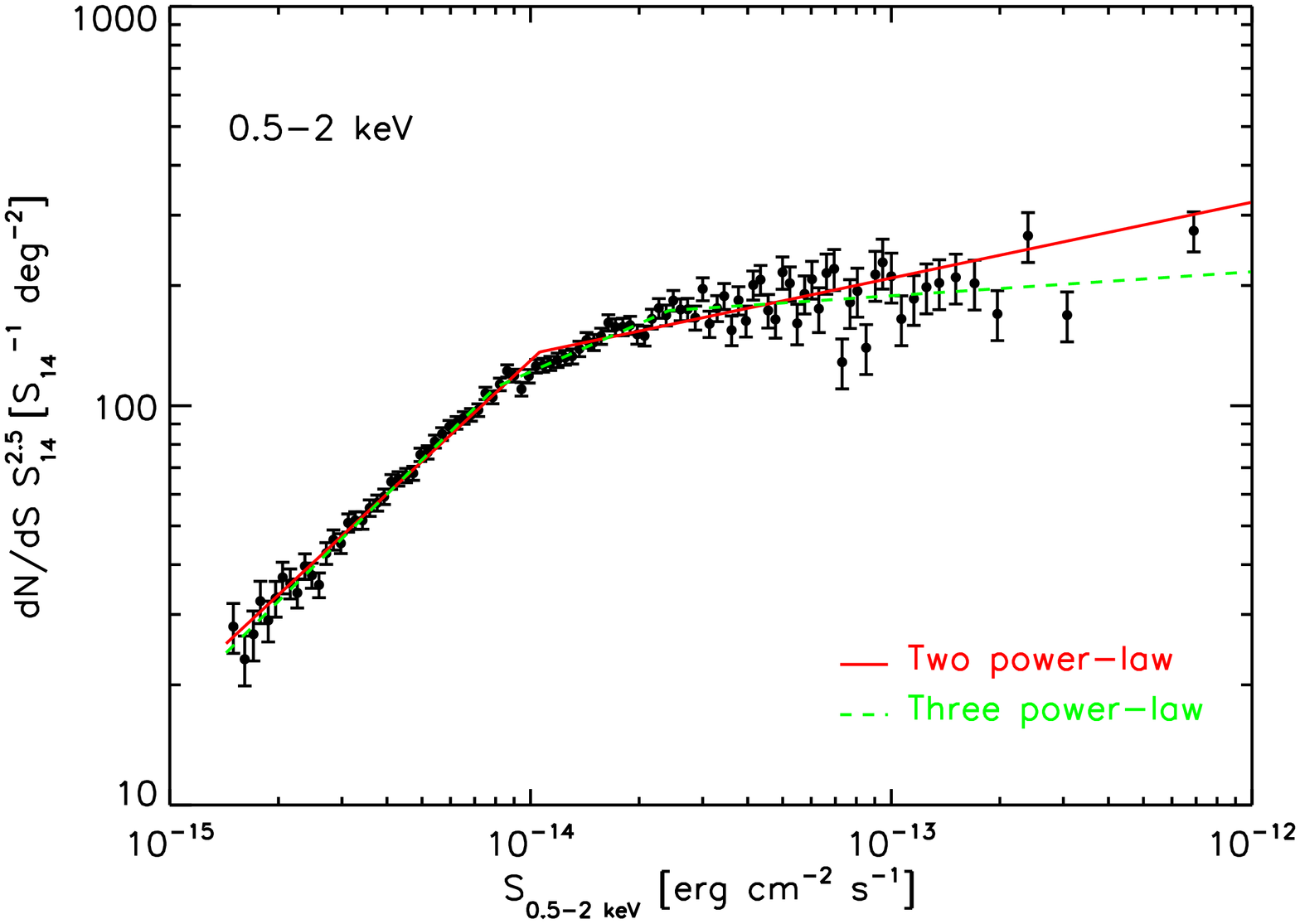}
   \includegraphics[angle=90,width=0.5\textwidth]{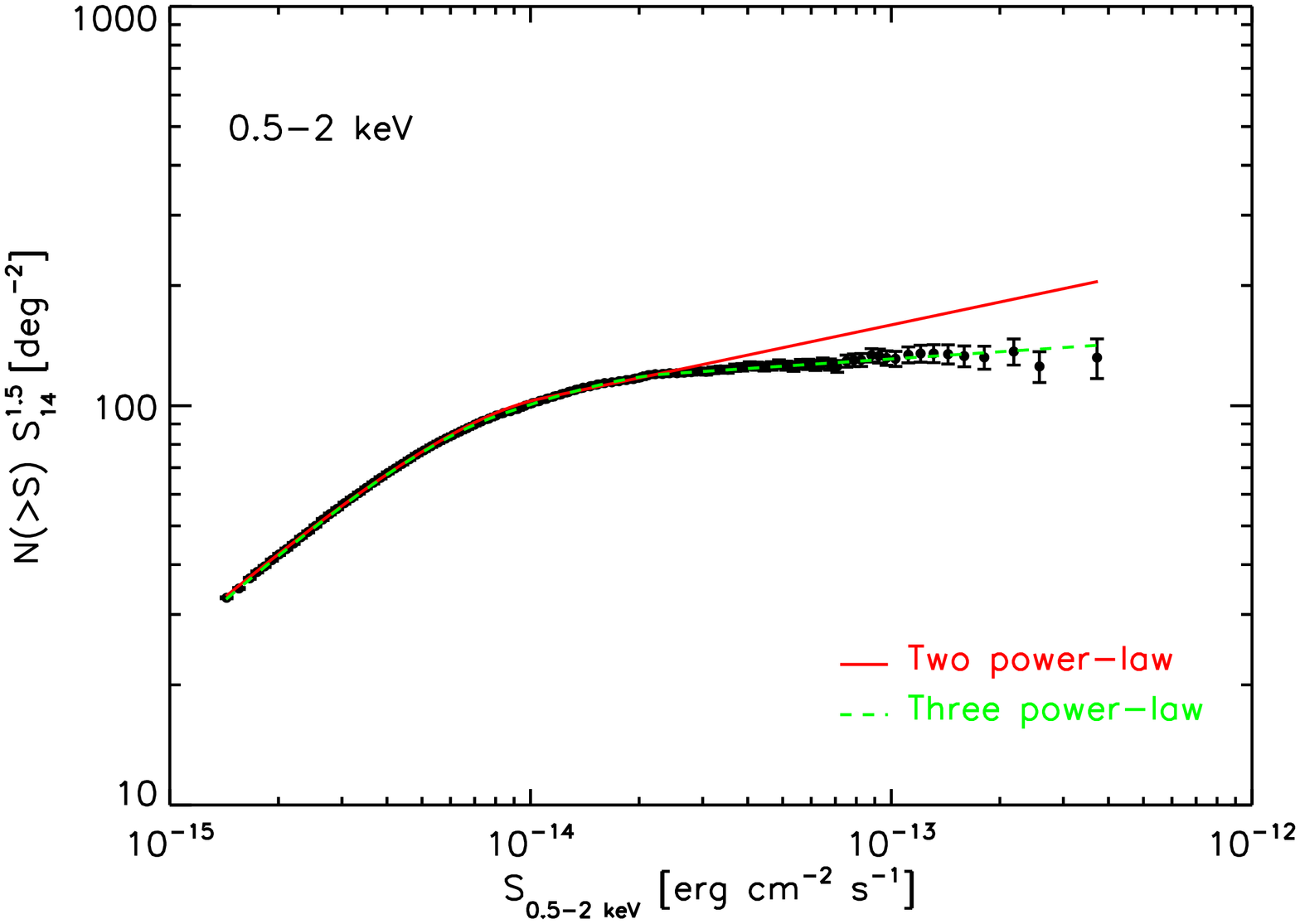}}
   \hbox{
   \includegraphics[angle=90,width=0.5\textwidth]{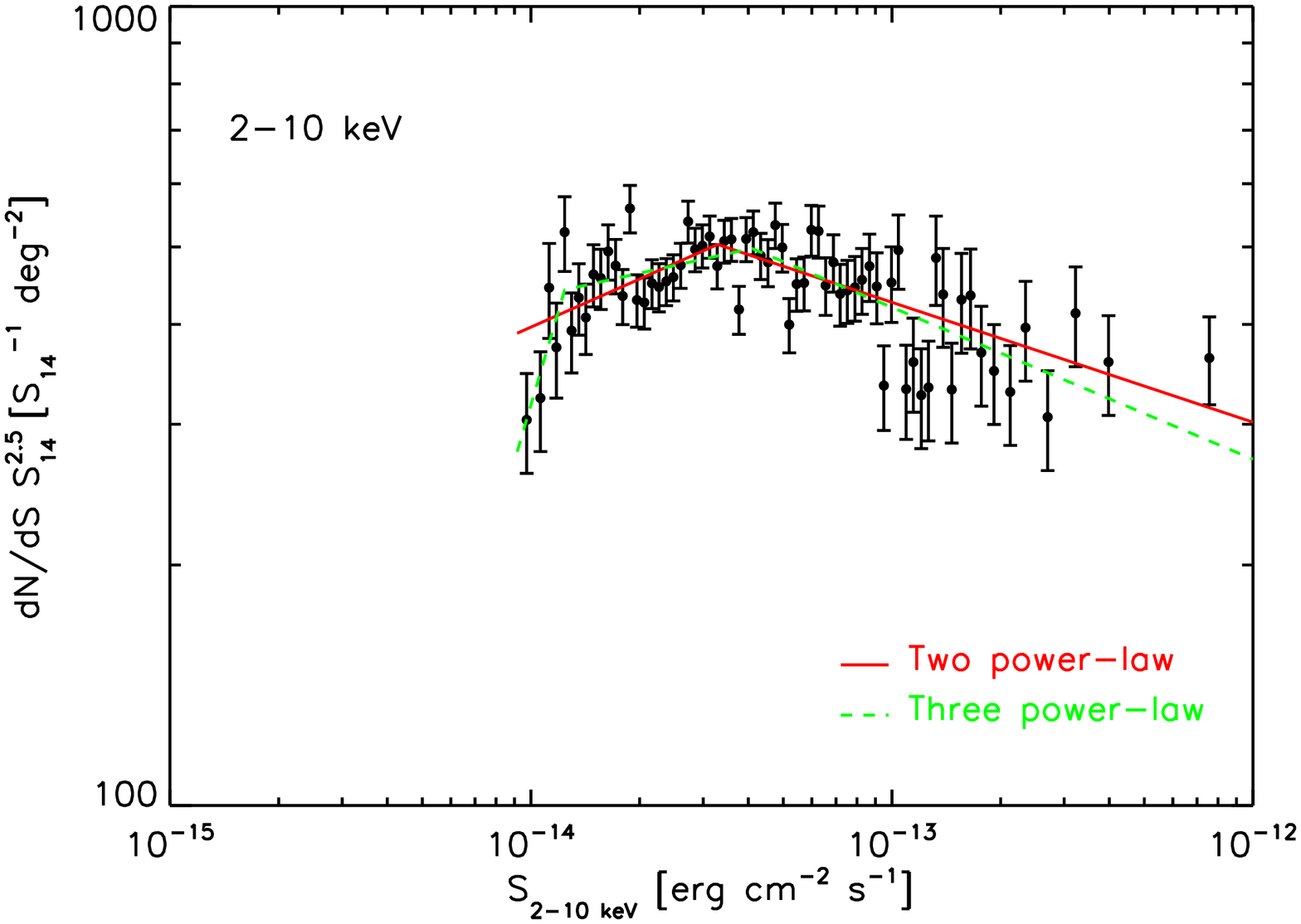}
   \includegraphics[angle=90,width=0.5\textwidth]{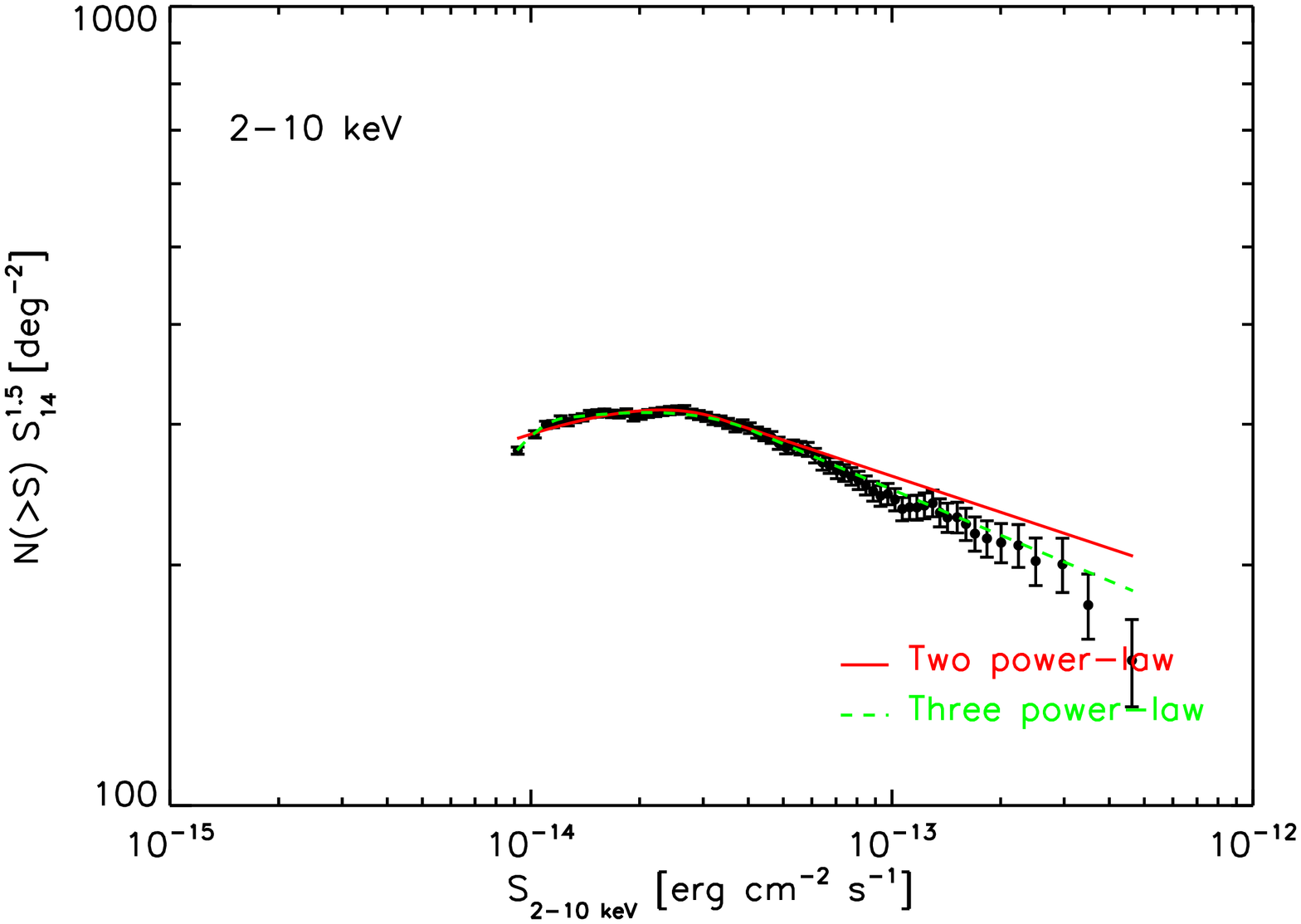}}
   \caption{Source count distributions in both a normalised differential (left) and normalised 
     integral form (right) for sources 
     detected in the 0.5-2 keV and 2-10 keV bands. 
     The lines show the results of the fitting of the data using a model with two (solid) and three (dashed)
       power-laws (see Sec.~\ref{ml_fit}).
     Error bars correspond to 1$\sigma$ confidence. } 
              \label{soft_hard_dists}%
    \end{figure*}

\section{The source counts}

\subsection{Calculation of source count distributions}
\label{number_counts_representation}
In X-ray astronomy, it is traditional to use the integral form to show the shape of source count distributions. However here we use both 
binned differential and integral representations. Differential counts have the advantage that the data 
points are independent, which makes it easier to see changes in the underlying shape of the distribution.
\begin{enumerate}
\item {\it Differential source counts}: The number of sources per unit flux and unit sky area, $n(S)$, is obtained as
\[
n(S_j)= {dN \over dS\,d\Omega}={\sum_{i=1}^{i=m}\,{1 \over \Omega_i} \over \Delta S_j}
\]
where $m$ is the number of sources in bin j with assigned flux $S_j$, $\Omega_i$ is the sky coverage (in deg$^2$) of source $i$ in the 
bin and $\Delta S_j$ is the bin width. The corresponding 1$\sigma$ error bars due to Poissonian statistics are calculated as $n(S_j)/\sqrt m$. $S_j$ refers to the weighted mean of the fluxes of the sources in the bin, 
$S_j=\sum_{i=1}^{i=m}\,w_i \times S_i$, where $S_i$ are the individual source fluxes and the weights, $w_i$, are determined as:

\[
w_i={{1 \over \Omega_i} \over \sum_{i=1}^{i=m}\,{1 \over \Omega_i}}
\]
$S_j$ is a better representation of the centroid of the bin than the mean of the fluxes of the sources in the bin,
especially at bright fluxes where the number of sources per bin is much smaller and 
the impact of the binning is most acute.
\\
\item {\it Integral source counts}: The number of sources per unit sky area with flux higher than $S$, $N(>S)$, is 
defined as 
\[
N(>S_j)=\sum_{i=1}^{i=M}\,{1 \over \Omega_i} 
\]
where the sum is over all sources with flux $S_i>S_j$ and $S_j$ is the flux of the faintest object in the bin. 
In this case error bars are assigned as $N(>S_j)/\sqrt(M)$, based on Poissonian statistics (but are correlated from 
bin to bin), where $M$ is the total number of sources with $S_i>S_j$. 
\end{enumerate}

   \begin{figure*}
   \centering
   \hbox{
   \includegraphics[angle=90,width=0.5\textwidth]{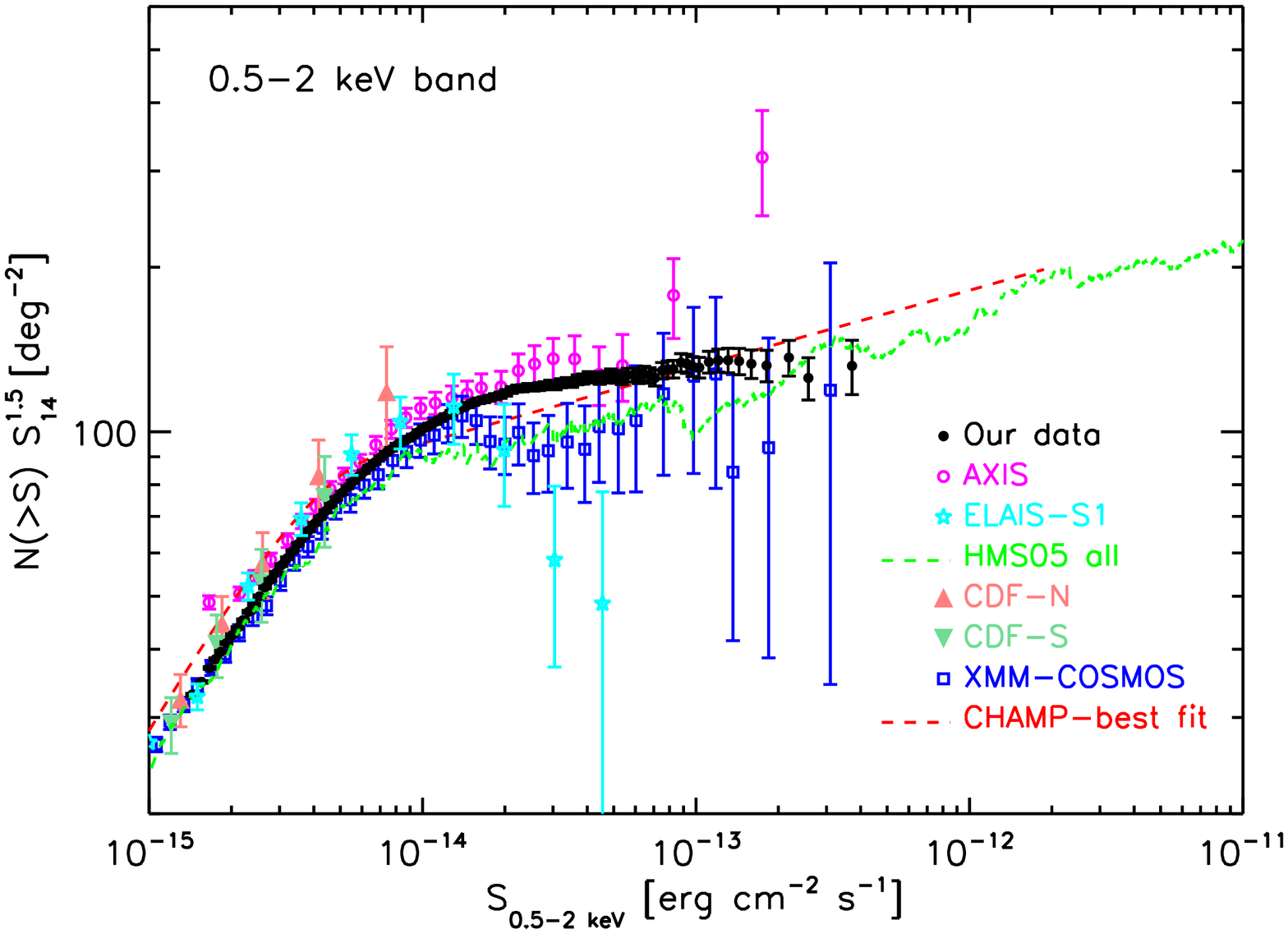}
   \includegraphics[angle=90,width=0.5\textwidth]{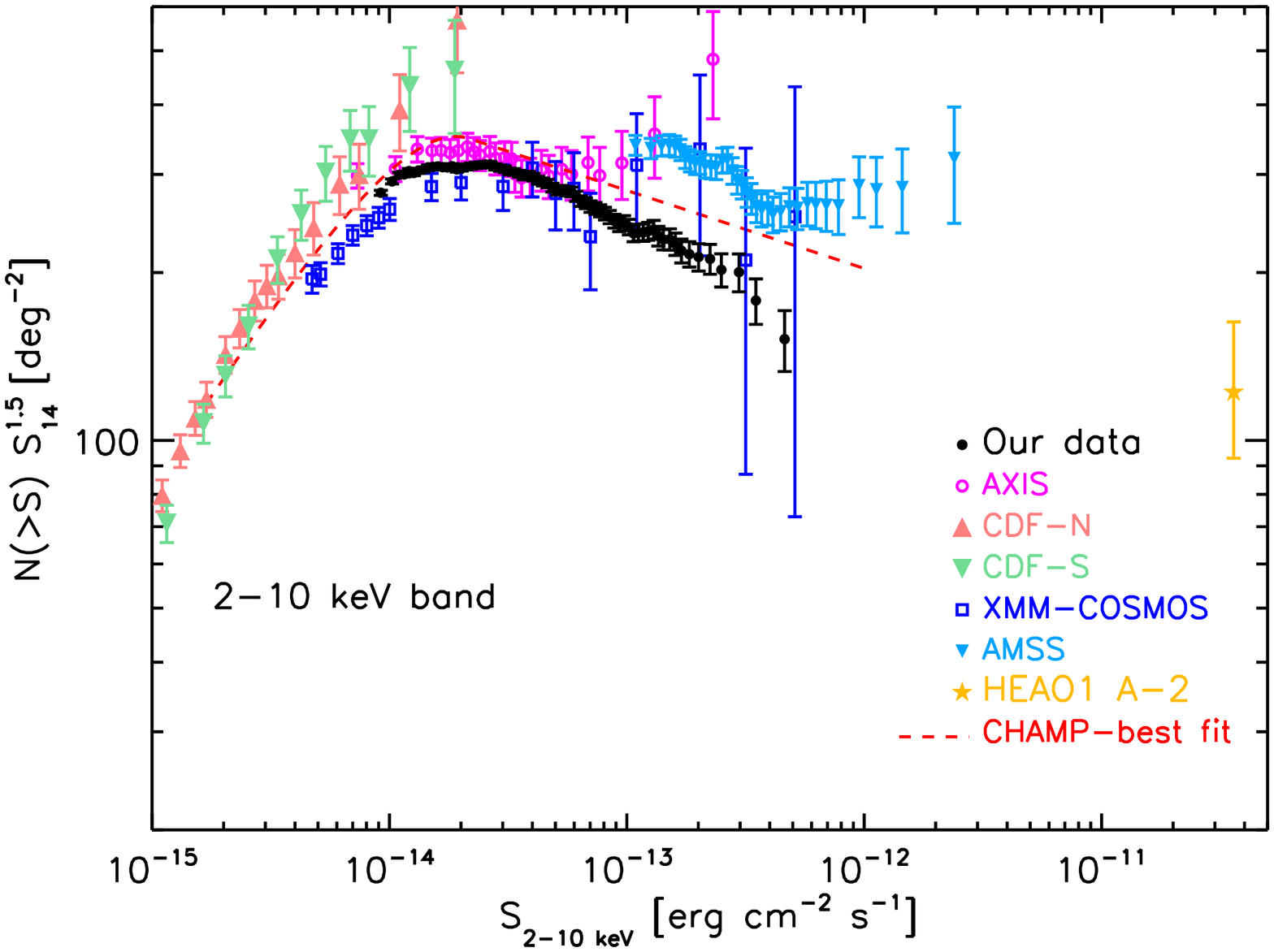}
   }
   \caption{Comparison of the 0.5-2 keV (left) and 2-10 keV (right) normalised integral source count distributions (filled circles) with a set of 
     representative results from previous surveys. Error bars correspond to 1$\sigma$ confidence.
   }
              \label{comp_soft_hard}%
    \end{figure*}

We apply a normalisation to both the differential and integral distributions, $n(S_j)\times (S_j/10^{-14})^{2.5}$ 
and $N($$>$$S_j)\times (S_j/10^{-14})^{1.5}$ respectively. The advantage of using a normalised representation is that 
it highlights deviations from the standard Euclidean form of the counts (with the Euclidean slope corresponding to a horizontal line in 
the normalised representation). 
We have defined the bin sizes of our distributions to have at least 50 sources per bin and a minimum 
bin size of 0.02 in log units of flux. Note that later the unbinned differential source counts are fitted 
with power-law models (see Sec.~\ref{ml_fit} for details).

When comparing source count distributions in different energy bands it is necessary to rescale the axes 
of at least one of the distributions (to that of the other energy band). If the flux rescaling factor is $\alpha$, the effect 
is to shift points along the x axis by a factor $\alpha$ and also up the y axis by a factor $\alpha^{1.5}$ (due to the normalisation). 
The energy band scaling factors used in this work are listed in Table~\ref{table:4}. The values are normalised to unit flux in 
the 0.5-10 keV band. 

\begin{table}
  \caption{Scaling factors used to convert fluxes to different energy bands.}
\label{table:4}      
\centering                          
\begin{tabular}{c c c c c c c c c c}        
\hline\hline                 
Energy band &  $\alpha=S_b/S_{\rm 0.5-10\,keV}$    \\
(1) & (2) \\
\hline                        
0.5-1  & 0.15\\
1-2    & 0.16\\
2-4.5  & 0.29\\
4.5-10 & 0.39 \\
0.5-2  & 0.32\\
2-10   & 0.68\\
2-8    & 0.56\\
2-12   & 0.79\\
4.5-7.5 & 0.24 \\
5.0-10  & 0.35 \\
0.5-10  & 1.0\vspace{0.1cm}\\
\hline                                   
\end{tabular}

(1) Energy band definition (in keV). 
(2) Flux scaling factor normalised to a unit flux in the 0.5-10 keV band, where $S_b$ is the flux in 
the band and $S_{\rm 0.5-10\,keV}$ is the flux in the 0.5-10 keV band. The values were obtained 
from an unabsorbed power-law spectrum of $\Gamma$=1.9 below 2 keV and $\Gamma$=1.6 above 2 keV.
\end{table}

Hereafter we will use the term `steeper' source counts to refer to  
those distributions having a greater numerical index for the slope, while `flatter' distributions will be those having 
smaller values of $|\Gamma|$. 

\subsection{The broad band source counts}
\label{dists_sh}
In Fig.~\ref{soft_hard_dists} we show the differential and integral source count distributions derived in the 'standard' 0.5-2 keV and 2-10 keV
bands. A comparison with those obtained using data taken directly from the {\tt 2XMM} catalogue is 
presented in Appendix~\ref{2xmm_vs_mydet}.

Our survey provides tight constraints on the X-ray source counts over more than 2 decades of X-ray flux 
in both energy bands. In Fig.~\ref{soft_hard_dists} the solid and dashed lines show the results of the fitting to the 
differential source counts with a power-law with two and three components (see Sec.~\ref{ml_fit} for details).
A break in the source count distributions is obvious in both bands, although in the 2-10 keV 
band the measurements do not go deep enough to properly define the shape of the distribution below the break. 
The cumulative angular density of sources in the broad energy bands at different fluxes is given in Table~\ref{table:7}. 

\begin{table}
  \caption{The cumulative angular density of sources in the broad bands.}
\label{table:7}      
\centering                          
\begin{tabular}{c c c c c c c c c c c c c c c c c c c c }        
\hline\hline                 
Flux &  ${\rm N(>S)}$ & ${\rm N}$&  ${\rm N(>S)}$ & ${\rm N}$\\
& 0.5-2 keV & & 2-10 keV \\
(1) & (2) & (3) & (4) & (5) \\
\hline
 -14.84   &  605.7$\pm$ 3.4$^*$&   31837  &        -           &      -  \\
 -14.70   &  474.0$\pm$ 2.7  &   31465    &        -           &      -  \\
 -14.43   &  287.4$\pm$ 1.7  &   27944    &        -           &      -  \\
 -14.40   &  268.4$\pm$ 1.6  &   27119    &        -           &      -  \\
 -14.10   &  132.9$\pm$ 1.0  &   16715    &        -           &      -  \\
 -14.04   &  111.2$\pm$ 0.9  &   14283    &  315.6$\pm$ 3.2$^*$&    9431 \\
 -13.85   &   65.9$\pm$ 0.7  &    8674    &  181.2$\pm$ 1.9    &    8911 \\
 -13.80   &   57.0$\pm$ 0.7  &    7521    &  155.6$\pm$ 1.7    &    8628 \\
 -13.50   &   21.7$\pm$ 0.4  &    2871    &  54.0 $\pm$ 0.7    &    5443 \\
 -13.20   &    8.0$\pm$ 0.2  &    1057    &  17.0 $\pm$ 0.4    &    2172 \\
 -12.90   &    3.0$\pm$ 0.2  &     395    &  5.3  $\pm$ 0.2    &     704 \\
 -12.60   &    1.0$\pm$ 0.1  &     138    &  1.6  $\pm$ 0.1    &     213 \\
 -12.30   &    0.4$\pm$ 0.1  &      53    &  0.4  $\pm$ 0.1    &      55 \\
\hline                        
\end{tabular}

(1) Energy band flux in log units.
(2) Cumulative angular density of sources in units of ${\rm deg^{-2}}$ above a given flux in the 0.5-2 keV energy band.
(3) Number of sources above given flux in the 0.5-2 keV energy band.
(2) Cumulative angular density of sources in units of ${\rm deg^{-2}}$ above a given flux in the 2-10 keV energy band.
(3) Number of sources above given flux in the 2-10 keV energy band.
$^*$ Cumulative angular density of sources at the flux limits of our survey in the 0.5-2 keV and 2-10 keV bands.
\end{table}

We have compared our results with previous findings from deep and shallow representative surveys (see Fig.~\ref{comp_soft_hard}).  
In the 0.5-2 keV band, the form of the source counts below $\sim$${\rm 10^{-14}\,erg\,cm^{-2}\,s^{-1}}$ 
has been determined previously from both medium-deep XMM-{\it Newton} (ELAIS-S1: Puccetti et al.~\cite{Puccetti06}, 
{\tt XMM-COSMOS}: Cappelluti et al.~\cite{Cappelluti07}, An XMM-{\it Newton} International Survey 
({\tt AXIS}): Carrera et al.~\cite{Carrera07}) and {\it Chandra} surveys (CDF-N and CDF-S: Bauer et al.~\cite{Bauer04}, 
{\tt Champ}: Kim et al.~\cite{Kim07}). Fig.~\ref{comp_soft_hard} (left) also shows the data from Hasinger et al.~(\cite{Hasinger05}, HMS05) which is a 
compilation of results from various {\tt ROSAT}, XMM-{\it Newton} and {\it Chandra} surveys.  
In general the present measurements are in good agreement with the published results.

At bright 0.5-2 keV fluxes, where the number of sources included in the surveys is much lower, there are still uncertainties 
in the shape of the source count distributions.
We note that in the flux range ${\rm 2\times10^{-14}-2\times10^{-13}\,erg\,cm^{-2}\,s^{-1}}$  
our distribution tends to lie above the results from the {\tt XMM-COSMOS} and {\tt Champ} surveys. 
One important difference in our analysis (and also in the {\tt AXIS} survey) is that our source count distributions 
include both point and (modestly) extended sources, while the distributions from the {\tt XMM-COSMOS} and {\tt Champ} surveys are 
for point sources only. Indeed if we exclude the sources detected in our 
analysis as extended, then a better agreement between our results and the {\tt XMM-COSMOS} survey is 
obtained, although the shape of the distribution is still somewhat steeper than the one from the {\tt Champ} survey. 

   \begin{figure*}
   \centering
   \hbox{
   \includegraphics[angle=90,width=0.45\textwidth]{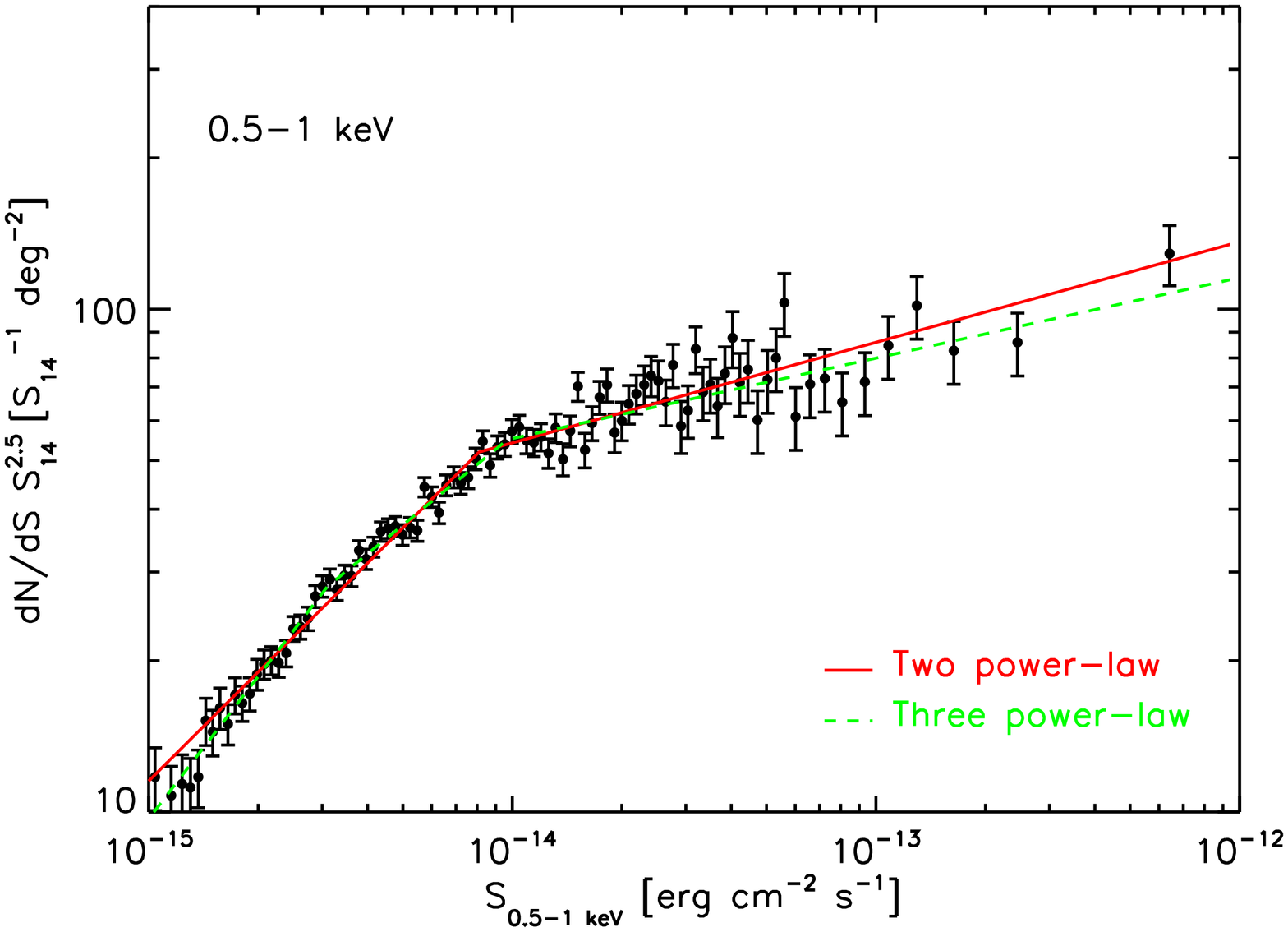}
   \includegraphics[angle=90,width=0.45\textwidth]{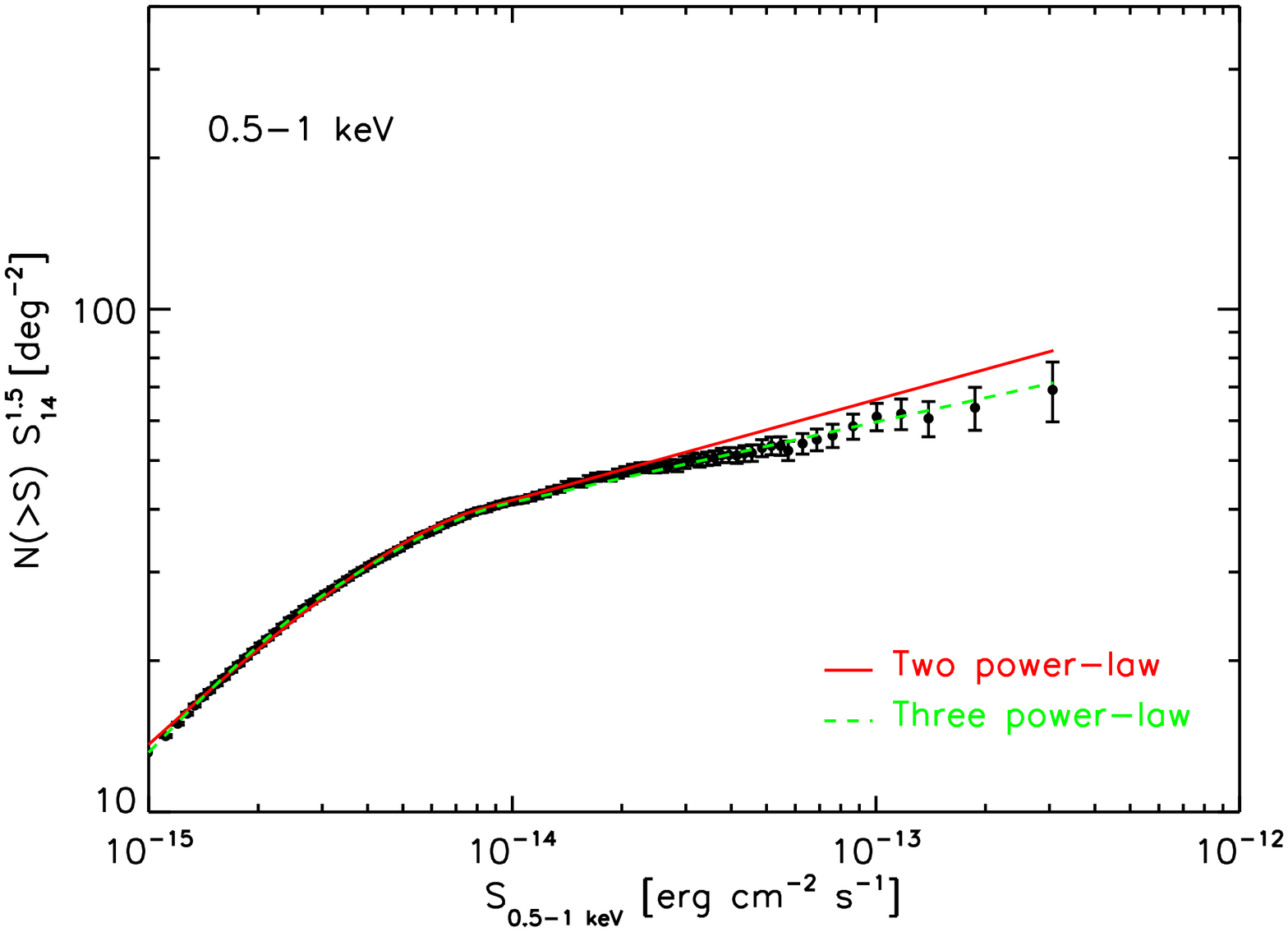}}
   \hbox{
   \includegraphics[angle=90,width=0.45\textwidth]{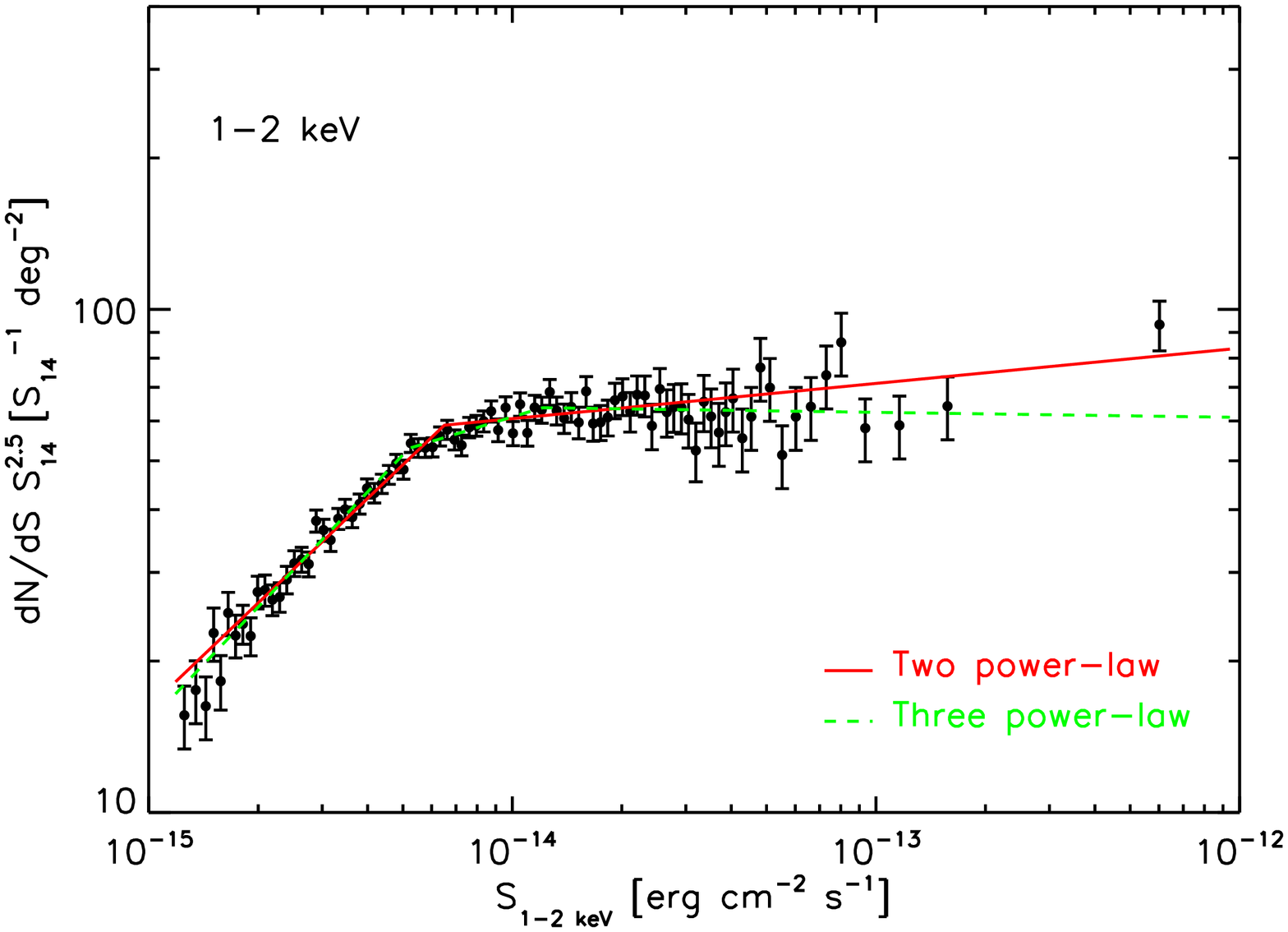}
   \includegraphics[angle=90,width=0.45\textwidth]{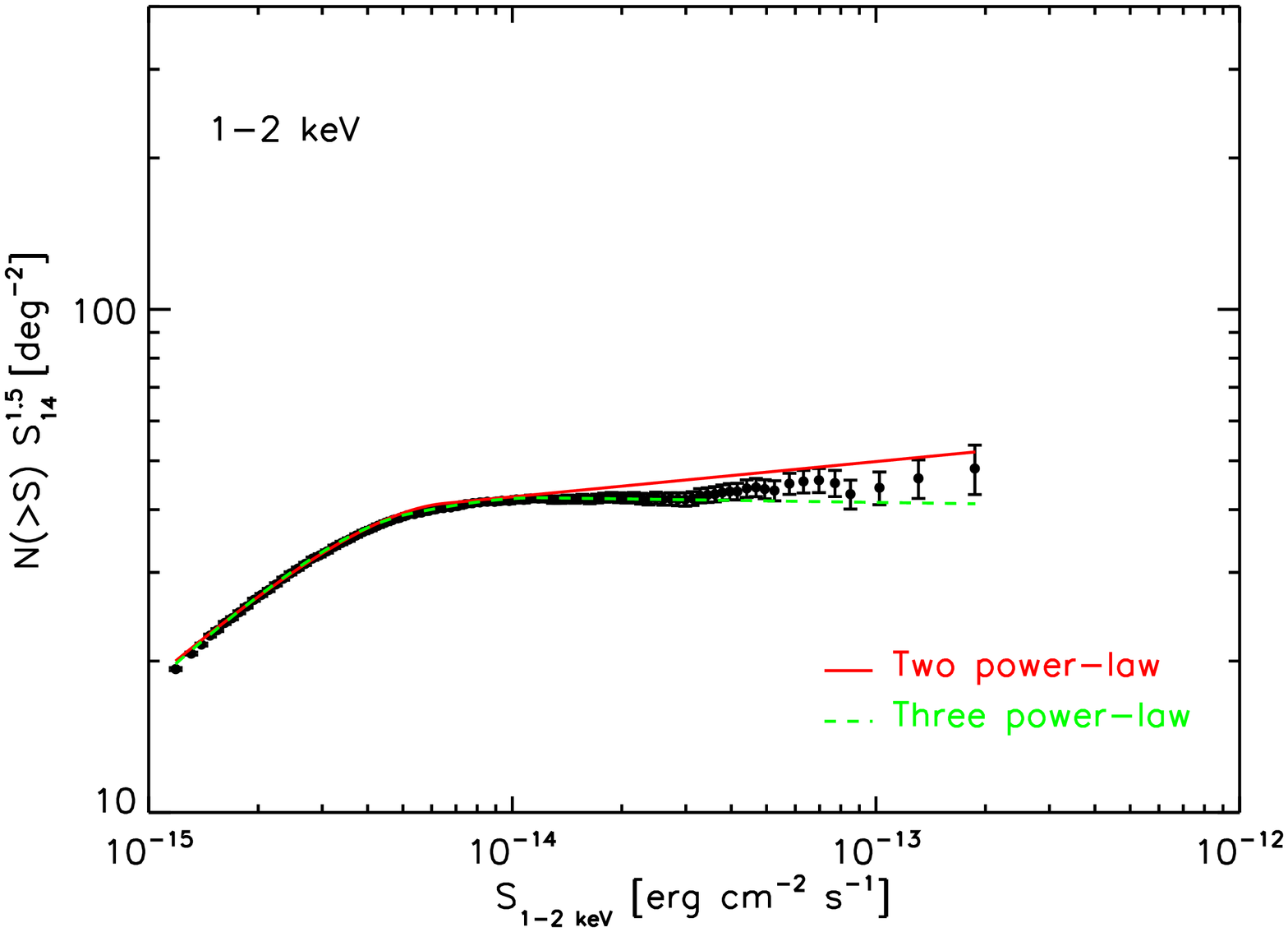}}
   \hbox{
   \includegraphics[angle=90,width=0.45\textwidth]{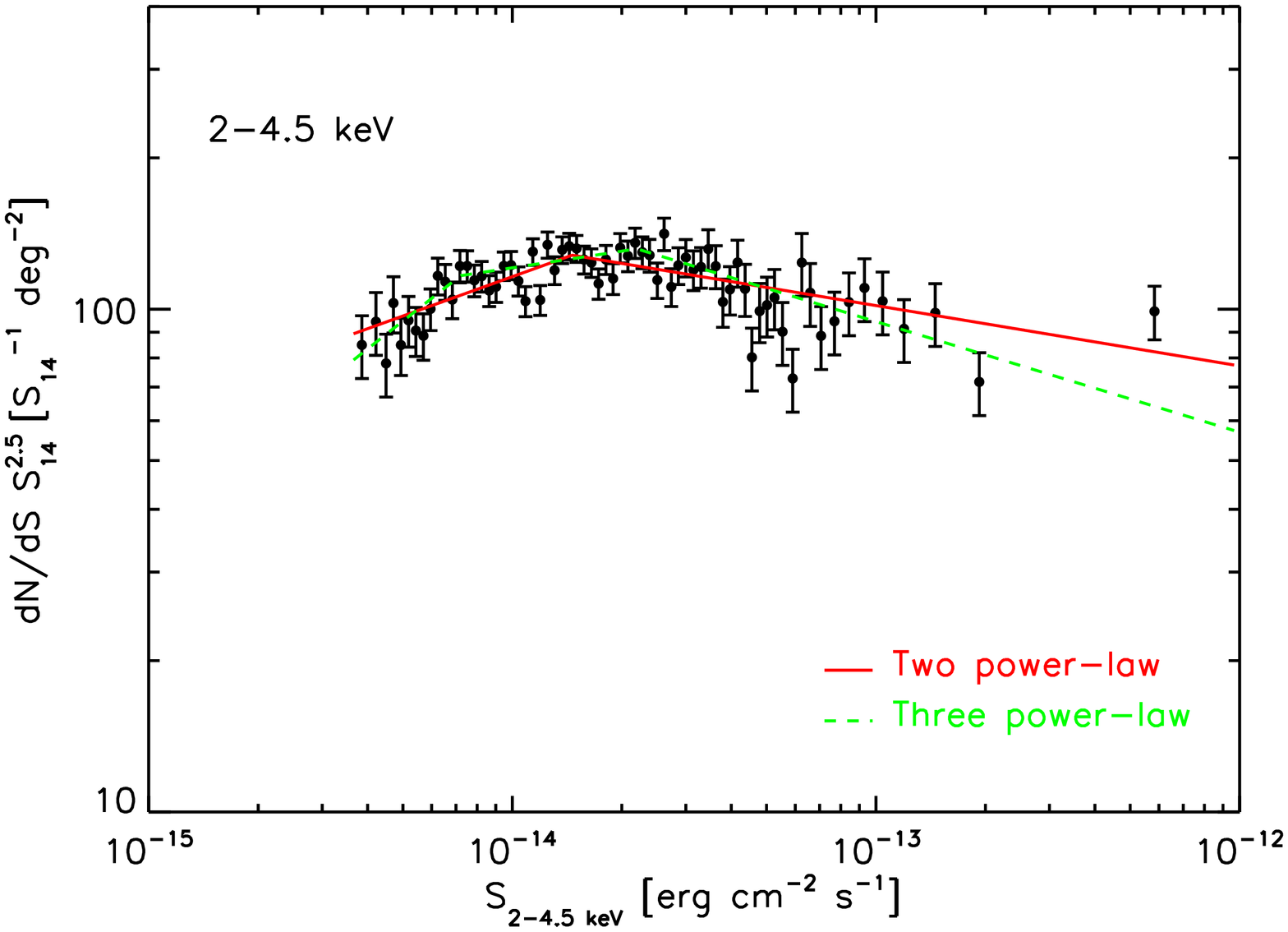}
   \includegraphics[angle=90,width=0.45\textwidth]{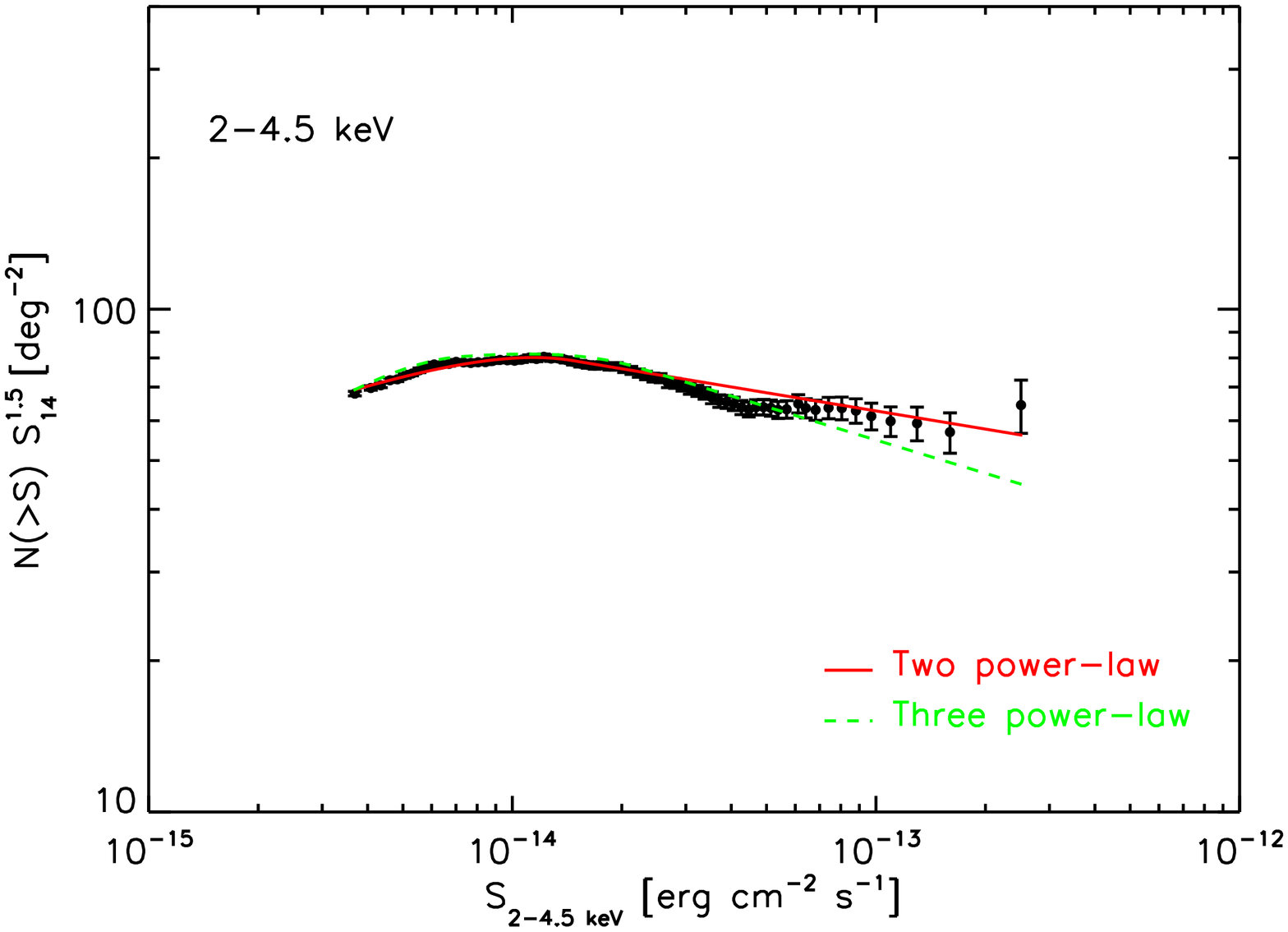}}
   \hbox{
   \includegraphics[angle=90,width=0.45\textwidth]{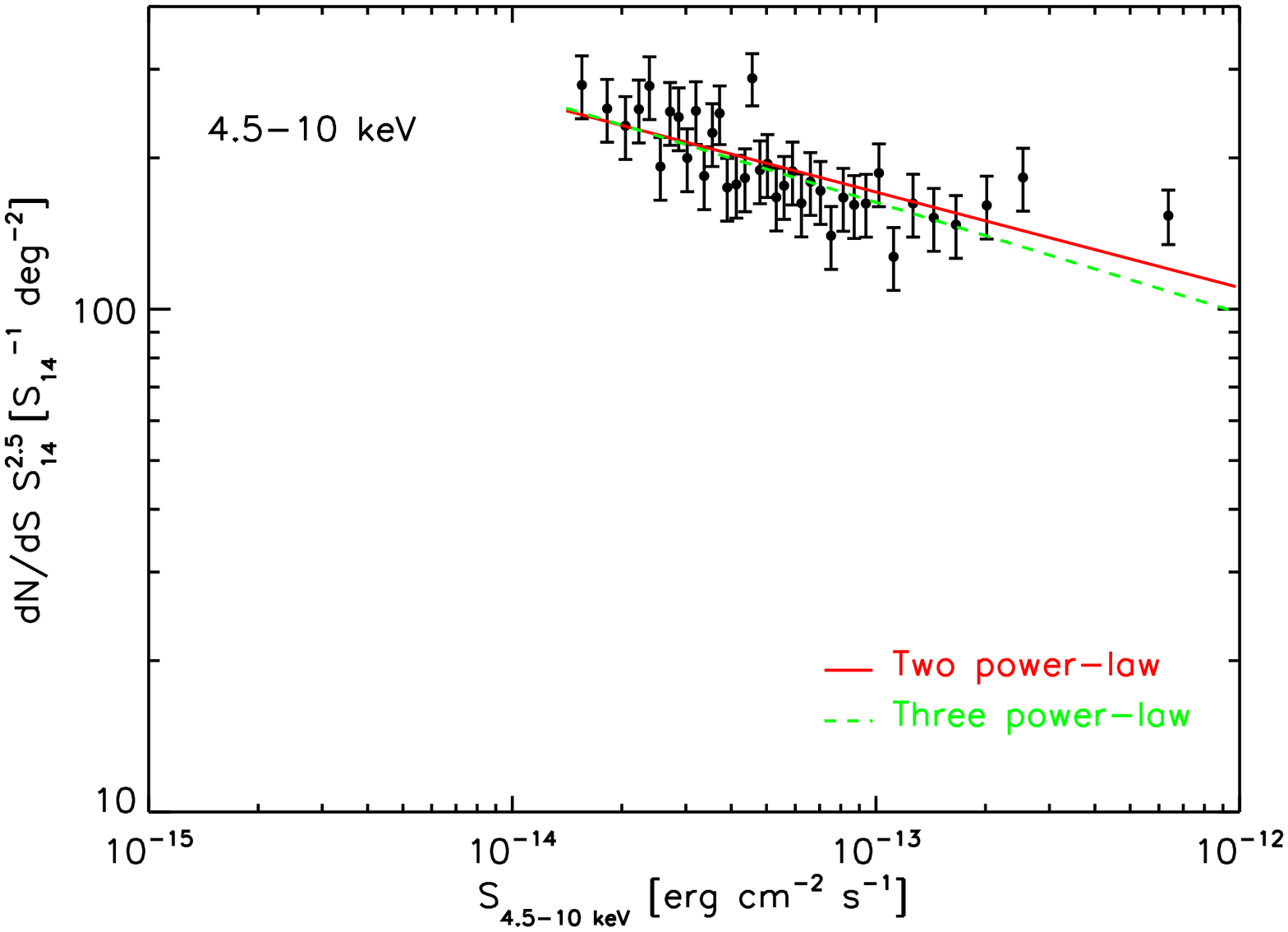}
   \includegraphics[angle=90,width=0.45\textwidth]{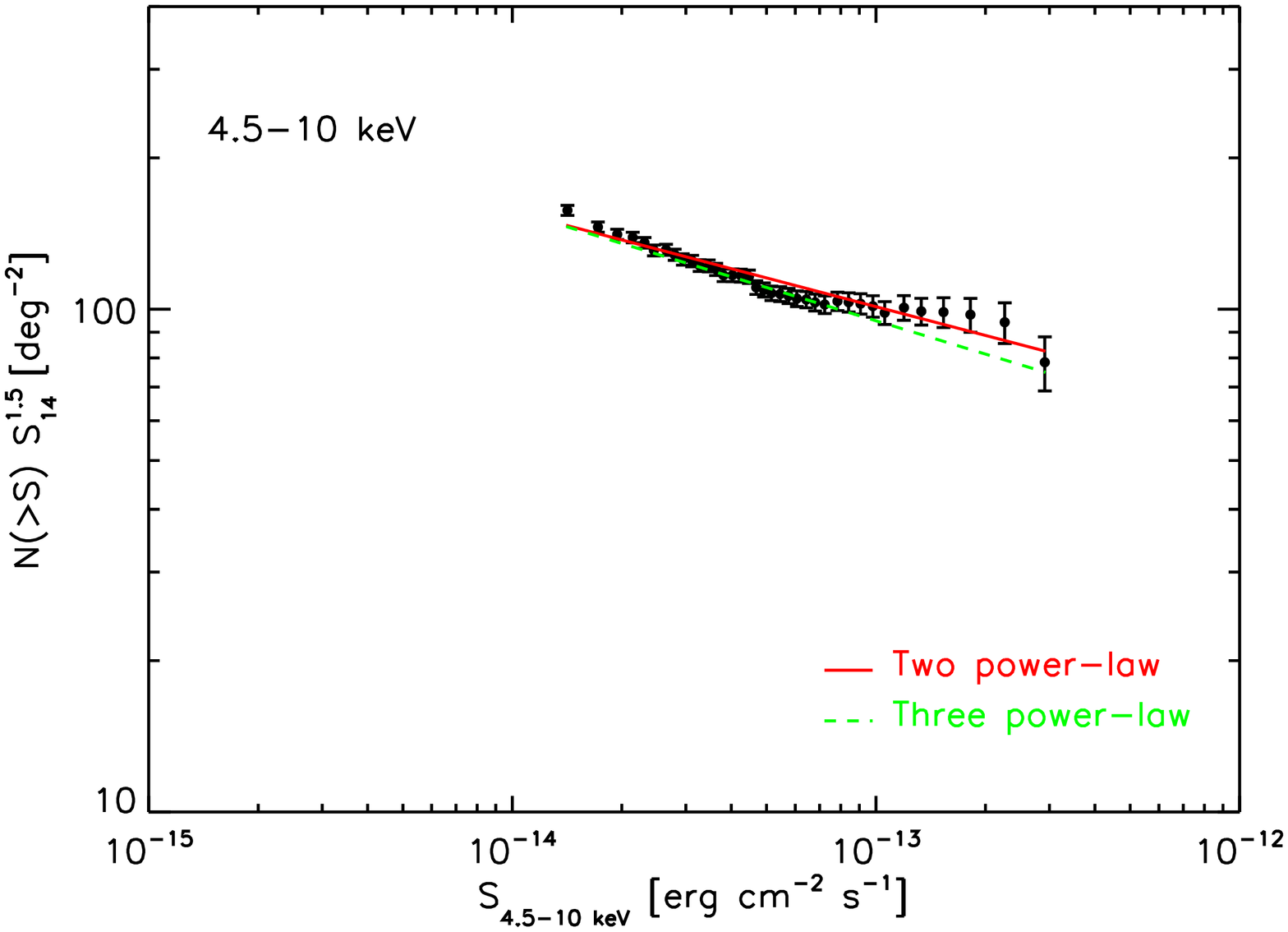}}
   \caption{Source count distributions in both normalised differential (left) and normalised 
     integral form (right) for sources detected in the 0.5-1 keV, 1-2 keV, 2-4.5 keV and 4.5-10 keV bands. 
     The lines show the results of the fitting of the data using a model with two (solid) and three (dashed)
       power-laws (see Sec.~\ref{ml_fit}).
     Error bars correspond to 1$\sigma$ confidence. } 
              \label{narrow_dists}
    \end{figure*}

\begin{table*}
  \caption{Cumulative angular density of sources in the narrow energy bands.}
\label{table:8}      
\centering                          
\begin{tabular}{c c c c c c c c c c c c c c c c c c c c }        
\hline\hline                 
Flux &  ${\rm N(>S)}$ & ${\rm N}$&  ${\rm N(>S)}$ & ${\rm N}$&  ${\rm N(>S)}$ & ${\rm N}$&  ${\rm N(>S)}$ & ${\rm N}$\\
& 0.5-1 keV & & 1-2 keV & & 2-4.5 keV && 4.5-10 keV \\
(1) & (2) & (3) & (4) & (5) & (6) & (7) & (8) & (9)  \\
\hline
 -15.00  &  417.6$\pm$ 2.9$^*$  &   20694  & -              &      -   &  -              &     -   & -              &      -\\
 -14.93 &   363.0$\pm$ 2.5  &   20604  &470.7$\pm$ 3.2$^*$  &   21185  &  -              &     -   & -              &      -\\
 -14.84 &   315.8$\pm$ 2.2  &   20394  &404.4$\pm$ 2.8  &   21059  &  -              &     -   & -              &      -\\
 -14.70  &  239.4$\pm$ 1.7  &   19429  &300.3$\pm$ 2.1  &   20338  &  -              &     -   & -              &      -\\
 -14.43 &   132.9$\pm$ 1.1  &   14856  &157.6$\pm$ 1.2  &   16004  & 302.4$\pm$ 3.1$^*$  &    9564 & -              &      -\\
 -14.40  &  122.3$\pm$ 1.0  &   14054  &144.5$\pm$ 1.2  &   15212  & 272.5$\pm$ 2.8  &    9523 & -              &      -\\
 -14.10  &   56.1$\pm$ 0.7  &    7291  & 58.2$\pm$ 0.7  &    7452  & 110.6$\pm$ 1.2  &    7980 & -              &      -\\
 -14.04  &   46.3$\pm$ 0.6  &    6071  &47.1$\pm$ 0.6  &    6125   &  89.6$\pm$ 1.0  &    7325 & -              &      -\\
 -13.85  &   26.5$\pm$ 0.4  &    3501  &24.9$\pm$ 0.4  &    3286   &  46.8$\pm$ 0.7  &    5022 & 92.2$\pm$ 2.1$^*$  &    1895\\ 
 -13.80  &   22.8$\pm$ 0.4  &    3013  & 21.1$\pm$ 0.4  &    2796  &  39.0$\pm$ 0.6  &   4407  & 73.8$\pm$ 1.7  &    1867\\
 -13.50  &    8.9$\pm$ 0.3  &    1183  &  7.5$\pm$ 0.2  &     989  &  12.3$\pm$ 0.3  &   1598  & 22.1$\pm$ 0.6  &    1432\\
 -13.20  &    3.4$\pm$ 0.2  &     453  &  2.9$\pm$ 0.1  &     380  &   4.0$\pm$ 0.2  &    534  &  6.6$\pm$ 0.2  &     731\\
 -12.90  &    1.4$\pm$ 0.1  &     187  &  1.0$\pm$ 0.1  &     134  &   1.3$\pm$ 0.1  &    175  &  2.3$\pm$ 0.1  &     293\\
 -12.60  &    0.6$\pm$ 0.1  &      73  &  0.4$\pm$ 0.1  &      55  &   0.5$\pm$ 0.1  &      66  & 0.7$\pm$ 0.1  &      89\\
 -12.30  &    0.1$\pm$ 0.1  &      18  &  0.1$\pm$ 0.1  &      16  &   0.1$\pm$ 0.1  &      15  & 0.2$\pm$ 0.1  &      27\\
\hline                        
\end{tabular}

(1) Energy band flux in log units.
(2) Cumulative angular density of sources in units of ${\rm deg^{-2}}$ above given flux in the 0.5-1 keV energy band.
(3) Number of sources above given flux in the 0.5-1 keV energy band.
(4) Cumulative angular density of sources in units of ${\rm deg^{-2}}$ above given flux in the 1-2 keV energy band.
(5) Number of sources above given flux in the 1-2 keV energy band.
(6) Cumulative angular density of sources in units of ${\rm deg^{-2}}$ above given flux in the 2-4.5 keV energy band.
(7) Number of sources above given flux in the 2-4.5 keV energy band.
(8) Cumulative angular density of sources in units of ${\rm deg^{-2}}$ above given flux in the 4.5-10 keV energy band.
(9) Number of sources above given flux in the 4.5-10 keV energy band.
$^*$ Cumulative angular density of sources at the flux limit of our survey in the various narrow energy bands.
\end{table*}
 
In the 2-10 keV band a detailed comparison is made more difficult by the fact that 
the effective area of the X-ray detectors typically varies substantially across the band, which may introduce 
systematic errors in the comparison of the results from different instruments. 
For example, because of the low effective area of the X-ray detectors on-board {\it Chandra} above 
$\sim$8 keV, published {\it Chandra} results are limited to the 2-8 keV band. In Appendix~\ref{2xmm_vs_mydet} we compare the source count 
distributions in the 2-10 keV and 2-8 keV energy bands for our sources.
There is in general a good agreement between these distributions although 
the slope of the source counts below the break is marginally flatter for the distribution in the 2-8 keV band.
In addition the results from the {\tt XMM-COSMOS} survey in the 2-10 keV band are based on source detection in 
the 2-4.5 keV energy band, and hence their analysis could be missing a population of sources with very hard 
X-ray spectra. Additional discrepancies between source counts may be explained by the different spectral 
assumption used in their construction. As we explained in Sec.~\ref{flx_cal}, this 
could affect the measured normalisation of the source count distributions by up to $\sim$20\%.
Finally if we adopt a conservative 10\% estimate on the absolute flux calibration of the EPIC pn camera, then 
a $\sim$15\% uncertainty in the normalisation of the source counts obtained with XMM-{\it Newton} might still 
be present (Stuhlinger et al.~\cite{Stuhlinger08}).

Despite these caveats the overall agreement between most surveys in the 2-10 keV energy band 
is better than 10\% at fluxes ${\rm \le 10^{-13}\,erg\,cm^{-2}\,s^{-1}}$.
At brighter fluxes, the largest discrepancy is found when comparing with 
the results from the ASCA Medium Sensitivity Survey (AMSS, Ueda et al.~\cite{Ueda05}) 
which has a normalisation $\sim$20-30$\%$ higher than ours. A higher normalisation 
on the {\tt ASCA} source counts compared with previous XMM-{\it Newton} surveys has been already 
reported (Cowie et al.~\cite{Cowie02}). 
Cross calibration effects need to be taken into account when comparing results from different missions and could explain the 
observed discrepancies. Snowden~(\cite{Snowden02}) investigated the cross calibration between different missions, including 
{\tt ASCA} and XMM-{\it Newton} and found an agreement between {\tt ASCA} and XMM-{\it Newton} fluxes at the $\sim$10\% level. Since this analysis, changes in EPIC-pn 2-10 keV fluxes associated with improvements in the calibration have been $\sim$1-2\%, and 
therefore the $\sim$10\% discrepancy between {\tt ASCA} and XMM-{\it Newton} EPIC-pn still holds. We conclude 
that the observed discrepancy with respect to {\tt ASCA} source counts cannot be explained as a cross calibration 
effect alone. 

We also note that the recently published source count distributions for sources detected by XMM-{\it Newton} in the Lockman Hole 
field in the 0.5-2 keV and 2-10 keV energy bands are also consistent with our results within 
1 to 2-$\sigma$ at fluxes ${\rm \le 10^{-14}\,erg\,cm^{-2}\,s^{-1}}$ (Brunner et al.~\cite{Brunner08}).
In order to compare our source count distributions with previous results at fluxes brighter than 
those sampled by our analysis we compare with the ROSAT All-Sky Survey data in the 0.5-2 keV band 
(HMS05, Fig.~\ref{comp_soft_hard} left) and the HEAO1 A-2 all-sky survey in the 2-10 keV band for AGN-only sources 
(Piccinotti et al.~\cite{Piccinotti82},  Fig.~\ref{comp_soft_hard} right).
The extrapolation to brighter fluxes of our source counts, shows that our distributions lie marginally below 
these results. This discrepancy can be explained as being due to the fact that surveys using pointed observations 
(such as this survey) may be biased against bright sources because the targets of the observations have to be excluded 
from the analysis.

\subsection{The narrow band source counts}
\label{narrow_band_dists}
If we compare the 0.5-2 keV and 2-10 keV source counts we see that there is a strong dependence of 
the shape of the distributions on the energy band. 
A similar trend is found when we compare the source count distributions 
in our narrow energy bands, 0.5-1 keV, 1-2 keV, 2-4.5 keV and 4.5-10 keV 
(see Fig.~\ref{narrow_dists}): the source count distributions become steeper both at faint and bright fluxes 
as we move to higher energies. The cumulative angular density of sources in the narrow energy bands at different fluxes is given in Table~\ref{table:8}. 

   \begin{figure}
   \centering
   \hspace{-0.3cm}\includegraphics[angle=90,width=0.5\textwidth]{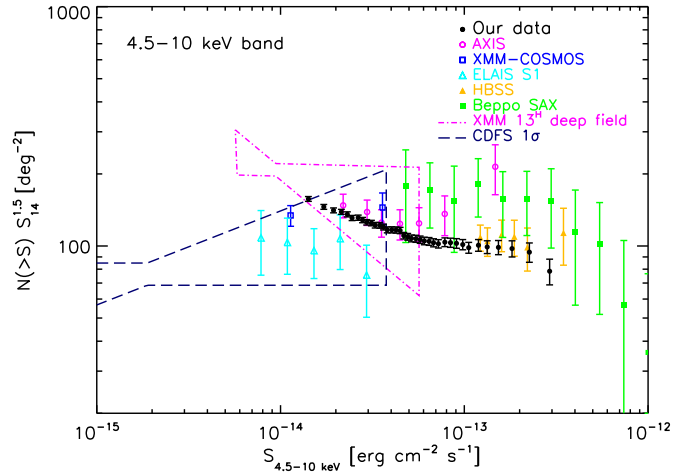}
   \caption{Comparison of the measured normalised integral 4.5-10 keV source count distribution (filled circles) with a sample of representative results from 
   previous surveys. Error bars correspond to 1$\sigma$ confidence.}
     
              \label{comp_uhard}%
    \end{figure}

In Fig.~\ref{comp_uhard} we show the source count distribution in the highest energy range sampled in our 
analysis, namely 4.5-10 keV. 
We also include some representative results from previous surveys for comparison: in the 5-10 keV 
band the results from ELAIS-S1 (Puccetti et al.~\cite{Puccetti06}), CDF-S 1$\sigma$ (Rosati et al.~\cite{Rosati02}), 
{\tt XMM-COSMOS} (Cappelluti et al.~\cite{Cappelluti07}) and 
the XMM-{\it Newton} ${\rm 13^H}$ field (Loaring et al.~\cite{Loaring05}); in the 4.5-7.5 keV energy band 
the XMM-{\it Newton} Hard Bright Serendipitous Survey (HBSS, Della Ceca et al.~\cite{Ceca04}) and 
{\tt AXIS} (Carrera et al.~\cite{Carrera07}); in the 4.5-10 keV band 
the Beppo SAX data from Fiore et al.~(\cite{Fiore01}).
In order to convert the fluxes from these surveys to the 4.5-10 keV energy band we assumed that the spectra of the 
sources can be well represented by a power-law of slope $\Gamma$=1.6. 
The corresponding factors used to convert fluxes to the 4.5-10 keV energy band are listed in Table~\ref{table:4}.

At energies $\gtrsim$4.5 keV there is still a lack of strong observational constraints in the shape of the 
source count distribution, as large discrepancies in the results for both the shape and normalisation of 
the distribution are evident. In some cases this amounts to $>$30\%. Because the effective area of the X-ray detectors 
at these energies is relatively small, the number of sources involved in these analyses is correspondingly limited.
In addition, a $\sim$10-20\% uncertainty in the normalisation 
can arise due to the uncertainty in the absolute flux calibration of the instruments and the 
spectral shape chosen in the count rate to flux conversion (see Sec.~\ref{flx_cal}).
The latter, however, cannot fully explain the observed discrepancies in the results as most of the surveys compute
their fluxes using the same spectral index. We note that the measurements of Beppo SAX (Fiore et al.~\cite{Fiore01}) are 
systematically higher than the results based on XMM-{\it Newton} data. The disagreement with the Beppo SAX results 
was already noted by Della Ceca et al.~(\cite{Ceca04}), who suggested that an offset in the Beppo SAX absolute 
flux calibration of $\sim$30\% could explain the discrepancy in the results.

Thanks to our study we can now constrain the shape and normalisation of the source counts above 4.5 keV over a reasonably broad range of 
flux, from ${\rm \sim10^{-14}\,erg\,cm^{-2}\,s^{-1}}$ to  ${\rm \sim 3\times 10^{-13}\,erg\,cm^{-2}\,s^{-1}}$.   
Note that in the 4.5-10 keV band no break in the source count distribution is detected 
down to the limiting flux of our survey, $\sim$${\rm 1.4\times 10^{-14}\,erg\,cm^{-2}\,s^{-1}}$. 
This is consistent with the results from deeper X-ray surveys which suggest that the break in the 
source counts at energies above $\sim$4.5 keV must occur at fluxes $\lesssim$5-8${\rm \times10^{-15}\,erg\,cm^{-2}\,s^{-1}}$ 
(see e.g. Loaring et al.~\cite{Loaring05}, Brunner et al.~\cite{Brunner08}, Georgakakis et al.~\cite{Georgakakis08}).

   \begin{figure}[!t]
   \centering
   \hbox{
   \hspace{-0.3cm}\includegraphics[angle=90,width=0.5\textwidth]{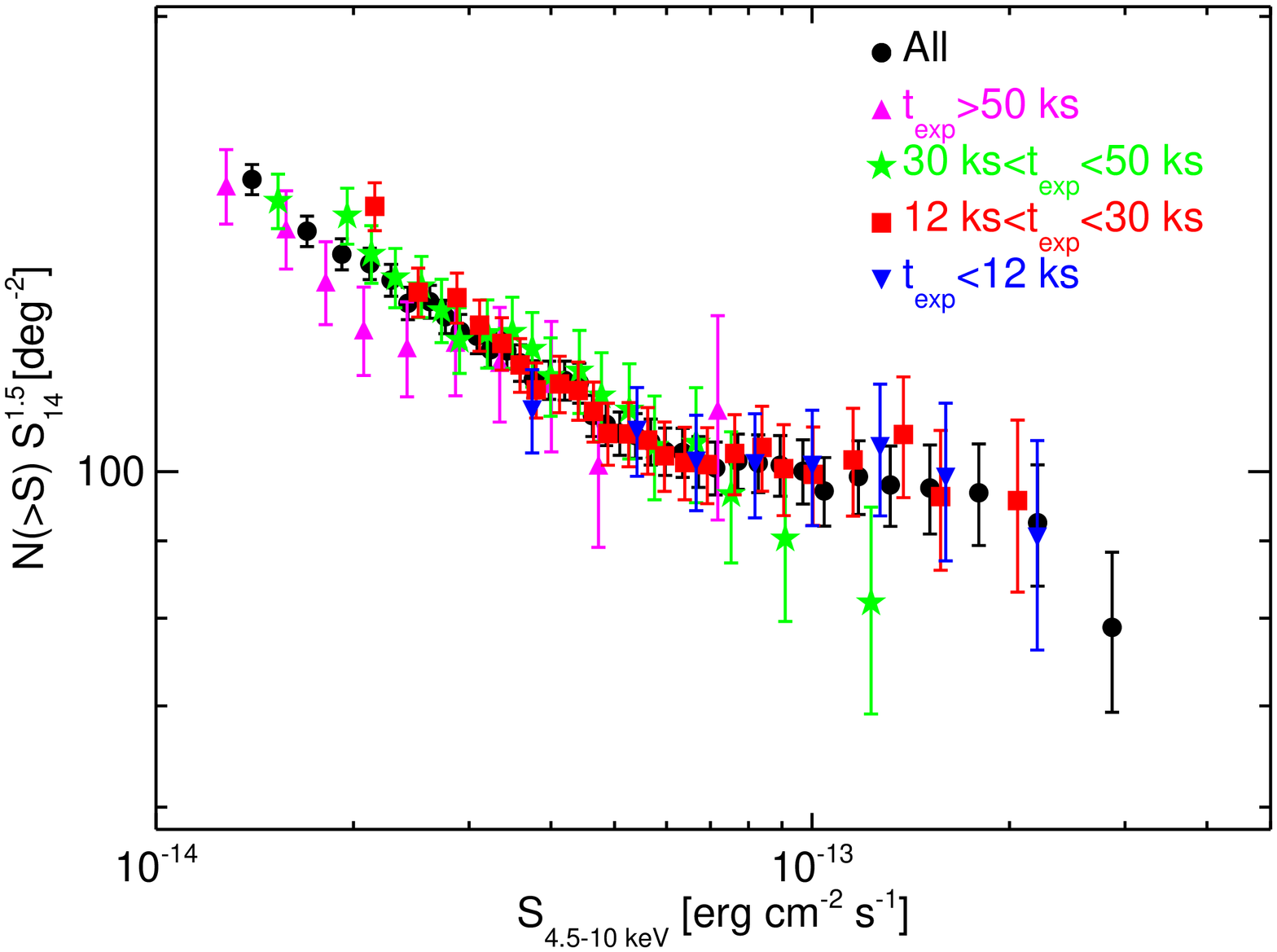}}
   \hbox{
   \hspace{-0.3cm}\includegraphics[angle=90,width=0.5\textwidth]{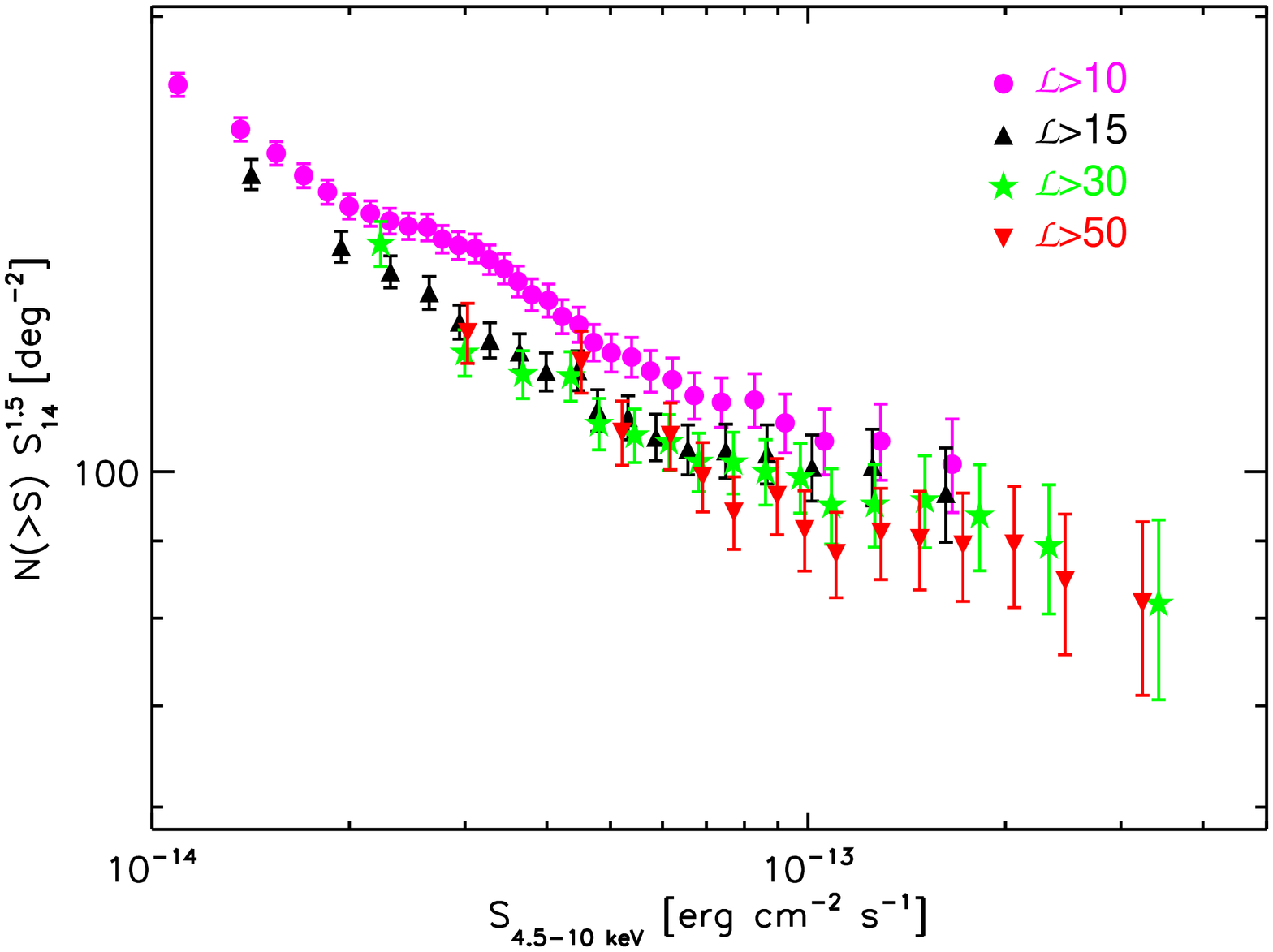}}
   \caption{Top: Dependence of the 4.5-10 keV normalised source count distribution in integral form 
     on the exposure time of the observations. Bottom: Comparison of our 4.5-10 keV normalised source count distribution 
     in integral form ($\mathcal{L}$$\ge$15) with the distributions obtained if lower/higher detection likelihood thresholds are 
     used instead.
     Error bars correspond to 1$\sigma$ confidence.}
              \label{dists_vs_texp}%
    \end{figure}

\begin{table}
  \caption{Results of the maximum likelihood fits to our source count distributions with a broken power-law model.}
\label{table:3}      
\centering                          
\begin{tabular}{c c c c c c c c c c}        
\hline\hline                 
Energy band &  ${\rm \Gamma_b}$     &  ${\rm \Gamma_f}$ & $S_{\rm b}$ & $K$ \\
(keV)   &  &  & (${\rm 10^{-14}\,cgs}$) &  (${\rm deg^{-2}}$) & \\
(1) & (2) & (3) & (4) & (5) \\
\hline                        
0.5-1        & $2.30_{-0.01}^{+0.02}$ & $1.78_{-0.02}^{+0.01}$ & $0.81_{-0.01}^{+0.01}$ & $71.3_{-11.6}^{+1.5}$ \vspace{0.05cm}\\ 
1-2          & $2.43_{-0.01}^{+0.01}$ & $1.81_{-0.01}^{+0.01}$ & $0.65_{-0.01}^{+0.01}$ & $112.3_{-1.0}^{+2.2}$ \vspace{0.05cm}\\
2-4.5        & $2.62_{-0.02}^{+0.02}$ & $2.24_{-0.03}^{+0.04}$ & $1.46_{-0.07}^{+0.15}$ & $72.6_{-12.1}^{+4.7}$ \vspace{0.05cm}\\
4.5-10$^{a}$ & $2.69_{-0.03}^{+0.03}$ & ----                 & ----                 & $264.8_{-14.7}^{+15.8}$ \vspace{0.25cm} \\

0.5-2       & $2.31_{-0.01}^{+0.01}$ & $1.66_{-0.02}^{+0.01}$ & $1.06_{-0.01}^{+0.13}$ & $124.9_{-18.6}^{+1.8}$ \vspace{0.05cm}\\
0.5-2$^{b}$ & $2.29_{-0.55}^{+0.01}$ & $1.63_{-0.33}^{+0.01}$ & $1.00_{-0.01}^{+0.01}$ & $131.9_{-45.5}^{+1.1}$ \vspace{0.05cm} \\
2-10        & $2.65_{-0.02}^{+0.02}$ & $2.30_{-0.03}^{+0.05}$ & $3.29_{-0.08}^{+0.15}$ & $84.4_{-6.1}^{+3.3}$ \vspace{0.05cm}\\
2-10$^{a}$  & $2.54_{-0.01}^{+0.01}$ & ----                   & ----                   & $482.6_{-9.2}^{+10.8}$ \vspace{0.05cm}\\
2-10$^{b}$  & $2.55_{-0.02}^{+0.03}$ & $1.19_{-0.01}^{+0.01}$ & $0.78_{-0.10}^{+0.01}$ & $709.3_{-18.0}^{+110.6}$ \vspace{0.1cm}\\
\hline                                  
\end{tabular}

(1) Energy band definition (in keV). (2) Power-law slope above the flux break.
(3) Power-law slope below the flux break.
(4) and (5) Flux break (in units of ${\rm 10^{-14}\,erg\,cm^{-2}\,s^{-1}}$) and normalisation in each band.
Errors are 1$\sigma$ uncertainty.
$^a$ Best fit parameters from using a single power-law.
$^b$ Best fit parameters from fitting our data together with the data from the CDF-N and CDF-S.
\end{table}

\begin{table*}
  \caption{Results of the $\chi^2$ fits to our source count distributions with a model with three power-laws.}
\label{table:5}      
\centering                          
\begin{tabular}{c c c c c c c c c c}        
\hline\hline                 
Energy band &  ${\rm \Gamma_b}$     &  ${\rm \Gamma_i}$  &  ${\rm \Gamma_f}$ & $S_{\rm b}^{\rm b}$ & $S_{\rm b}^{\rm f}$ &$K$ & F-test \\
(keV)   &  &  & & (${\rm 10^{-14}\,cgs}$) &  (${\rm 10^{-14}\,cgs}$) & (${\rm deg^2}$)  & prob. (\%)\\
(1) & (2) & (3) & (4) & (5) & (6) & (7) & (8)\\
\hline                        
0.5-1 & $2.34_{-0.02}^{+0.02}$ & $1.91_{-0.03}^{+0.02}$ & $1.56_{-0.06}^{+0.06}$ & $0.97_{-0.07}^{+0.06}$ & $0.31_{-0.02}^{+0.04}$  &$57.6_{-3.7}^{+4.9}$ & $\>$99.99 \vspace{0.05cm}\\
1-2 & $2.51_{-0.02}^{+0.02}$ & $2.28_{-0.04}^{+0.04}$ & $1.74_{-0.03}^{+0.03}$ & $1.21_{-0.20}^{+0.15}$ & $0.52_{-0.01}^{+0.04}$ & $47.9_{-7.6}^{+12.4}$ & $\>$99.99  \vspace{0.05cm}\\
2-4.5 & $2.72_{-0.03}^{+0.03}$ & $2.39_{-0.03}^{+0.04}$ & $1.93_{-0.22}^{+0.13}$ & $2.2_{-0.16}^{+0.19}$ & $0.72_{-0.08}^{+0.06}$  &$40.4_{-5.9}^{+5.9}$ & $\>$97.3  \vspace{0.05cm}\\
4.5-10& $2.72_{-0.03}^{+0.03}$ & -  & - & - & - & $270.7_{-16.2}^{+17.2}$ \\
0.5-2 & $2.44_{-0.02}^{+0.03}$ & $2.10_{-0.04}^{+0.03}$ & $1.61_{-0.02}^{+0.03}$ & $2.42_{-0.22}^{+0.19}$ & $0.80_{-0.06}^{+0.04}$  &$46.0_{-4.0}^{+6.1}$ & $\>$99.99 \vspace{0.05cm} \\
2-10 & $2.69_{-0.03}^{+0.02}$ & $2.40_{-0.04}^{+0.01}$ & $0.96_{-0.51}^{+0.40}$ & $4.09_{-0.69}^{+0.22}$ & $1.24_{-0.13}^{+0.04}$  &$60.2_{-3.4}^{+6.9}$ & $\>$94.3 \vspace{0.1cm}\\
\hline                                   
\end{tabular}

(1) Energy band definition (in keV). (2), (3) and (4) Power-law slopes at the 
brightest, intermediate and fainter fluxes respectively. 
(5) and (6) Flux breaks (in units of ${\rm 10^{-14}\,erg\,cm^{-2}\,s^{-1}}$) at bright 
and faint fluxes. (7) Normalisation of the model in each band. (8) F-test significance of 
improvement of the quality of the fit with respect to a broken power-law model.
Errors are 1$\sigma$ uncertainty.
\end{table*}

\subsection{Confusion, bias and other systematic effects in source counts}
\label{biases}
We have not corrected our source count distributions for biases associated with the source detection procedure such as 
Eddington bias, source confusion or spurious detections. 
It is therefore important to quantify the potential effect of these biases on our results. 

First we note that there is excellent agreement between our results and previous surveys 
which have gone substantially deeper and hence are less susceptible to bias effects at flux thresholds relevant 
to the current survey (e.g. Bauer et al.~\cite{Bauer04}, Cappelluti et al.~\cite{Cappelluti07}). 
This suggests that our source count distributions are not strongly affected by source detection biases. 

Source confusion occurs when two or more sources fall in a single resolution element of the detector, and depends on 
the sky density of sources and the size of the telescope beam. As shown in Loaring et al.~(\cite{Loaring05}), the 
XMM-{\it Newton} confusion limit is reached at a source density of $\sim$2000 deg$^{-2}$, corresponding to fluxes ${\rm <10^{-16}\,erg\,cm^{-2}\,s^{-1}}$ 
in all energy bands. This is well below the flux limits reached by our survey (see Table~\ref{table:1}). Therefore we expect the effect of source confusion on our source count 
distributions to be negligible. Furthermore, due to the relatively high threshold in detection likelihood used in our 
analysis ($\mathcal{L}$$\ge$15), we expect the fraction of spurious detections in our samples 
to be low. From Fig. 6 in Loaring et al.~(\cite{Loaring05}), a detection likelihood $\mathcal{L}$=10
corresponds to a $\sim$2.6\% fraction of spurious detections in the 5-10 keV band, i.e. the 
expected number of spurious sources per XMM-{\it Newton} pointing in this energy band is $\sim$1.1.
Because we have increased the detection likelihood threshold by 5, the number of spurious sources in our survey in the 4.5-10 keV 
band is effectively reduced by $e^{-5}$=$6.7\times10^{-3}$. Therefore for our selected threshold in significance of 
detection, $\mathcal{L}$$\ge$15, the expected number of spurious detections per field in our 4.5-10 keV band is 
$\sim7\times10^{-3}$ and the total number of spurious detections in our 1129 fields is $\sim$8.3. As we have 
1895 objects in the 4.5-10 keV band the fraction of spurious detections in this band is $\le$0.44\%. 
Note however that the computation of the fraction of spurious detections for $\mathcal{L}$=15 from an extrapolation of the results 
for $\mathcal{L}$=10 is probably too conservative, and the real fraction is probably marginally higher ($\sim$1-2\%, see Watson et al.~\cite{Watson08}). 
The fraction of spurious detections is largest in the 4.5-10 keV band so we can conclude that the fraction of 
spurious detections is $\lesssim$2\% in all our energy bands.

The Eddington bias (i.e. a systematic offset in the number of detected sources at a given flux) 
depends on the uncertainty in the measured fluxes and the shape of the source count distributions (Eddington~\cite{Eddington13}). The steeper the 
source counts the stronger the Eddington bias. 
As shown in Loaring et al.~(\cite{Loaring05}) if a less-conservative detection limit is used 
($\mathcal{L}$$\ge$6-8), the Eddington bias can increase the measured source counts by up to $\sim$23\% at the 
faintest fluxes. In the present study we expect Eddington bias effects to be most important in the 4.5-10 keV band, since the background is 
higher (and therefore for a source at a given flux the corresponding 
statistical error is higher) and due to the fact that the source count distribution is steeper for this energy band 
(see Sec.~\ref{narrow_band_dists}). On this basis we have focused our attention on the impact of the Eddington bias in the measured 
source count distributions in the 4.5-10 keV band. 

Because we use a set of observations having a broad range of exposure times our source counts are expected to be affected by 
Eddington bias over a broad range of flux. We 
have investigated how the shape of our source count distribution changes if we use observations with 
different exposure times. 
We divided our 1129 observations into four 
groups having different ranges of exposures and calculated source count distributions for each 
subsample. The degree of Eddington bias is not directly related to 
the exposure time as within each field the flux limit increases for larger offaxis angles due to the vignetting. However, 
we have defined the range of exposure times within each group broadly enough to account for the variation of the flux 
limit within each observation. No obvious changes in the shape of the distributions are seen 
suggesting that Eddington bias effects must be rather small (see Fig.~\ref{dists_vs_texp} top).

We also have compared source count distributions obtained when one increases the threshold in 
the significance of detection of sources. The results of this test are shown in Fig.~\ref{dists_vs_texp} (bottom). 
For comparison we have also included the distribution obtained using a 
threshold in the significance of detection $\mathcal{L}$=10. At $\mathcal{L}$=10 
we see deviations of up to 10 percent with respect to the other curves, suggesting 
that Eddington bias would have an influence if we had adopted this likelihood threshold for our study. In contrast, the 
results are entirely consistent for likelihood thresholds of 15 and larger, implying that  Eddington bias is smaller 
than our statistical errors at the detection threshold that we have adopted for this work. 

\begin{table*}[!ht]
  \caption{Intensity of the X-ray background contributed by our sources in the various energy bands.}
\label{table:6} 
\centering      
\begin{tabular}{c c c c c c c c c c}
\hline\hline                 
Energy band &  $S_{\rm min}$  &  $S_{\rm max}$  &  $I_{\rm CXRB}(S_{\rm min}\le S\le S_{\rm max})$ & $I_{\rm CXRB}$ & $f_{\rm CXRB}$\\
(keV)   &  (${\rm cgs}$) &  (${\rm cgs}$) & (${\rm 10^{-12} cgs\,deg^{-2}}$) & (${\rm 10^{-12} cgs\,deg^{-2}}$) \\
(1) & (2) & (3) & (4) & (5) & (6)\\
\hline                        
0.5-1  & $9.9\times10^{-16}$ & $10^{-12}$ & 2.4 & 2.96   & 0.81   \vspace{0.05cm}\\
       & $10^{-12}$ & & 0.32 &    & 0.11   \vspace{0.05cm}\\

1-2    & $1.2\times10^{-15}$ & $10^{-12}$ & 2.5 & 4.50*  & 0.55   \vspace{0.05cm}\\
       & $10^{-12}$ & & 0.12 &    & 0.03   \vspace{0.05cm}\\

2-4.5  & $3.7\times10^{-15}$ & $10^{-12}$ & 3.3 & 7.80   & 0.42   \vspace{0.05cm}\\
       & $10^{-12}$ & & 0.08 &    & 0.01   \vspace{0.05cm}\\

4.5-10 & $14 \times10^{-15}$ & $10^{-12}$ & 2.8 & 12.4   & 0.22   \vspace{0.05cm}\\
       & $10^{-12}$ & & 0.14 &    & 0.01   \vspace{0.25cm}\\

0.5-2  & $1.4\times10^{-15}$ & $10^{-12}$  & 4.9 & 7.50* & 0.65   \vspace{0.05cm}\\
       & $10^{-12}$ & & 0.48 &    & 0.06   \vspace{0.05cm}\\

2-10   & $9.0\times10^{-15}$ & $10^{-12}$  & 8.0 & 20.2* & 0.39   \vspace{0.05cm}\\
       & $10^{-12}$ & & 0.39 &    &  0.02  \vspace{0.05cm}\\
\hline                                   
\end{tabular}

(1) Energy band definition (in keV). (2) and (3) Minimum and maximum flux used in the integration.
(4) Intensity of the X-ray background contributed by our sources. Note that the quoted values include 
the contribution from both clusters of galaxies and stars.
(5) Total X-ray background intensity. Errors are 1$\sigma$ uncertainty. The values indicated with an asterisk 
are from Moretti et al.~(\cite{Moretti03}). These values were used to estimate the CXRB intensity in the various energy bands
assuming a power-law model of $\Gamma$=1.4 (see Sec.~\ref{ixrb} for details).
(6) Fraction of X-ray background resolved by our sources.
\end{table*}

\subsection{Maximum likelihood fitting to the source counts}
\label{ml_fit}
\subsubsection{Two power-law fitting model}
\label{ml_fit_new_model}
Number counts below $\sim$10 keV can be well fitted by 
broken power-law shapes with the break in the distributions occurring at fluxes ${\rm \sim10^{-15}-10^{-14}\,erg\,cm^{-2}\,s^{-1}}$ 
(see e.g. Cowie et al.~\cite{Cowie02}, Moretti et al.~\cite{Moretti03}, Ueda et al.~\cite{Ueda03}, Bauer et al.~\cite{Bauer04}, 
Carrera et al.~\cite{Carrera07}).
We have fitted the unbinned differential source count distributions using
the parametric, unbinned maximum likelihood method described in Carrera et al.~(\cite{Carrera07}). In Eq. 2 of that
paper, a Poisson term was added to the 'usual' maximum likelihood expression to take into
account the difference between the observed number of sources in each sample and the
expected number given the model parameters being fitted.
This analysis is equivalent to the one used in Marshall et al.~(\cite{Marshall83}) to fit the
luminosity function of quasars. However, the analysis presented in Carrera et al.~(\cite{Carrera07})
also takes into account both the errors in the fluxes of the sources and the
changing sky area with flux.

The model adopted to represent the shape of the distributions is a broken power-law,
\[
{dN \over dS d\Omega}=\left\{ \begin{array}{ll}
{K \over S_{\rm b}}\,({S \over S_{\rm b}})^{-\Gamma_{\rm f}} & S\le S_{\rm b}  \vspace{0.1cm} \\

{K \over S_{\rm b}}\,({S \over S_{\rm b}})^{-\Gamma_{\rm b}} & S> S_{\rm b}
\end{array}
\right.
\]
This model has four independent parameters: the break flux $S_{\rm b}$, the normalisation $K$ and the slopes of the 
differential counts at bright ($\Gamma_{\rm b}$) and faint ($\Gamma_{\rm f}$) fluxes. 

The results of the fits are summarised in Table~\ref{table:3} 
while the best fit broken power-law models are represented as solid lines in Fig.~\ref{soft_hard_dists} and Fig.~\ref{narrow_dists}.
Table~\ref{table:3} also lists the best fit parameters obtained when fitting the current 
measurements simultaneously with the CDF-N and CDF-S counts in both the 0.5-2 keV and 2-10 keV energy bands.

In the 0.5-2 keV band we see that the broken power-law model provides only a modest 
fit to the data above its break suggesting that the curvature of the source counts in the 0.5-2 keV band cannot be well 
represented by a simple broken power-law model.
Although in the 2-10 keV band the model seems to provide a better representation of the shape of the source counts, 
this is only achieved by allowing a break at a 
flux ${\rm >3\times10^{-14}\,erg\,cm^{-2}\,s^{-1}}$, 
well above the value of ${\rm \sim10^{-14}\,erg\,cm^{-2}\,s^{-1}}$ typically found in deeper 
surveys (see e.g. Cappelluti et al.~\cite{Cappelluti07}). We interpret this 
as an indication that the curvature of the source counts also in the 2-10 keV band cannot be well reproduced by the broken power-law model.
For the narrow energy bands the broken power-law model provides a 
somewhat better although far from perfect fit to all the data sets (see Fig.~\ref{narrow_dists}).

   \begin{figure*}
   \centering
   \hbox{
   \hspace{-2cm}\includegraphics[angle=-90,width=0.50\textwidth]{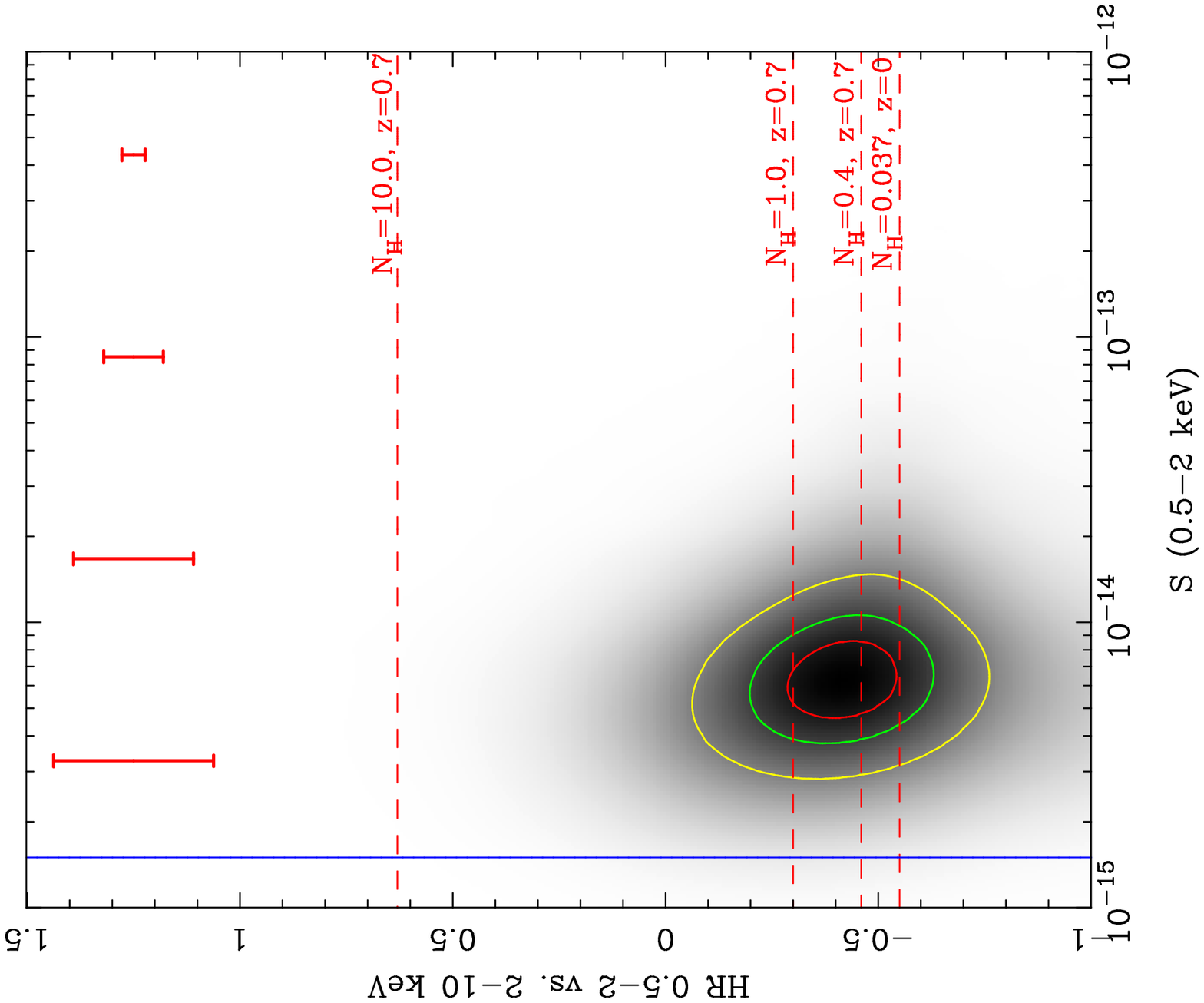}
   \includegraphics[angle=-90,width=0.50\textwidth]{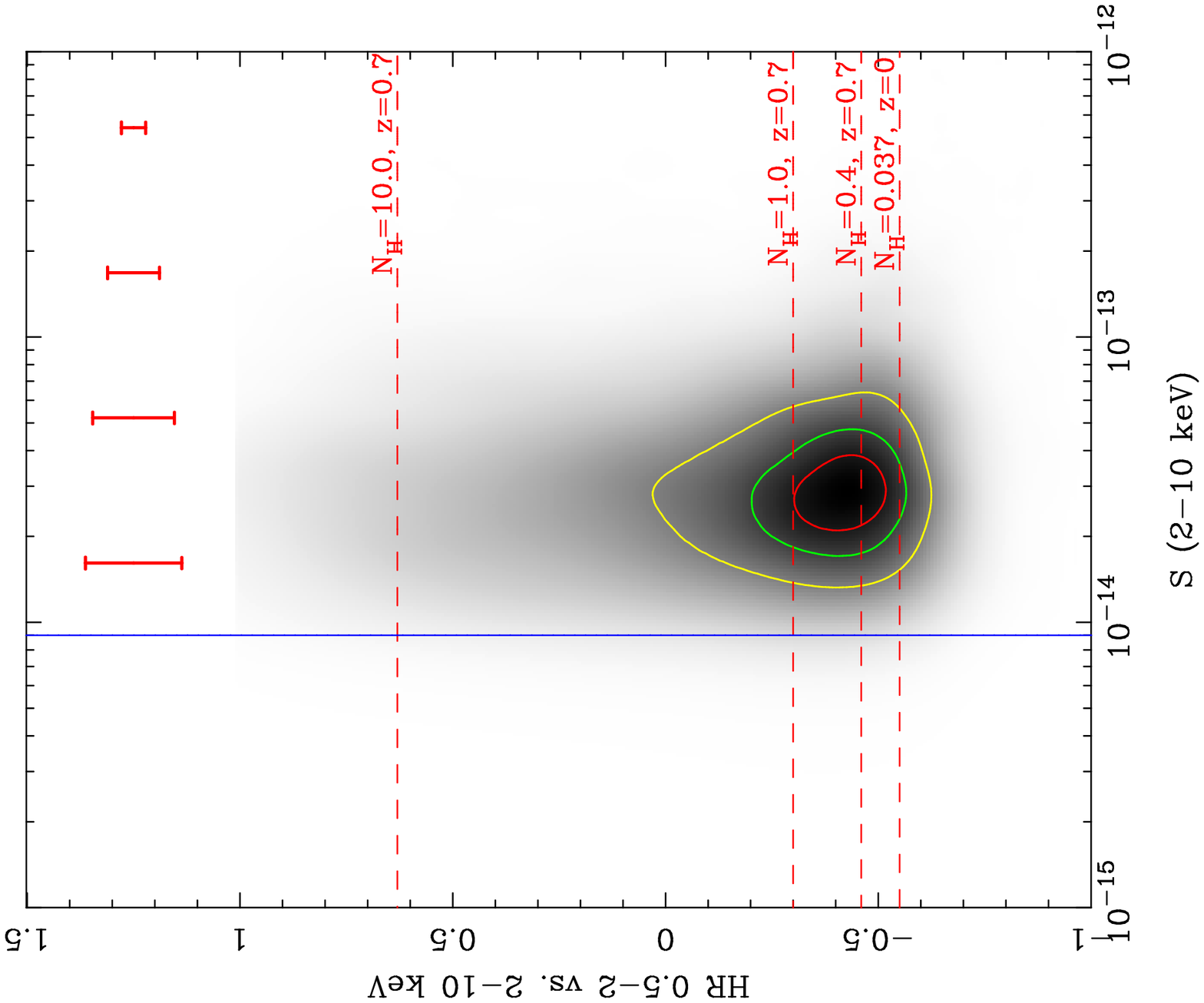}}
   \caption{Distributions of the X-ray colour vs. flux for sources detected in our broad energy bands. The 
     contours show the 90\%, 75\% and 50\% level of peak intensity. Flux limits in the band are shown with 
   a vertical solid line. The X-ray colour for a power-law spectral model with $\Gamma$=1.9 and 
   different values of redshift and rest-frame absorption (${\rm N_H}$ in units of ${\rm 10^{22}\,cm^{-2}}$) 
   are represented with horizontal lines. 
   The error bars at the top of the plots indicate the mean error in the X-ray colour 
   at different fluxes.
   }
              \label{hr_vs_flx_broad_bands}%
    \end{figure*}

\subsubsection{Three power-law fitting model}
\label{ml_fit_new_model}
In order to improve the quality of the fits to our source counts we performed a $\chi^2$ 
fitting to the binned differential distributions using a model with three power-law components. 
This model has six independent parameters: the break fluxes at bright and faint fluxes, 
$S_{\rm b}^{\rm b}$ and $S_{\rm b}^{\rm f}$; the normalisation $K$ at the bright flux break and the slopes at 
bright ($\Gamma_{\rm b}$), intermediate ($\Gamma_{\rm i}$) and faint ($\Gamma_{\rm f}$) fluxes. 
A summary of the results of the fitting are given in Table~\ref{table:5}. Column 8 lists 
the F-test significance of improvement of the fits relative to a model with two power-laws.

The results for the 0.5-1 keV, 1-2 keV and 0.5-2 keV spectral regimes confirm that the curvature of the 
source counts in these energy bands cannot be well reproduced with the standard broken power-law model. The same result
probably applies also to the source counts in the 2-4.5 keV and 2-10 keV energy bands,
although in this case the lower significance of improvement in the fits is probably due to the fact that 
our survey is not deep enough at these energies to provide strong constraints on the shape of the distributions 
below the region of downward curvature in the counts.

   \begin{figure*}[!t]
   \centering
   \hbox{
   \includegraphics[angle=90,width=0.50\textwidth]{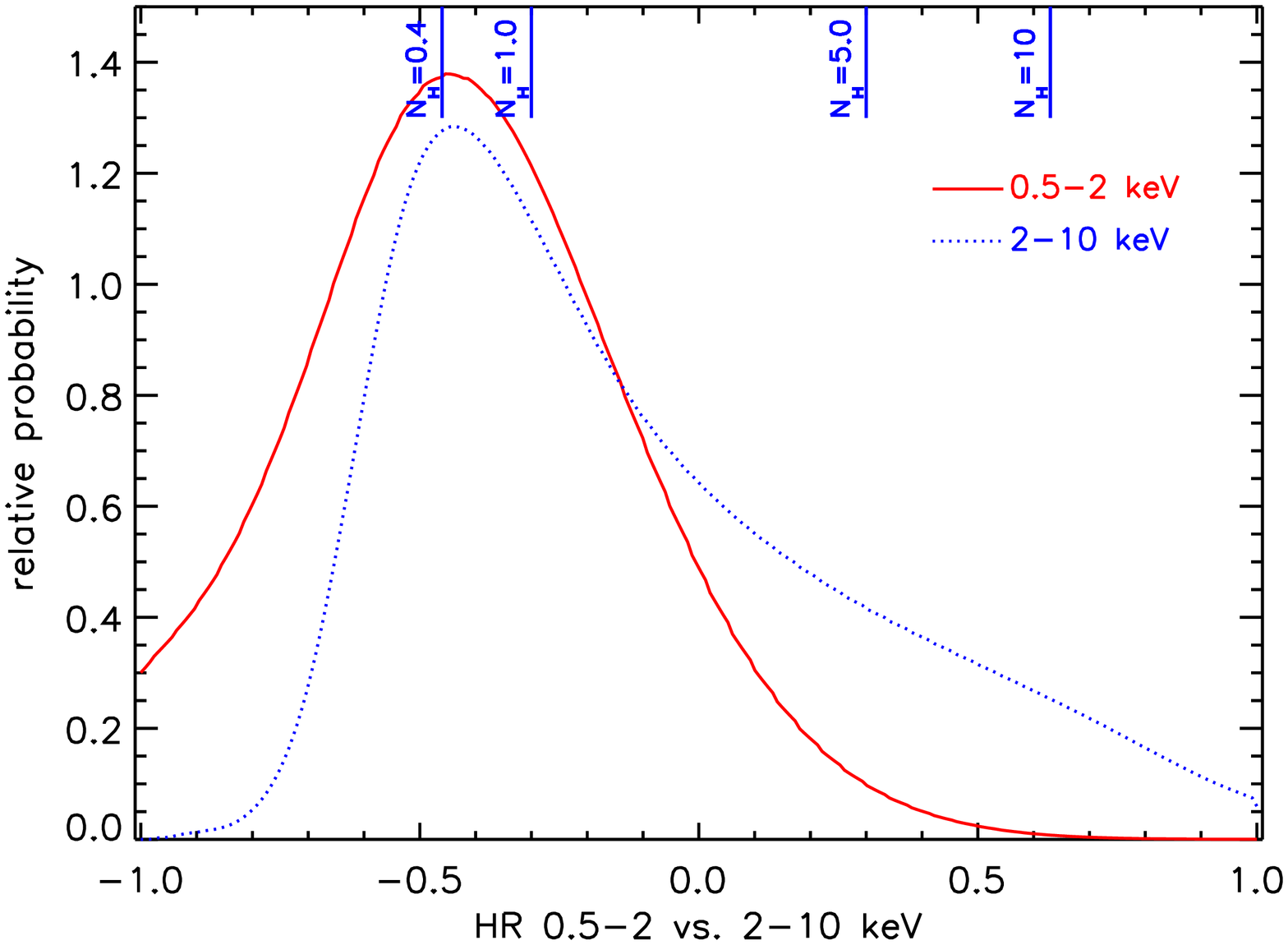}
   \includegraphics[angle=90,width=0.50\textwidth]{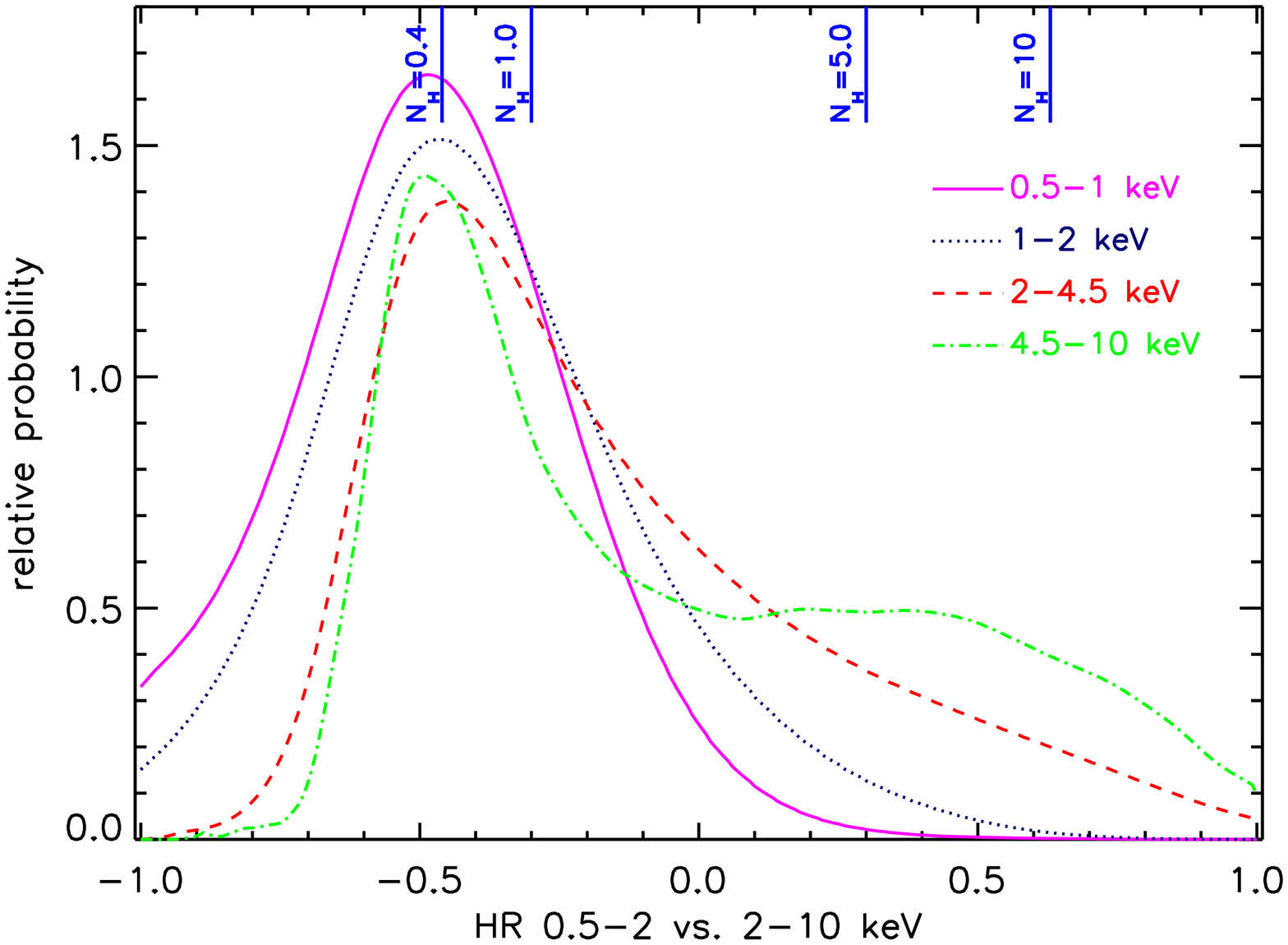}}
   \hbox{
   \includegraphics[angle=90,width=0.50\textwidth]{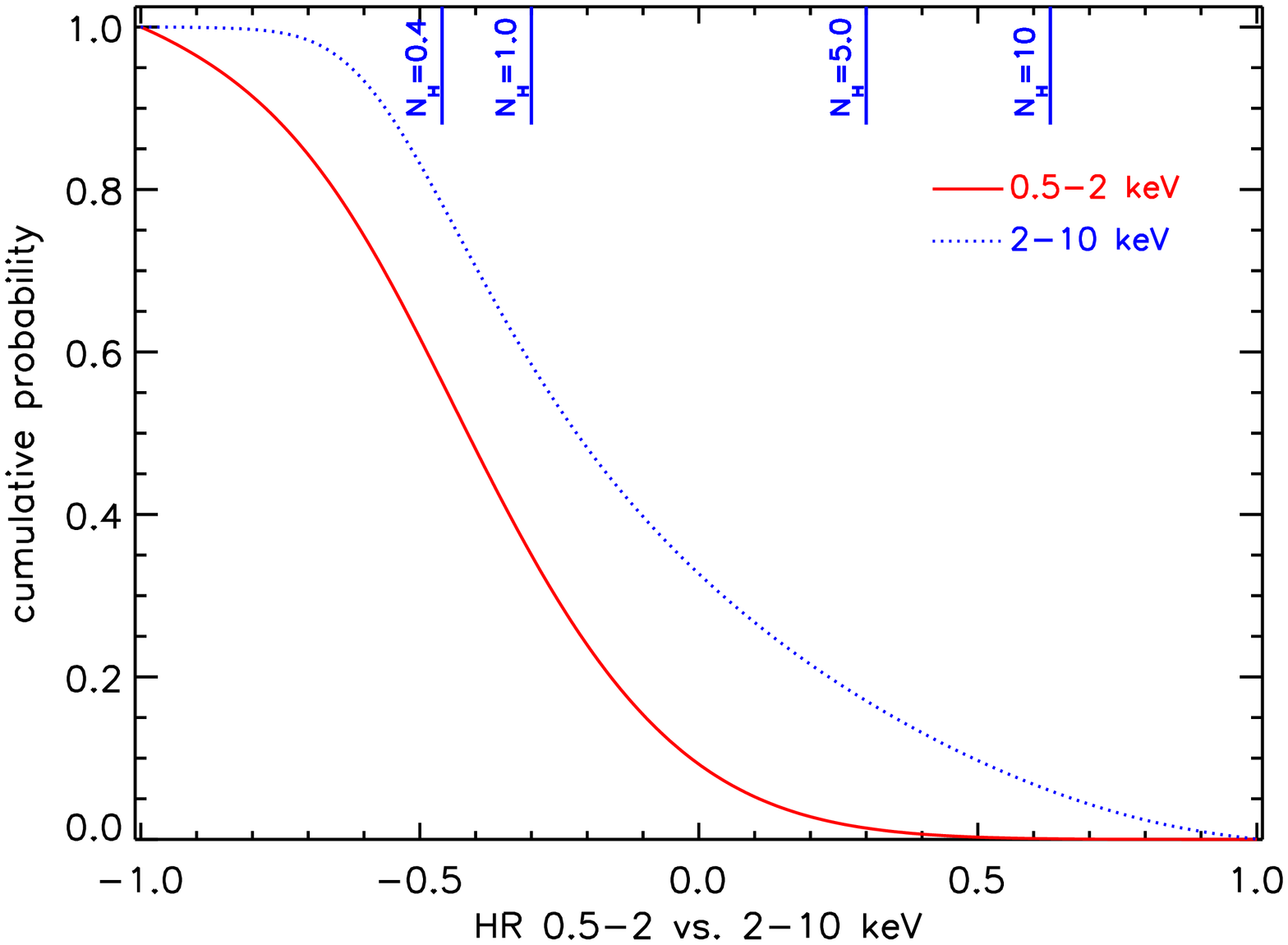}
   \includegraphics[angle=90,width=0.50\textwidth]{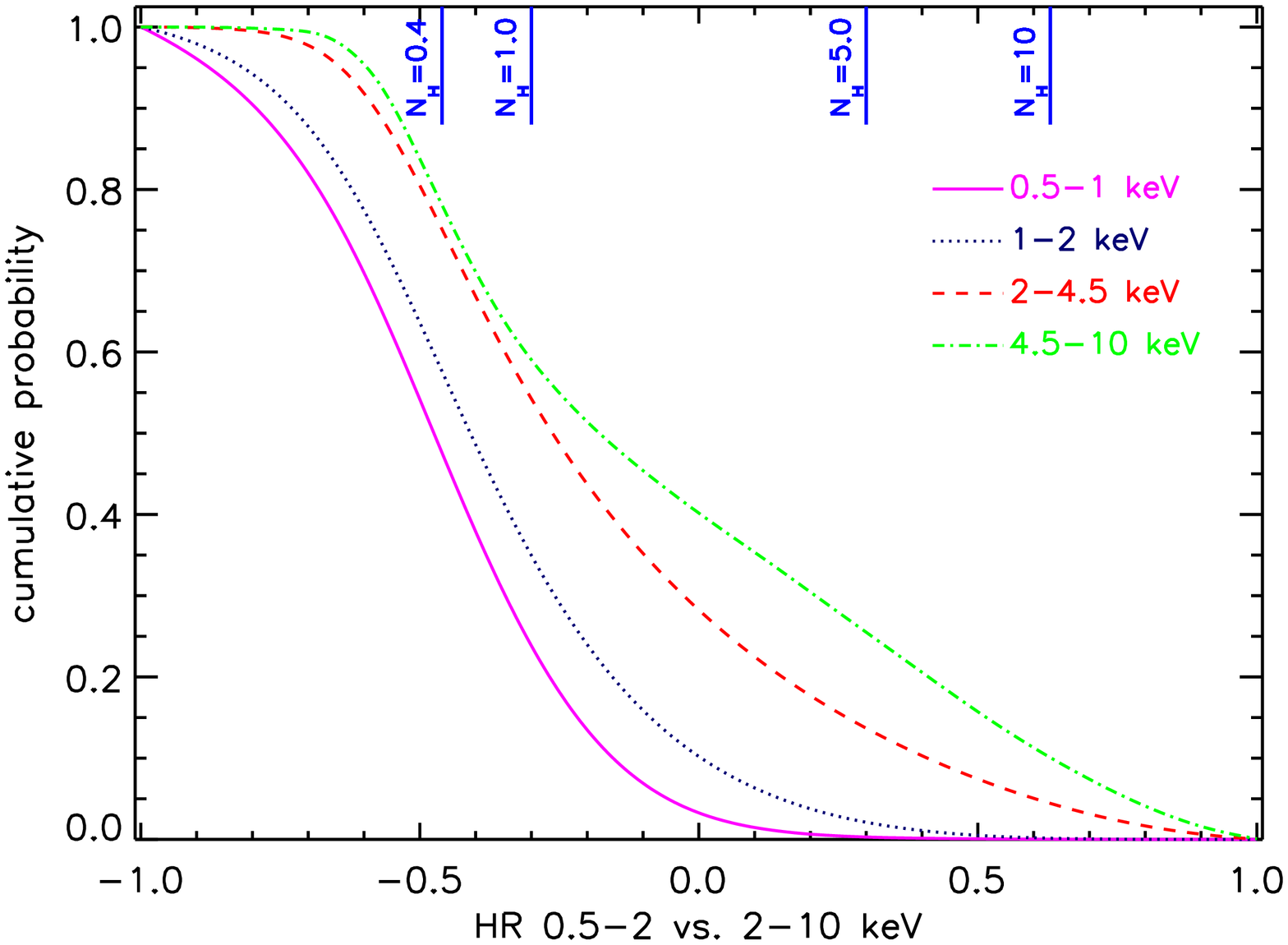}}
   \caption{
     Top: Probability density distributions of the X-ray colour for sources detected in different energy bands.     
       Bottom: Cumulative distributions of the X-ray colour for sources detected in different energy bands.
       The vertical solid lines indicate the X-ray colour for an object at redshift z=0.7 with a power-law
       spectrum of photon index $\Gamma$=1.9 and different amounts of rest-frame absorption (in units of 
       ${\rm 10^{22}\,cm^{-2}}$).}
              \label{hr_vs_flx_hist}%
    \end{figure*}

\subsection{Contribution to the cosmic X-ray background}
\label{ixrb}
We have used the best fit parameters of our source count distributions (three power-law model, see Table~\ref{table:5})
to estimate the intensity contributed by our sources to the cosmic X-ray background in the various energy bands.
The results are presented in Table~\ref{table:6}. 
Here we use the CXRB intensity measurements obtained by Moretti et al.~(\cite{Moretti03}) in the 1-2 keV and 
2-10 keV bands, which are consistent within 1$\sigma$ with the more recent measurements of 
the X-ray background intensity obtained by De Luca \& Molendi~(\cite{De Luca04}) and Hickox \& Markevitch~(\cite{Hickox06}). 
In order to compute the values in our energy bands we assumed a spectral model with power-law $\Gamma$=1.4, which 
is known to be an appropriate representation of the CXRB spectrum at energies above 2 keV (Lumb et al.~\cite{Lumb02}). The shape of 
the CXRB spectrum at energies below $\sim$1 keV is rather uncertain although source stacking analyses suggest a marginally 
softer spectrum (e.g. Streblyanska et al.~\cite{Streblyanska04}). We decided to use the same value of $\Gamma$
down to 0.5 keV and therefore the value of the CXRB intensity in the 0.5-1 keV band reported in Table~\ref{table:6}
could be potentially underestimated.
In Table~\ref{table:6} we also list the values of the CXRB intensity contributed by sources with fluxes above 
${\rm 10^{-12}\,erg\,cm^{-2}\,s^{-1}}$. These values were obtained by integrating the best fit model of our source count distributions. We 
estimate that the uncertainty in the values reported in Table~\ref{table:6} is dominated 
by systematics (e.g. those associated with spectral assumptions, flux conversions and instrumental calibrations), i.e. not source statistics, and 
therefore uncertainties in our measurements of the CXRB intensity should be $\lesssim5\%$. 

The source count distributions obtained by Moretti et al.~(\cite{Moretti03}) in the 
0.5-2 keV and 2-10 keV energy bands have been frequently used to estimate the contribution from bright sources to the X-ray 
background (e.g. Worsley et al.~\cite{Worsley04}, Worsley et al.~\cite{Worsley05}). However, it has already been pointed out that 
the Moretti et al.~(\cite{Moretti03}) bright end slopes might be too steep 
suggesting that bright-end corrections for the CXRB intensity could be an underestimate (see Worlsley et al.~\cite{Worsley05}).
We also found that the bright end slope of our source counts is flatter than the value reported in Moretti et al.~(\cite{Moretti03}) ($2.82_{-0.09}^{+0.07}$). 
However in the 2-10 keV band our bright end slope is marginally flatter although still compatible with 
the Moretti et al.~(\cite{Moretti03}) value at less than 2$\sigma$ ($2.57_{-0.08}^{+0.10}$). 

At energies below $\sim$2 keV our sources contribute more than $\sim$60\% of the CXRB intensity, while the fraction reduces to 
$\sim$40\% at higher energies. We also note that there is an important decline in the fraction of CXRB resolved by our sources 
as a function of energy, especially across the 2-10 keV bandpass, where the value goes down from 
$\sim$40\% in the 2-4.5 keV band to just $\sim$22\% above 4.5 keV.

\section{X-ray spectral properties of the sources}
\label{discussion}
Our analysis has revealed that the shape of the source count distributions 
becomes substantially steeper as we move to higher X-ray energies. In order to understand the origin of this effect 
we have investigated the overall properties of the X-ray source populations detected in different energy bands 
at the fluxes sampled by our analysis. Fig.~\ref{hr_vs_flx_broad_bands} shows in the form of density 
distributions\footnote{In order to account for the uncertainty in both the measured 
flux and X-ray colour, we added the probability density distribution of the 
X-ray colour-flux of each individual source.
This distribution was defined as a 2-d Gaussian centred at the measured value of the X-ray 
colour and flux and with dispersion equal to the corresponding 1$\sigma$ errors of the parameters.}
the X-ray colour distribution of sources detected in the 0.5-2 keV and 2-10 keV energy bands as a function of flux.
The contours overlaid indicate the 90\%, 75\% and 50\% levels of the peak intensity while the flux limit in the band 
is shown by a vertical solid line. The error bars at the top of the plots indicate the mean error in the X-ray colour 
at different fluxes.
The X-ray colour for each source was obtained as the normalised ratio of the count rates 
in the energy bands 0.5-2 keV and 2-10 keV. 
In the same figure we also show for comparison the X-ray colour for a power-law model 
of photon index $\Gamma$=1.9 at redshift z=0 and z=0.7 (the majority of type-2 AGN identified in surveys 
at intermediate fluxes have redshifts z$\lesssim$0.7-0.8, see Barcons et al.~\cite{Barcons07}, 
Caccianiga et al.~\cite{Caccianiga07}, Della Ceca et al.~\cite{Ceca08}) and various amounts of rest-frame 
absorption (${\rm N_H\,in\,units\,of\,10^{22}\,cm^{-2}}$). 
The peak of the distribution of X-ray colours corresponds to the X-ray colour typical for a power-law spectral model with 
low observed X-ray absorption. This result seems to hold for 
sources detected in both the 0.5-2 keV and 2-10 keV energy bands and at all fluxes sampled by our analysis although there seems to 
be a larger contribution from sources with hard X-ray colours in the 2-10 keV band.

We have computed the probability density distributions of the X-ray colour 
for sources detected in our broad energy bands 
by projecting the distributions in Fig.~\ref{hr_vs_flx_broad_bands} onto the y axis.
These distributions together with the corresponding cumulative distributions are shown in Fig.~\ref{hr_vs_flx_hist}.
Both distributions peak at a similar X-ray colour, however at energies $\ge$ 2 keV sources with very soft X-ray colours become 
less important while sources with hard X-ray colours become more important. 
In Fig.~\ref{hr_vs_flx_hist} (top right) we show the corresponding probability density distributions for sources detected in 
the narrow energy bands. These distributions show the same energy dependent trends as observed in our broad energy bands.
Because the typical errors in the X-ray colour of the sources are $\Delta$HR$\le$0.1-0.2 in all energy bands 
they cannot alone account for the observed large dispersion in the distribution of X-ray colours.

The decrease in the apparent fraction of sources with very soft X-ray colours at high energies can be explained by 
a significant decrease in the contribution from non-AGN sources such as stars and clusters of galaxies 
as we move to higher energies. On the other hand, in the harder band we are less biased against absorbed 
sources and hence we expect more absorbed sources to be detected at these energies. In order to estimate the contribution
from absorbed objects from the observed X-ray colour distributions, we used as a 
dividing line between unabsorbed and absorbed AGN the value of the X-ray colour for a source with spectral slope $\Gamma$=1.9 
at redshift z=0.7 and rest-frame absorption ${\rm N_H=4\times10^{21}\,cm^{-2}}$. 
According to Caccianiga et al.~(\cite{Caccianiga07}), the separation between optically absorbed (type-2) and optically 
unabsorbed (type-1) AGN corresponds to an optical extinction ${\rm A_V\sim}$2 mag. This value, assuming a 
Galactic ${\rm A_V/N_H}$ relation, corresponds to a column density ${\rm N_H=4\times10^{21}\,cm^{-2}}$. 
For a threshold corresponding to a power-law spectrum with rest-frame absorption ${\rm N_H=4\times 10^{21}\,cm^{-2}}$ at redshift z=0.7 
we find that the fraction of 'hard' sources 
is 0.46, 0.57, 0.74 and 0.77 at 0.5-1 keV, 1-2 keV, 2-4.5 keV and 4.5-10 keV respectively. The 
corresponding values for the broad energy bands are 0.55 at 0.5-2 keV and 0.77 at 2-10 keV. 
These numbers should be considered as lower limits to the 
fraction of X-ray absorbed objects contributing to the AGN source population at different energies because if the sources 
are moderately absorbed but at higher redshift the signatures of absorption will be outside of the applicable 
energy bandpass, and hence 
these objects will exhibit the X-ray colours of unabsorbed AGN. 

   \begin{figure}
   \centering
    \hspace{-0.3cm}\includegraphics[angle=90,width=0.5\textwidth]{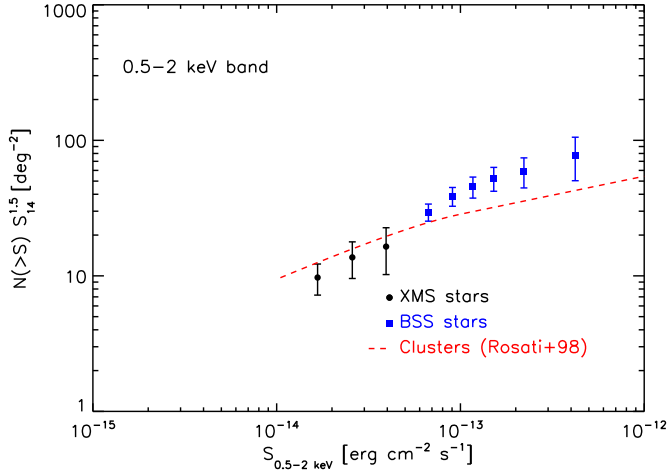}
   \caption{Normalised source count distribution in integral form for stars in the 0.5-2 keV band 
     from two XMM-{\it Newton} serendipitous surveys at high galactic latitudes (${\rm |b|>}$20{$^\circ$}): the XMM-{\it Newton} 
     Medium Sensitivity Survey ({\tt XMS}, circles) and the XMM-{\it Newton} bright survey ({\tt XBS}, squares). 
     The dashed line shows the best fit to the 0.5-2 keV source count distribution for clusters from Rosati et al.~(\cite{Rosati98}).
     Error bars correspond to 1$\sigma$ confidence.}
              \label{dist_stars}%
    \end{figure}

We can also compare the measured cumulative sky density of sources in the soft energy bands 
at a given flux with the values obtained in higher energy bands at the expected flux for the source. 
For example, assuming an unabsorbed power-law spectrum with $\Gamma$=1.9 at a flux 
of ${\rm 1.5\times10^{-14}\,erg\,cm^{-2}\,s^{-1}}$ in the 1-2 keV energy band\footnote{The flux was chosen to be 
high enough to guarantee that the expected flux in the energy bands at higher energies is well above the flux limit in the bands.}
we measure a cumulative sky density in the band of 23$\pm$1 deg$^{-2}$. The corresponding fluxes in the 
2-4.5 keV and 4.5-10 keV energy bands are ${\rm 1.9\times10^{-14}\,erg\,cm^{-2}\,s^{-1}}$ and 
${\rm 2\times10^{-14}\,erg\,cm^{-2}\,s^{-1}}$ respectively. At these fluxes the cumulative sky density of sources 
is 30$\pm$1 deg$^{-2}$ in the 2-4.5 keV band and 50$\pm$1 deg$^{-2}$ in the 4.5-10 keV band. 
The implied substantial increase in the sky density of objects as we move to higher energies suggests that 
a fraction of sources in the 4.5-10 keV band must have a spectrum harder than the assumed $\Gamma$=1.9 power-law. 
Furthermore, the substantial increase in the sky density from the 2-4.5 keV to the 4.5-10 keV band 
indicates that $\sim$40\% of the sources detected in the 4.5-10 keV band must have absorbing column densities high 
enough (${\rm \sim 10^{22} - 10^{23}\,cm^{-2}}$) to significantly reduce the observed flux in the 
2-4.5 keV band. This is consistent  with the shape of the probability distribution of X-ray colours 
presented in Fig.~\ref{hr_vs_flx_hist}. The higher efficiency of selection of type-2 AGN at energies $>$4.5 keV 
has already been reported (Caccianiga et al.~\cite{Caccianiga07}, Della Ceca et al.~\cite{Ceca08}).
 
The overall emission properties of the objects change with the energy band, but it is unclear whether this effect 
is entirely due to the fact that we are sampling a different spectral range at different 
energies or we are detecting an intrinsically different population of sources as we move to higher energies. 
The fraction of non-AGN sources decreases as we move to higher energies. This has an 
effect on both the distribution of the X-ray colours, as shown previously and, as we will see in Sec.~\ref{xrb_models},
on the shape of the source count distributions. At energies $\ge$2 keV we find that $\sim$5\% of the sources 
detected in the 4.5-10 keV band were not detected in the 2-4.5 keV band, however the majority of 4.5-10 keV sources 
were detected both in the 2-10 keV ($\ge$99\%) and 0.5-2 keV ($\ge$88\%) energy bands. 
Thus, we do not have any strong evidence that 
highly absorbed AGN with no soft flux dominate the source counts in the harder energy bands.
The changes in both the overall emission properties of the objects 
and the shape of the source counts at different energies are 
best explained by the varying mix of the population of objects at different fluxes and energies and the different 
sampling of the spectra of AGN. The dependence of the contribution from different populations of objects to the CXRB 
on the energy band was already noticed by Bauer et al.~(\cite{Bauer04}). However their analysis was based on relatively broad X-ray 
energy bands (0.5-2 keV and 2-8 keV). The use of narrower bands has allowed us to investigate 
the varying mix of different populations of objects and the relative contribution from absorbed and unabsorbed AGN 
throughout these bandpasses and to extend the study to higher energies (10 keV).

Although the overall emission properties of the objects in the 2-4.5 keV and 4.5-10 keV energy bands have changed 
substantially due to the strong increase in the fraction of absorbed objects at the highest energies, the slope of 
the source count distributions in these energy bands has not varied significantly, indicating that the cosmological evolution 
of the sources detected in the two energy bands must be quite similar.

\section{Implications for CXRB synthesis models}

\subsection{The contribution of stars and clusters}
\label{dists_stars_clusters}
Before comparing our source counts in different energy bands with the predictions from current 
CXRB synthesis models we need to account first for the contribution from non-AGN 
to our samples of X-ray sources. The two most important contributors are clusters of 
galaxies and stars, especially at bright fluxes and soft X-ray energies.

   \begin{figure*}
   \centering
   \hbox{
   \includegraphics[angle=90,width=0.5\textwidth]{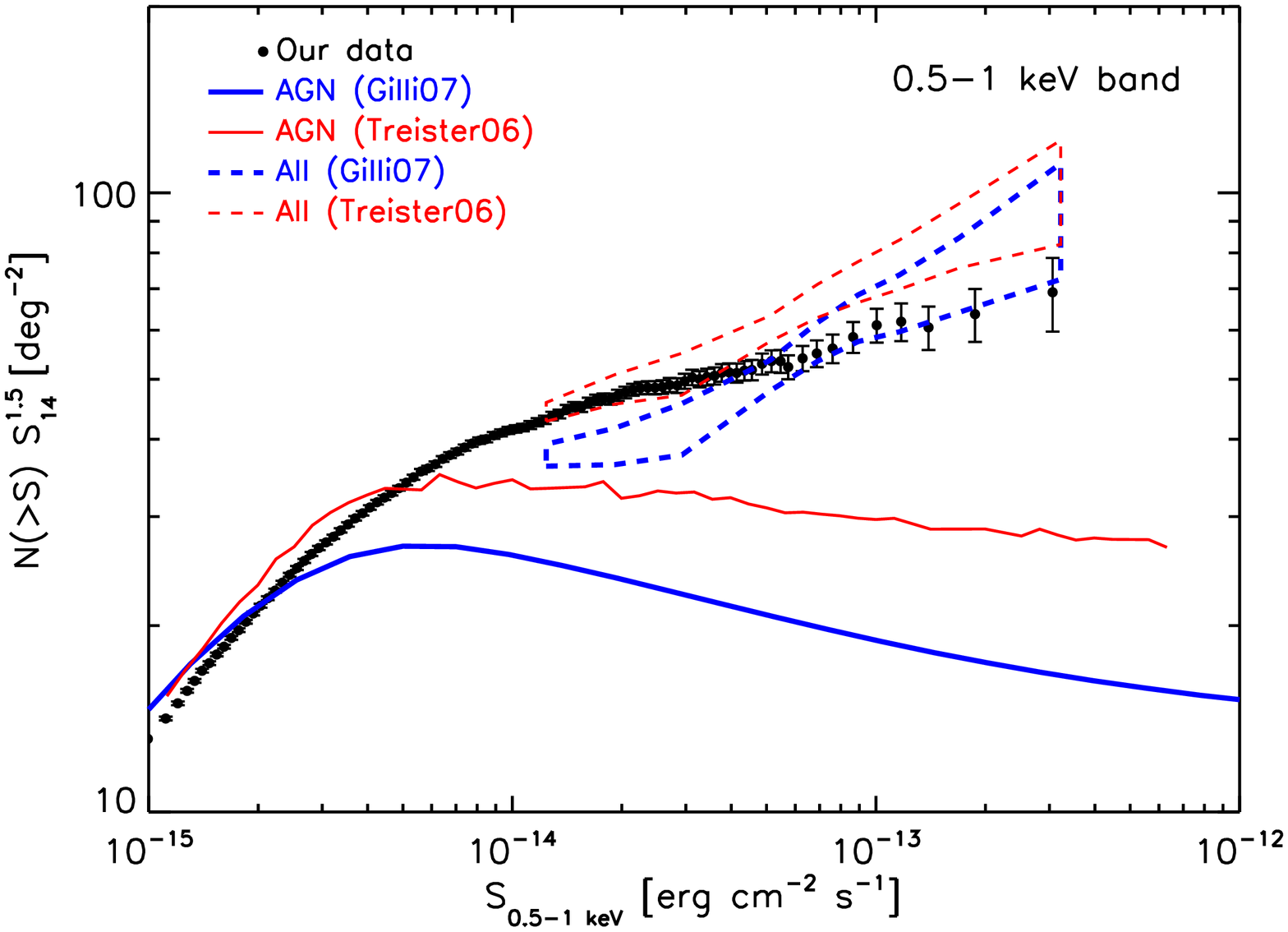}
   \includegraphics[angle=90,width=0.5\textwidth]{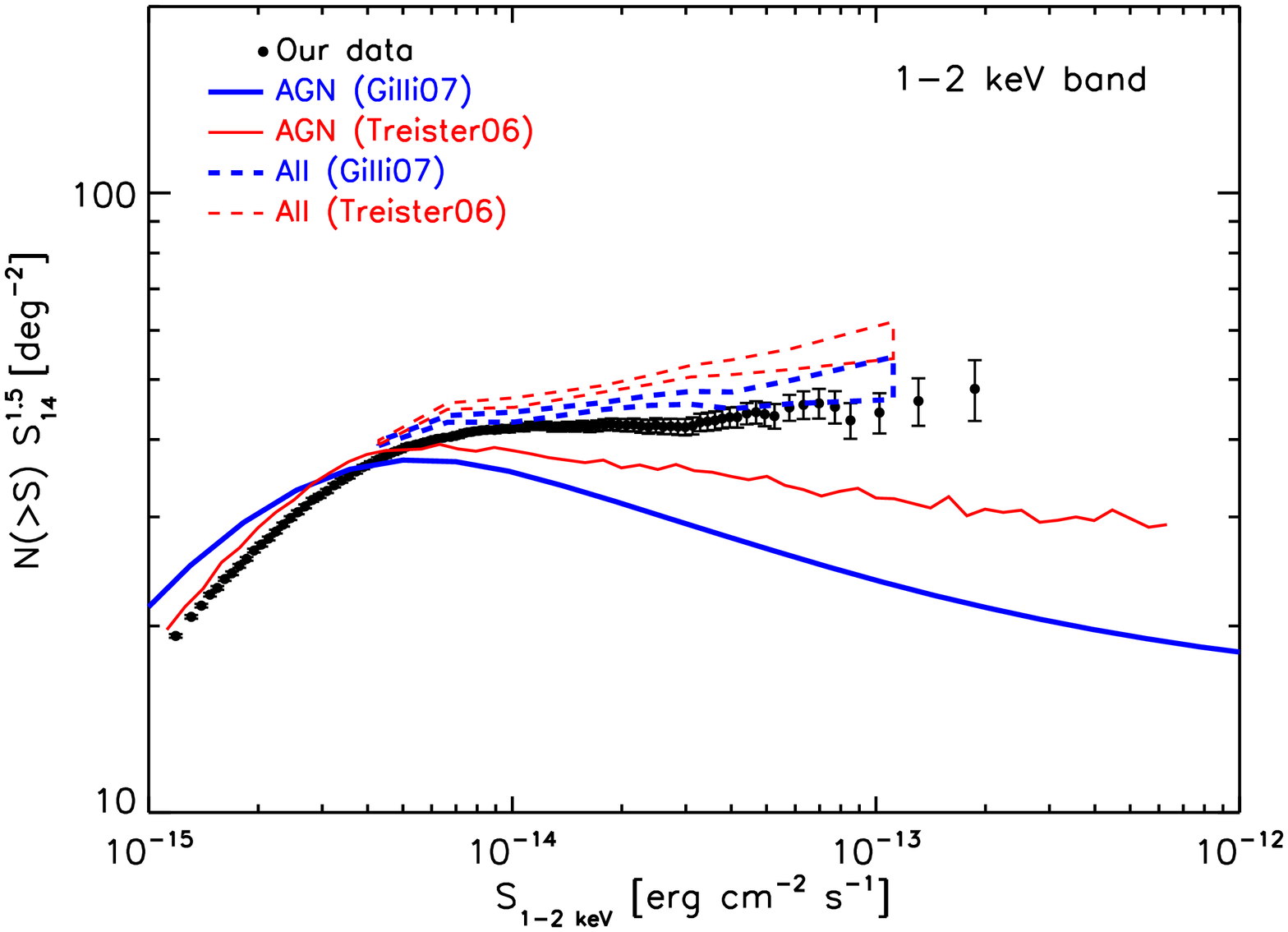}
   }
   \hbox{
   \includegraphics[angle=90,width=0.5\textwidth]{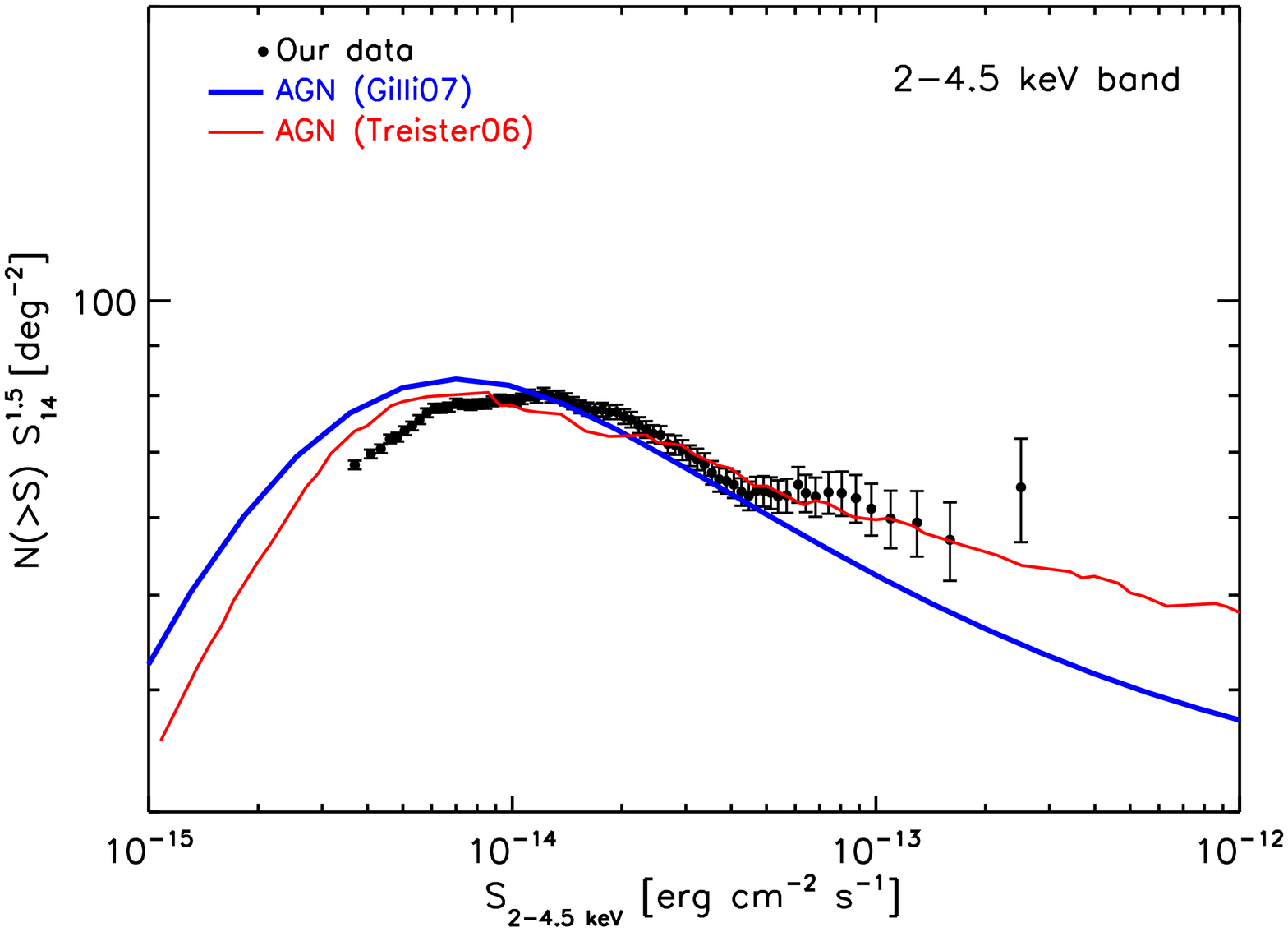}
   \includegraphics[angle=90,width=0.5\textwidth]{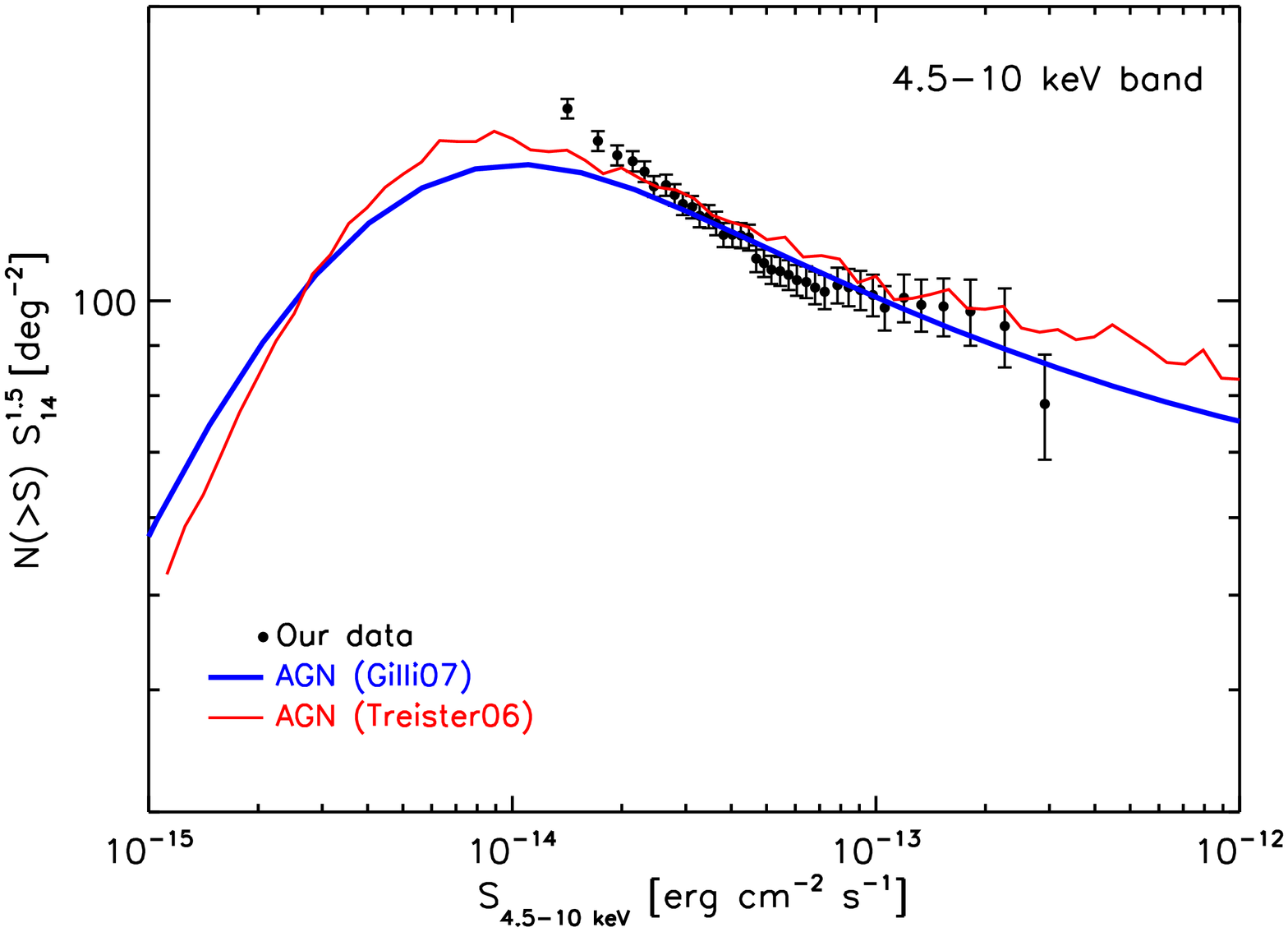}
   }
   \caption{Comparison of the normalised source count distributions in integral form in the 0.5-1 keV, 1-2 keV, 
     2-4.5 keV and 4.5-10 keV bands with the predictions from the synthesis models of the CXRB of 
     Treister \& Urry~(\cite{Treister06}) and Gilli et al.~(\cite{Gilli07}): AGN only (solid lines), 
     AGN+clusters+stars (dashed lines). 
     Error bars correspond to 1$\sigma$ confidence.}
              \label{xrb_data_narrow_bands}%
    \end{figure*}

   \begin{figure*}
   \centering
   \hbox{
   \includegraphics[angle=90,width=0.5\textwidth]{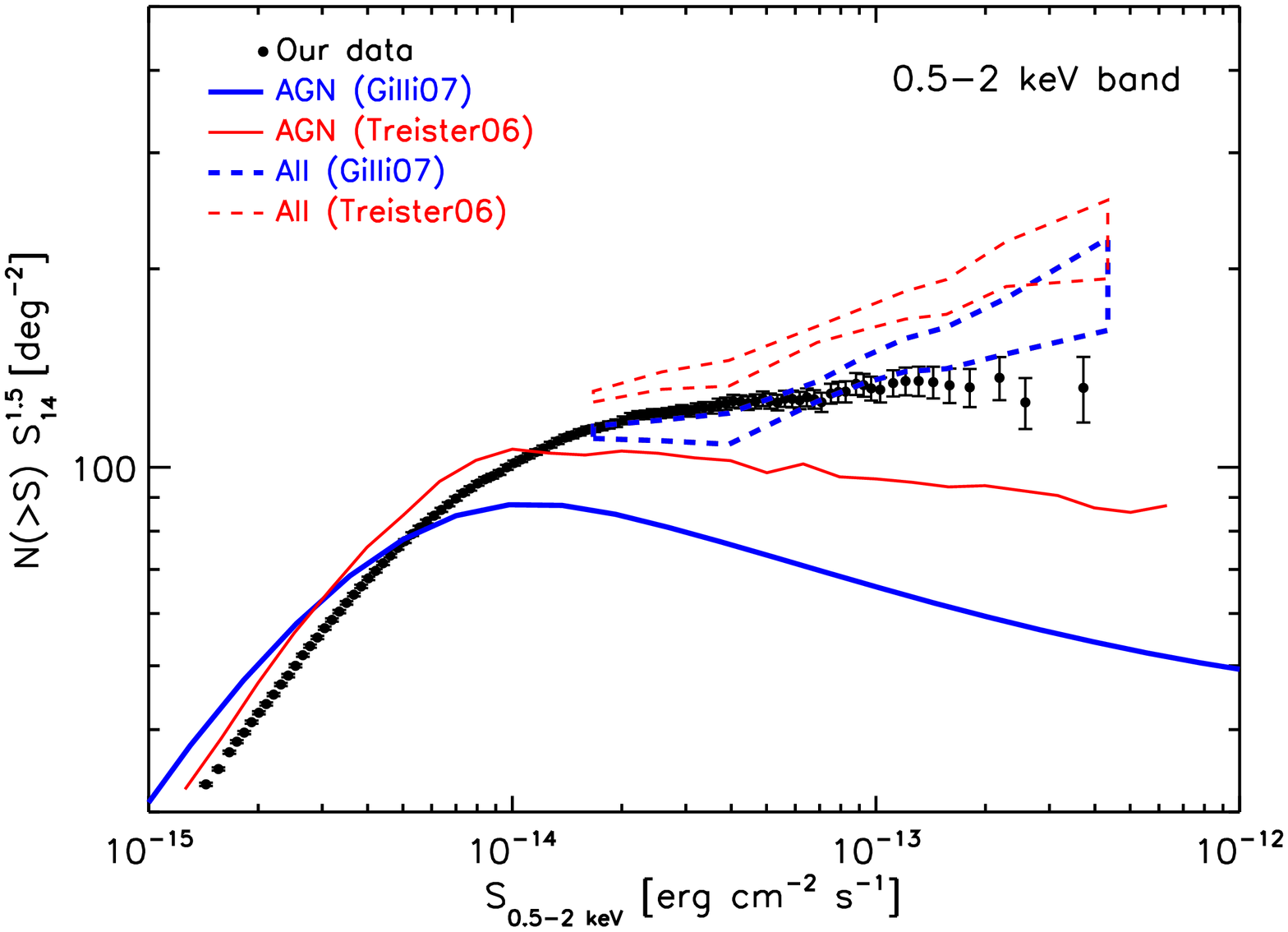}
   \includegraphics[angle=90,width=0.5\textwidth]{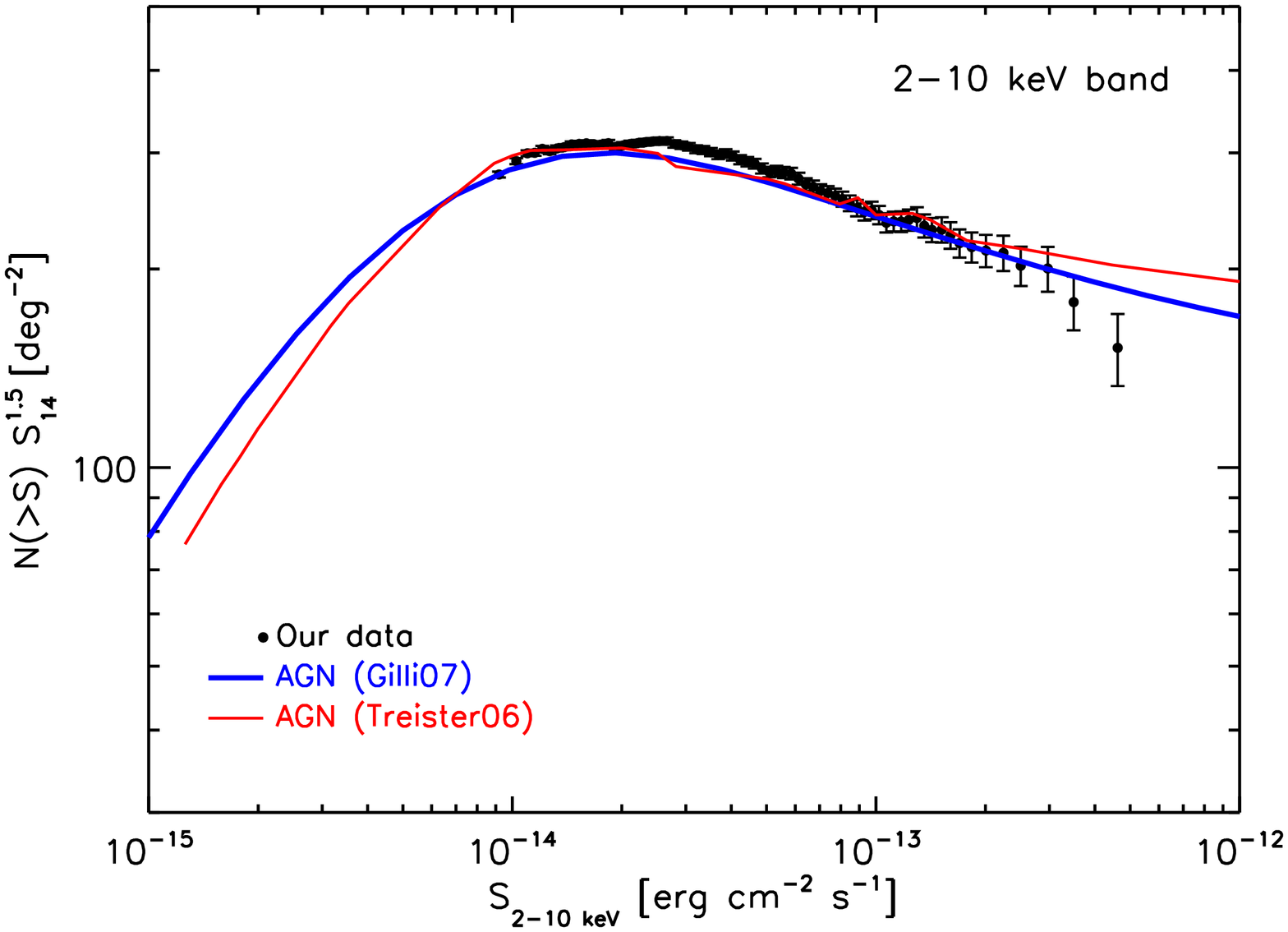}}
   \caption{Comparison of the normalised source count distributions in integral form in the 0.5-2 keV and 2-10 keV
     bands with the predictions from the synthesis models of the CXRB of 
     Treister \& Urry~(\cite{Treister06}) and Gilli et al.~(\cite{Gilli07}): AGN only (solid lines), 
     AGN+clusters+stars (dashed lines). 
     Error bars correspond to 1$\sigma$ confidence.}
              \label{xrb_data_broad_bands}%
    \end{figure*}

\begin{enumerate}
\item {\it Source counts for clusters}: In order to account for cluster contribution, we have used the 0.5-2 keV source 
count distribution for clusters from Rosati et al.~(\cite{Rosati98}), that covers 
the flux range from ${\rm \sim10^{-14}\,erg\,cm^{-2}\,s^{-1}}$ 
to ${\rm \sim10^{-12}\,erg\,cm^{-2}\,s^{-1}}$ (see Fig.~\ref{dist_stars}).
We have checked whether our source detection algorithm 
is able to detect all clusters at these fluxes (either as point like or extended). 
In order to do that we cross-correlated our sample of objects detected in the 0.5-2 keV 
band with the 50 clusters from the {\tt XMM-COSMOS} survey (Finoguenov et al.~\cite{Finoguenov07}) that lie in the area covered 
by our survey. We found that 18 of their clusters have been detected in our 0.5-2 keV band, 8 as extended and 10 as point like sources.  
We detected all {\tt XMM-COSMOS} clusters with 0.5-2 keV fluxes ${\rm \ge10^{-14}\,erg\,cm^{-2}\,s^{-1}}$ 
(5 as point like sources and 8 as extended sources), while only $\sim$13\%  {\tt XMM-COSMOS} clusters 
with 0.5-2 keV fluxes below ${\rm \sim10^{-14}\,erg\,cm^{-2}\,s^{-1}}$ are in our sample (all detected 
as point like objects). Although the numbers involved are very small, this test demonstrates that the assumption that 
all clusters with fluxes ${\rm \sim10^{-14}\,erg\,cm^{-2}\,s^{-1}}$ are included in our samples is reasonable.

We also need to estimate the contribution of clusters to the 0.5-1 keV and 1-2 keV energy bands (the contribution from clusters 
at energies $\ge$2 keV is negligible). Assuming that all clusters are also detected in these bands we have rescaled the 0.5-2 keV distribution 
to estimate the contribution to these bands. In order to convert the fluxes we have assumed that the mean spectrum 
of clusters can be well represented by a thermal spectrum with temperature=3 keV, abundance=1/3 and 
redshift=0.2 ({\tt apec} model in Xspec, Arnaud~\cite{Arnaud96})\footnote{From the bolometric luminosity-temperature 
relation, a cluster with a temperature $\sim$3 keV has a luminosity ${\rm \sim10^{43}\,erg\,s^{-1}}$. If the cluster 
is at a redshift of 0.2 the expected observed flux will be 
${\rm \sim3\times10^{-14}\,erg\,cm^{-2}\,s^{-1}}$, i.e. in the range of fluxes sampled by our survey. From the cluster luminosity 
function objects with very high temperature ($\sim$10 keV) are relatively rare, while for a flux limited survey the volume 
sampled for less luminous clusters (hence with low temperatures) is small. This implies that flux limited surveys are expected to be 
dominated by clusters with typical luminosities ${\rm \sim10^{43}-10^{44}\,erg\,s^{-1}}$ (i.e. with 
temperatures $\sim$3-4 keV). More luminous clusters (${\rm \sim10^{45}\,erg\,s^{-1}}$) will be relatively rare, and, 
at the flux level sampled by our survey, they will be at high redshift (z$\sim$1) and hence they will 
exhibit X-ray spectra very similar to the 
more common, less luminous objects (Henry et al.~\cite{Henry91}).}.
\\
\item {\it Source counts for stars}: We have calculated source count distributions for stars using the 
data from two XMM-{\it Newton} serendipitous surveys at high galactic latitudes ($|b|>20{^\circ}$): the XMM-{\it Newton} 
Medium Sensitivity Survey ({\tt XMS}; Barcons et al.~\cite{Barcons07}) and the XMM-{\it Newton} Bright Survey 
({\tt XBS}; Della Ceca et al.~\cite{Ceca04}, Caccianiga et al.~\cite{Caccianiga07}). 
The {\tt XMS} survey covers a total sky area of 3.33 deg$^2$ and it has spectroscopically identified all 
stars in the sample (15 objects) down to a 0.5-2 keV flux ${\rm \sim1.5\times10^{-14}\,erg\,cm^{-2}\,s^{-1}}$. 
On the other hand the {\tt XBS} covers a total sky area of 28.1 deg$^2$ and has a complete sample of stars (49) 
down to a 0.5-4.5 keV flux ${\rm \sim7\times10^{-14}\,erg\,cm^{-2}\,s^{-1}}$.
Stars at the fluxes sampled by our analysis have mainly low temperature thermal spectra (see e.g. Della Ceca et al.~\cite{Ceca04}, 
Lopez-Santiago et al.~\cite{Lopez-Santiago07}). Therefore 0.5-2 keV fluxes were scaled to our various energy bands assuming 
a thermal model with a temperature of 0.7 keV. The 0.5-2 keV source count distributions for {\tt XMS} and {\tt XBS} stars 
are shown in Fig.~\ref{dist_stars}. 

The contribution of stars to the 2-4.5 keV and 2-10 keV energy bands at high galactic latitudes is much less certain
because of the lack of data at these energies. According to the {\tt XBS} survey, the cumulative sky density of stars in the above  
bands at a flux of ${\rm 10^{-13}\,erg\,cm^{-2}\,s^{-1}}$ (where the survey is complete) is 0.21$\pm$0.09 deg$^{-2}$.    
We have used this value to estimate the effect of stars on the observed source count distributions in the two hard 
energy bands at bright fluxes (see Sec.~\ref{xrb_models}). Because stars have a thermal soft spectrum their 
contribution to our source counts should decrease 
significantly as we move to higher energies and above $\sim$4.5 keV their contribution is 
negligible (Barcons et al.~\cite{Barcons07}, Caccianiga et al.~\cite{Caccianiga07}).
\end{enumerate}

\subsection{Comparison with CXRB synthesis models}
\label{xrb_models}
In principle synthesis models of the CXRB incorporate all the available information about the mean spectral properties and cosmological evolution
of the sources. Therefore a comparison of our observational constraints with their predictions can give us some insight into
the origin of the observed trends in the source count distributions.
Conversely, any deviation of the measurements from the predictions, might indicate the need for refinement of the 
model assumptions.
 
We have compared our data with the predictions from the synthesis models of Treister \& Urry~(\cite{Treister06}) and 
Gilli et al.~(\cite{Gilli07}). 
These models have different recipes for the X-ray luminosity function and assume somewhat different distributions 
of X-ray absorption. One fundamental ingredient of these models, is the fraction of obscured AGN, $F$, its evolution and 
its dependence on luminosity. In the Treister \& Urry~(\cite{Treister06}) model, 
$F\propto(1+z)^{0.4}$, and its dependence on the X-ray luminosity is linear, from 
100\% at ${\rm L_X=10^{42}\,erg\,s^{-1}}$ to 0\% at ${\rm L_X = 3 \times 10^{46}\,erg\,s^{-1}}$.
On the other hand, in the Gilli et al.~(\cite{Gilli07}) model no evolution of $F$ is assumed and the dependence 
of $F$ on the luminosity is much flatter than in the Treister \& Urry ~(\cite{Treister06}) model. The significantly different assumption,
relating to the intrinsic absorption properties of AGN inherent in the two models would suggest that 
the predictions of the two models might differ, especially at low energies where the results are more affected by X-ray absorption effects. We see in 
Fig.~\ref{xrb_data_narrow_bands} that the predicted source counts from the two models are very similar above 4.5 keV, especially 
at the fluxes sampled by our survey. Differences between the two model predictions become more clear as we move to lower energies. 
The two most important differences are:
\begin{enumerate}
  \item At bright fluxes the Treister \& Urry~(\cite{Treister06}) model predicts a larger number density of AGN
    than the Gilli et al.~(\cite{Gilli07}) model. The effect becomes more important at low energies. For example, 
    at ${\rm \sim10^{-13}\,erg\,cm^{-2}\,s^{-1}}$ the Treister \& Urry~(\cite{Treister06}) model predicts 25-35\% 
    more AGN (depending 
    on the energy band) than the Gilli et al.~(\cite{Gilli07}) model.  
    \item At all fluxes the slope of the source counts is flatter in the 
      Treister \& Urry~(\cite{Treister06}) model than in the Gilli et al.~(\cite{Gilli07}) model 
      (in the sense that the implied value of $|\Gamma|$ is lower).
\end{enumerate}

The comparison of our measured source count distributions with the predictions from the two models is 
shown in Fig.~\ref{xrb_data_narrow_bands} and Fig.~\ref{xrb_data_broad_bands}. We show both 
the AGN-only predictions from the models (solid lines) and the predictions after adding to the AGN-only 
source counts (in the 0.5-1 keV, 1-2 keV and 0.5-2 keV bands) the contribution from non-AGN 
sources (clusters of galaxies and stars) as explained in Sec.~\ref{dists_stars_clusters}.

At energies below 2 keV and at bright fluxes (${\rm >10^{-14}\,erg\,cm^{-2}\,s^{-1}}$)
the measured source count distributions lie significantly above model predictions 
for AGN-only. This appears to be due to the fact that at these energies the contribution from non-AGN sources, mainly 
stars and clusters of galaxies, is not negligible. We estimated that the contribution from stars and clusters to the X-ray source 
population increases from $\sim$13-22\% at fluxes ${\rm \ge10^{-14}\,erg\,cm^{-2}\,s^{-1}}$ to 
$\sim$44-54\% at fluxes ${\rm \ge10^{-13}\,erg\,cm^{-2}\,s^{-1}}$ in the 0.5-2 keV band (the exact value depending on the CXRB 
model used to calculate the fractions). The net effect of stars and clusters on the shape of the source counts is that the bright slopes of 
the distributions are substantially flatter compared 
with the expectations from the models for AGN-only sources. 
Once we include the contribution from stars and clusters the predictions of the models are in better agreement 
with our results, although both models seem to overpredict the observed 
source counts by 20-30\% at the brightest fluxes sampled by our survey (although with the caveat that the 
exact contribution of stars and clusters of galaxies has some uncertainty).
Below 2 keV and at faint fluxes (${\rm <10^{-14}\,erg\,cm^{-2}\,s^{-1}}$) both models 
overpredict the source counts. The effect is more important from the comparison with the 
Gilli et al.~(\cite{Gilli07}) model, resulting in a discrepancy of the data with the model predictions $\sim$10-20\%. 

In the 2-4.5 keV band and at bright fluxes the Gilli et al.~(\cite{Gilli07}) model seems to underpredict the source 
counts. We have estimated the contribution from stars in this energy band at bright fluxes using the results 
from the {\tt XBS} survey as specified in Sec.~\ref{dists_stars_clusters} (${\rm 0.21\pm0.09\,deg^{-2}}$ 
at ${\rm 10^{-13}\,erg\,cm^{-2}\,s^{-1}}$). We find that the net effect of stars is to increase the source counts at 
${\rm 10^{-13}\,erg\,cm^{-2}\,s^{-1}}$ by $\sim$11\% with respect to the AGN-only distributions, 
obtaining a much better agreement of the model predictions 
with our data. On the other hand, at faint fluxes we find that both models 
overpredict the 2-4.5 keV source counts by 10-20\%, 
as we found from the comparison of the source counts at energies below 2 keV.

In the 2-10 keV and 4.5-10 keV energy bands the agreement of our data with the predictions from the models for 
AGN-only is better than 10\%. 
Only in the 4.5-10 keV band do the model predictions seem to slightly underpredict our source counts at the faintest 
fluxes sampled by our analysis. This effect could be explained if the break in the source counts in this energy band is located at 
lower fluxes than those predicted by the synthesis models as suggested from deeper X-ray 
surveys ($\lesssim$5-8${\rm \times10^{-15}\,erg\,cm^{-2}\,s^{-1}}$, e.g. Loaring et al.~\cite{Loaring05}, 
Brunner et al.~\cite{Brunner08}, Georgakakis et al.~\cite{Georgakakis08}). The effect of stars on the source counts in these energy bands is negligible. 
  
We find that the CXRB models overpredict by 10-20\% the source counts at energies below 4.5 keV and at faint fluxes. 
It is important to note that the models have been tuned to be in agreement with 
the {\it Chandra} Deep Field source counts at faint fluxes, but these appear to be only marginally higher ($\lesssim$10\%) 
than those estimated by our analysis. The results of the comparison suggest that 
the synthesis models might be overpredicting the number of faint absorbed AGN as has been reported in the 
past by previous surveys (e.g. Piconcelli et al.~\cite{Piconcelli02},~\cite{Piconcelli03}, 
Caccianiga et al.~\cite{Caccianiga04}). This would call for fine 
adjustment of some model parameters such as the obscured to unobscured AGN ratio and/or the details of 
the distribution of column densities at intermediate obscuration (${\rm N_H=10^{22}-10^{23}\,cm^{-2}}$) 
and the dependence of these on the X-ray luminosity and/or redshift (see e.g. Della Ceca et al.~\cite{Ceca08}).

\section{Summary and Conclusions}
\label{conclusions}
We have used the largest samples of X-ray selected sources available to date to 
provide strong observational constraints on the X-ray source count distributions over a broad range of fluxes 
and at different X-ray energies. Our source lists were built from 1129 XMM-{\it Newton} observations at high 
galactic latitudes, $|b|$$>20{^\circ}$, covering a total sky area of 132.3 deg$^2$. We have focused our study on 
four 'narrow bands' and the two 'standard' energy bands, 0.5-2 keV and 2-10 keV, where we have in excess of 30,000 sources.
Our data encompass roughly 3 decades of flux, from ${\rm \sim10^{-15}\,erg\,cm^{-2}\,s^{-1}}$ 
to ${\rm \sim10^{-12}\,erg\,cm^{-2}\,s^{-1}}$ at energies $\lesssim$2 keV and more than 2 decades in flux at 
higher energies, $\ge$2 keV, from ${\rm \sim10^{-14}\,erg\,cm^{-2}\,s^{-1}}$ to ${\rm \sim10^{-12}\,erg\,cm^{-2}\,s^{-1}}$. 
Our sources contribute more than $\sim$60\% of the CXRB intensity at energies below $\sim$2 keV and $\sim$40\% above $\sim$2 keV (although 
there is a marked decline in the fraction of CXRB resolved across the 2-10 keV bandpass).
Thanks to the large size of the samples employed, our 
results are not limited by cosmic variance effects or low counting statistics. For the first time we have been able to 
investigate how the changing population of X-ray sources, as we move to different energies, modifies the shape of the measured 
distributions. The main results are summarised below:
   \begin{enumerate}
     \item A comparison with previous representative surveys at the fluxes of interest shows overall a good 
       agreement.
       The largest discrepancies from the comparison were found at bright 
       fluxes ${\rm \ge10^{-14}\,erg\,cm^{-2}\,s^{-1}}$, where the 
       results of the majority of the surveys are limited by low counting statistics. Although cross-calibration issues between 
       missions might contribute to the scatter, especially at energies $\ge$2 keV, we have seen that it 
       cannot fully explain some of the observed discrepancies. We have also shown that 
       different spectral assumptions made in converting count rates 
       to fluxes can also introduce some additional scatter particularly above 2 keV, where the effective areas of the X-ray 
       detectors vary strongly with the energy. 

      \item Maximum likelihood fits to our distributions have been carried out using a broken power-law model.
	A break in the distributions is detected in all energy bands, except in the 4.5-10 keV band, where our survey
	does not go deep enough to detect the change in slope of the distribution. However, the results of the fits indicate that the 
	measured curvature of our source count distributions cannot be well fitted with a simple broken power-law shape. 
	A model with three power-law components provided a significantly better representation of the shape of our distributions 
	across the set of energy bands.

      \item We find that our source count distributions become significantly steeper both at high and low fluxes 
	as we move to higher energies, where we 
	have shown that the contribution from objects with hard X-ray colours becomes more important. We explain this 
	on the basis of a 
	varying mix of the population of objects at different fluxes and energies
	and the different sampling of the spectra of AGN. Stars and clusters of 
	galaxies become significantly less important as we move to higher energies, while 
	sources with hard X-ray colours become substantially more important at high energies. 
	We did not find any strong evidence that AGN with no soft flux dominate the source counts at the highest 
	energies sampled by our survey.

      \item We have compared our distributions with the predictions from the CXRB synthesis models from 
	Treister \& Urry~(\cite{Treister06}) and Gilli et al.~(\cite{Gilli07}) which assume different 
	absorption properties for the underlying population of AGN.
	Once we account for the contribution from clusters and stars at bright fluxes, the models seem to overpredict by 20-30\%
	the source counts at energies below 2 keV. However 
	our correction for the contribution from stars and clusters might contribute to the observed discrepancy.
	The two CXRB models predict different shapes for the 
	source count distributions at faint fluxes, however both models overpredict our source counts 
	(especially the Gilli et al.~\cite{Gilli07} model) by 10-20\% at energies below 4.5 keV and at faint fluxes. 
	This result suggest that the synthesis models 
	might overpredict the number of faint absorbed AGN. 
	On the other hand the models seem to underpredict our 4.5-10 keV source counts at the faintest fluxes sampled by 
	our analysis. This could be explained if, as suggested by deep X-ray surveys, 
	the break in the source counts in this energy band is located at 
	lower fluxes than those predicted by the models.
	The high statistical precision of source counts 
	will allow fine tuning of some model parameters such as 
	the obscured to unobscured AGN ratio and/or the details of the distribution of column densities 
	at intermediate obscuration 
	(${\rm N_H=10^{22}-10^{23}\,cm^{-2}}$).
   \end{enumerate}

\begin{acknowledgements}

\end{acknowledgements}
This work is based on observations obtained with XMM-{\it Newton}, an ESA science mission with instruments and 
contributions directly funded by ESA Member States and NASA. SM, and JAT acknowledge direct support from the 
UK STFC research council. FJC and JE acknowledge financial support by the Spanish Ministerio de Educaci\'on y Ciencia 
under project ESP2006-13608-C02-01. RDC and AC acknowledge financial support from MIUR, grants PRIN-MIUR 2006-02-5203 and from the Italian Space Agency (ASI) , grants n. I/088/06/0. We thank the referee D. Alexander for providing comments that improved this paper.

\begin{appendix} 

\section{Empirical sensitivity maps calculation}
\label{esens_calc}

\begin{figure*}
\centering
\hbox{
  \includegraphics[angle=-90,width=0.450\textwidth]{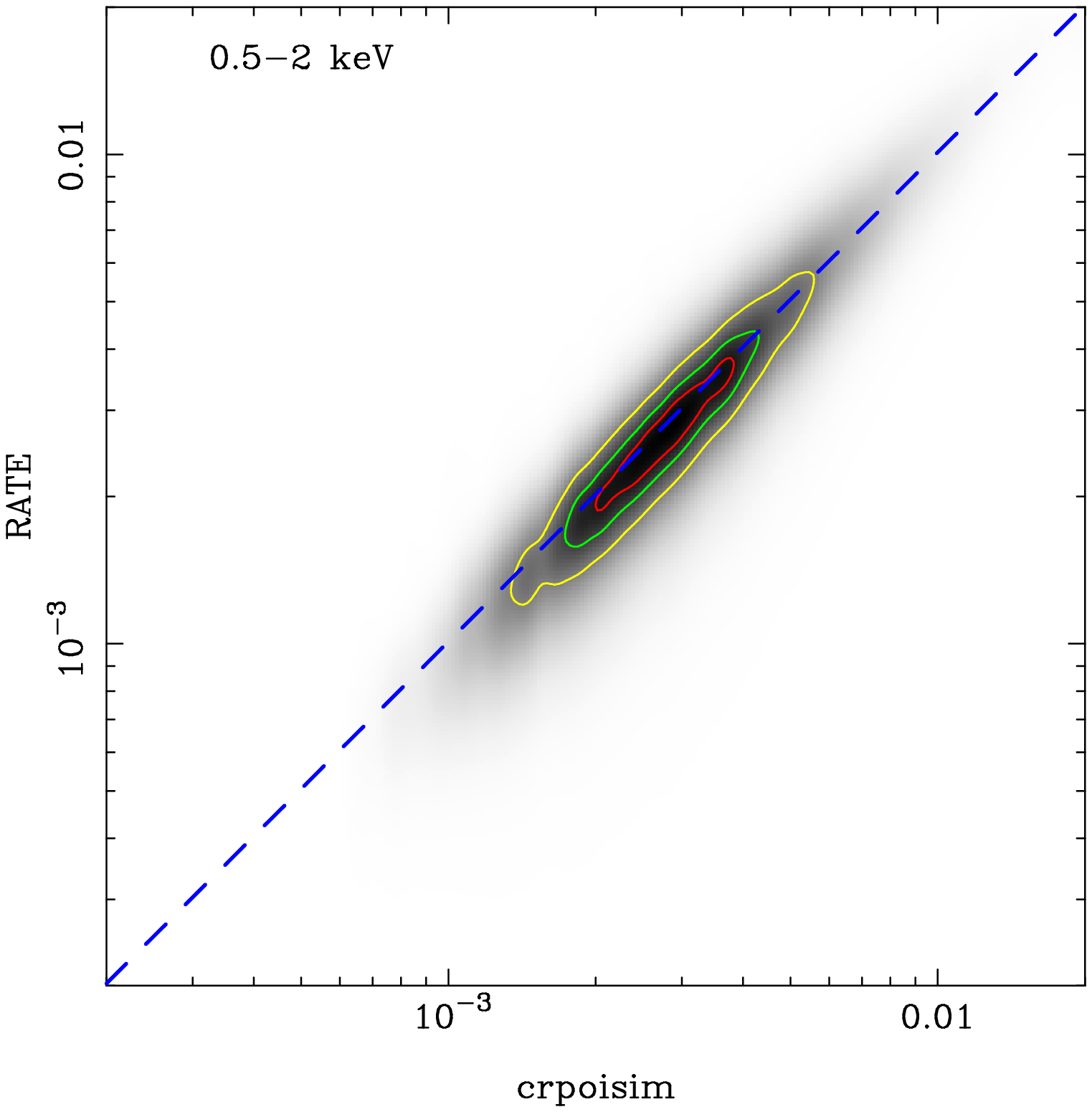}
   \includegraphics[angle=-90,width=0.450\textwidth]{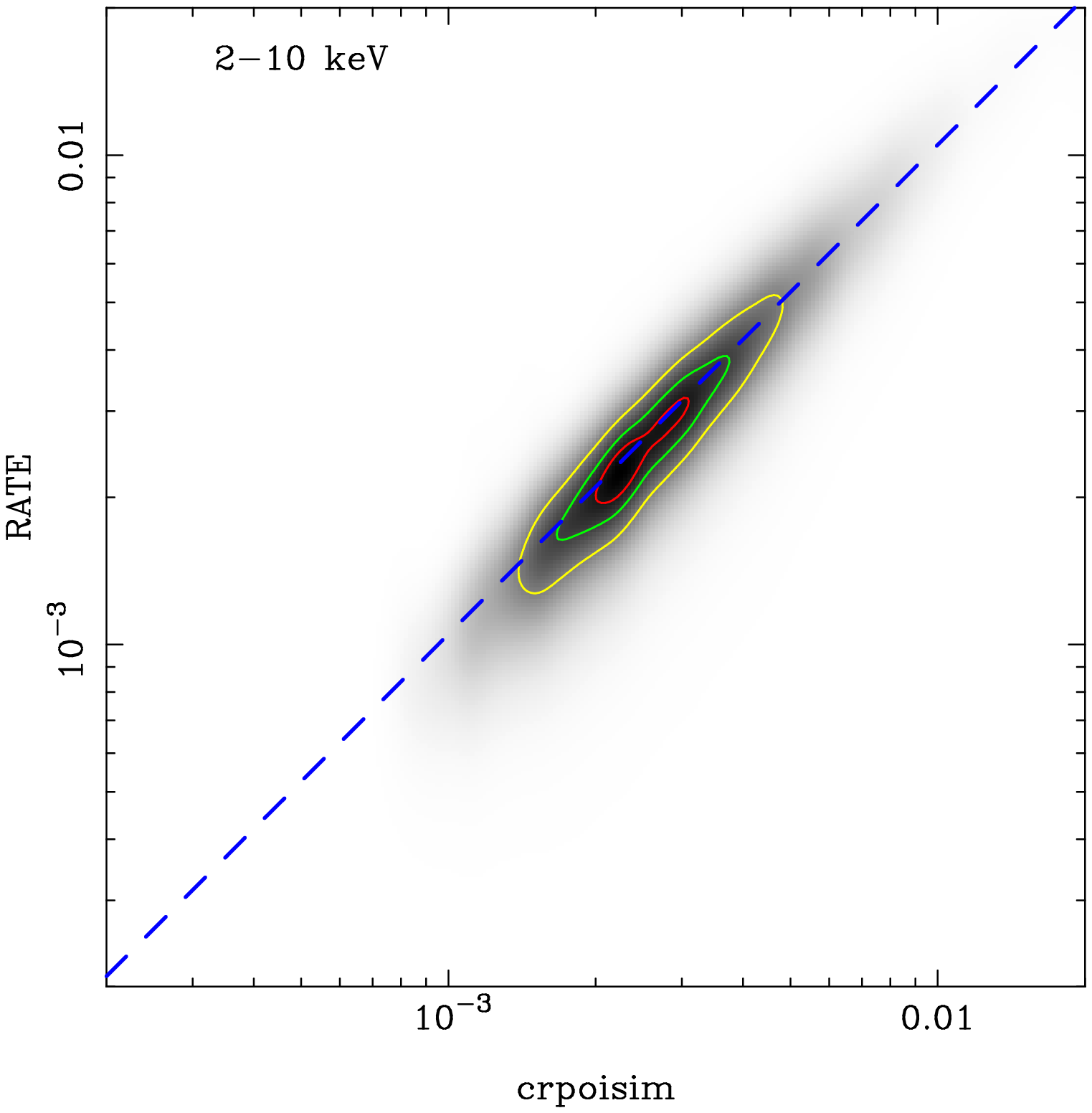}}
\hbox{
  \includegraphics[angle=-90,width=0.450\textwidth]{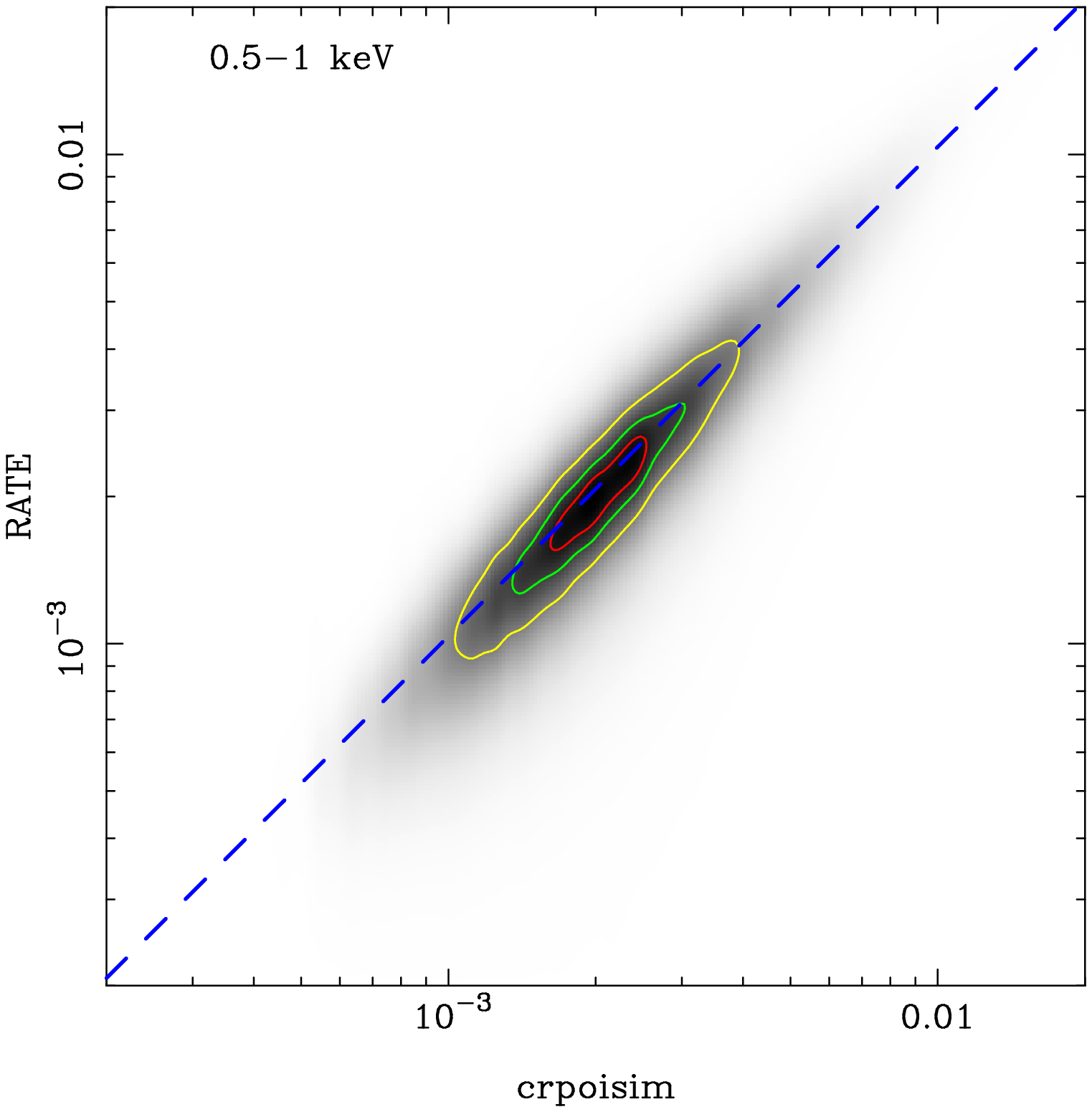}
   \includegraphics[angle=-90,width=0.450\textwidth]{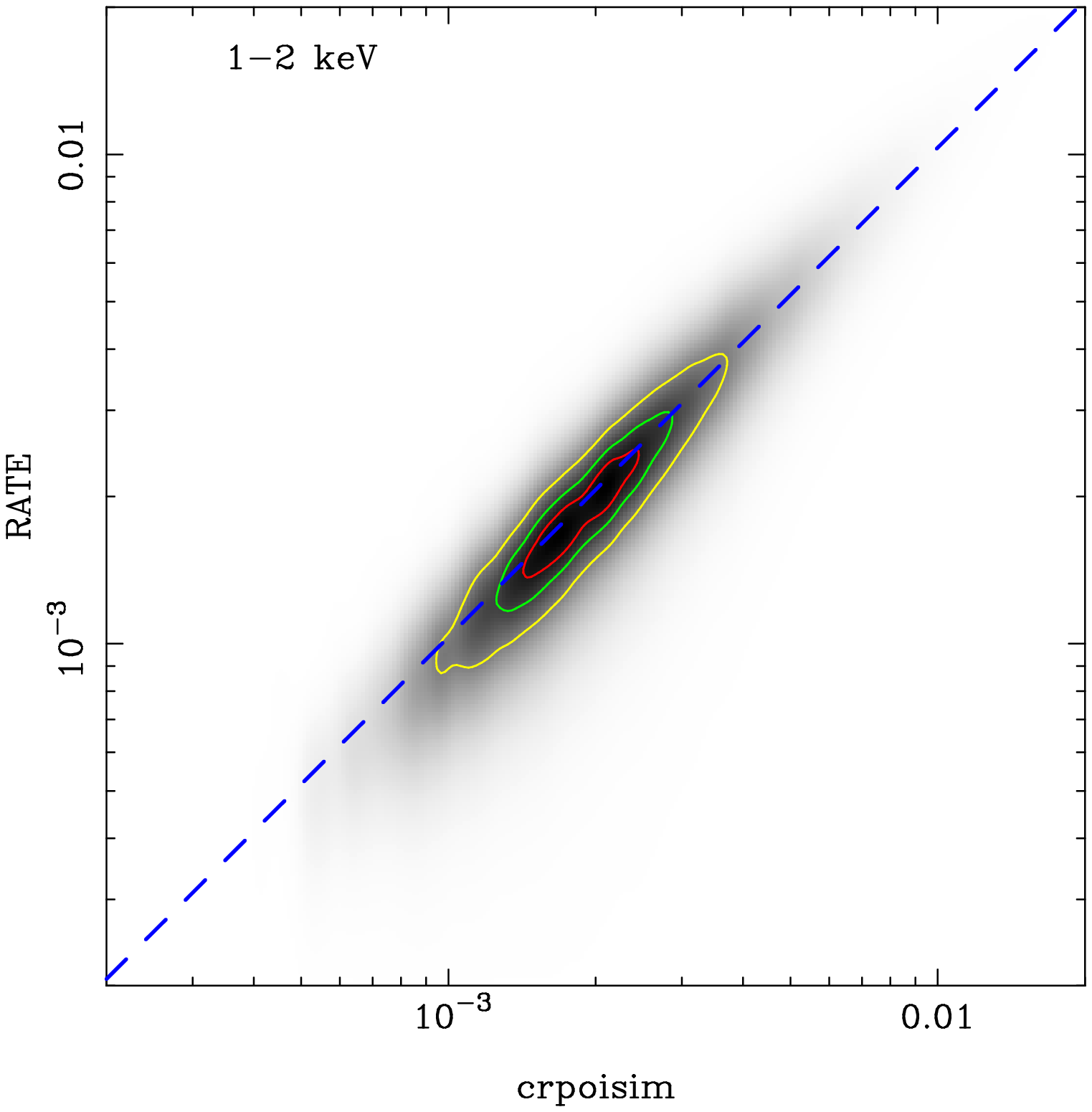}}
\hbox{
  \includegraphics[angle=-90,width=0.450\textwidth]{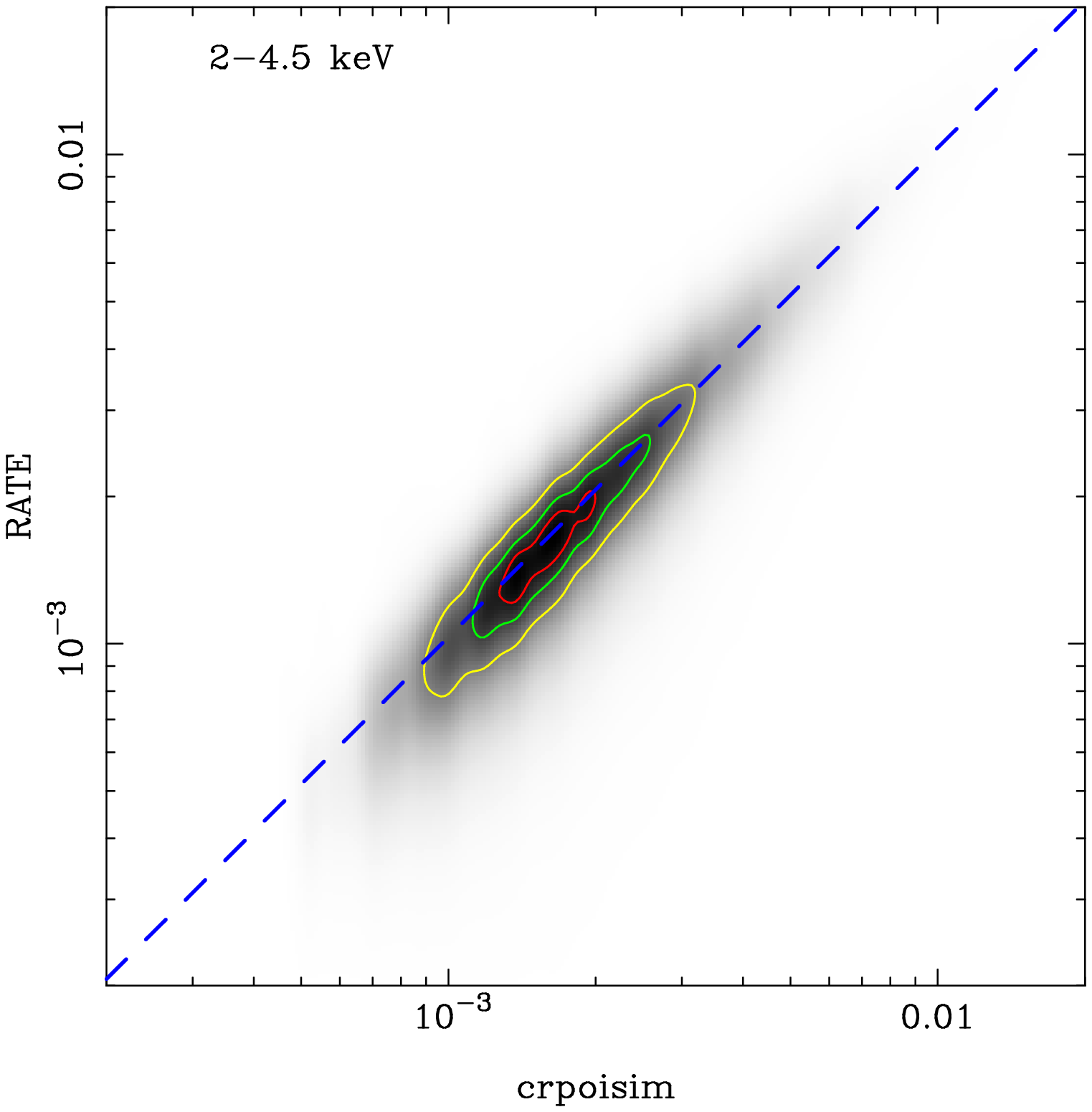}
   \includegraphics[angle=-90,width=0.450\textwidth]{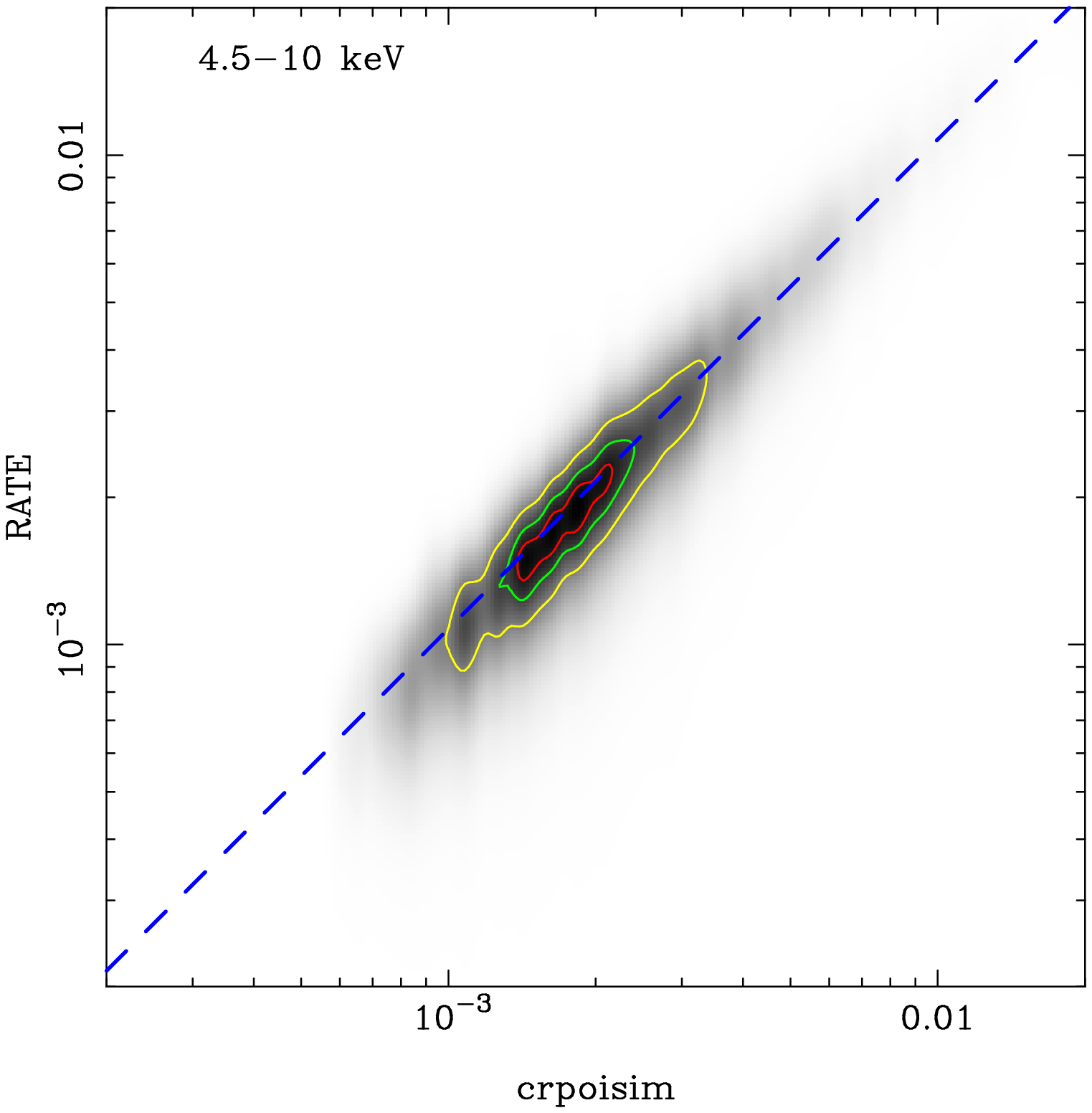}}
\caption{Count rates given by the {\tt emldetect} {\tt SAS} task vs pure Poissonian count rates (in units of cts s$^{-1}$). The 
  best fit to the observed distributions is also shown with a dashed line. The 
  contours show the 90\%, 75\% and 50\% level of peak intensity.}
\label{LI_plots}
\end{figure*}

\begin{table}
  \caption{Summary of the results of the linear fit to the distributions of {\tt emldetect} (RATE) vs Poissonian ($crpoisim$) 
count rates.}
\label{table:2}      
\centering                          
\begin{tabular}{c c c c c c }        
\hline\hline                 
Energy band      &  ${\rm N_{tot}}$ & LI & $\chi^2_{LI}$  & $\chi^2_0$ \\
(1) & (2) & (3) & (4) & (5) \\
\hline                        
0.5-1  & 16349 & $1.056_{-0.005}^{+0.005}$ & 4067  & 4402 \vspace{0.05cm} \\
1-2    & 17817 & $1.054_{-0.005}^{+0.005}$ & 3995  & 4323 \vspace{0.05cm} \\
2-4.5  &  8760 & $1.054_{-0.007}^{+0.008}$ & 1675  & 1824  \vspace{0.05cm}\\
4.5-10 &  1796 & $1.102_{-0.015}^{+0.016}$ & 276.5 & 395.1 \vspace{0.05cm}\\
0.5-2  & 22952 & $1.032_{-0.003}^{+0.004}$ & 6493  & 6692  \vspace{0.05cm}\\
2-10   &  8484 & $1.076_{-0.007}^{+0.007}$ & 1654  & 1983  \vspace{0.1cm} \\
\hline                                   
\end{tabular}

(1) Energy band definition (in keV). (2) Number of sources used in the fit. (3) Best fit slopes. 
(4) Best fit $\chi^2$ obtained when the slope is left free to vary.
(5) Best fit $\chi^2$ obtained when the slope is fix to a value of 1 (i.e. assuming that on average 
{\tt emldetect} and Poissonian count rates do not differ).
\end{table}

A detailed description of the method used in this work to obtain the sky coverage as a function of the 
X-ray flux is given in Appendix A in Carrera et al.~(\cite{Carrera07}). Briefly, 
assuming Poisson statistics hold, it is possible to determine, for a given detection likelihood, $\mathcal{L}$, 
the source detection threshold at each sky position.
The minimum count rate that a source must have in order to be detected at a certain position is given by solving the equation:
\[
-\log(P_{bgdim}(\ge(bgdim+crpoisim\times expim))=\mathcal{L} \label{eq:1}
\]
where $bgdim$ and $expim$ are respectively, the total background and mean exposure time within a circle of 5 pixel\footnote{The images were created with a 4 arcsec pixel side.} radius 
(the effective size of the FOV used to calculate the parameters for each object) at the source position. $Crpoisim$ is the 
minimum count rate that the object must have to be detected with a detection likelihood $\mathcal{L}$ at the source position. 
The {\tt SAS} task that creates the source lists we have used for our analysis, {\tt emldetect}, performs a maximum 
likelihood fit to the distribution of counts of the sources in the images convolved with the PSF of the telescope at the source positions.
Therefore, the count rate values resulting from {\tt emldetect} are not Poissonian. However, as it is shown in 
Appendix A of Carrera et al.~(\cite{Carrera07}), there exists a linear relationship between the Poissonian count rates, 
$crpoisim$, and those obtained by {\tt emldetect}, {\tt RATE}. 
The same linear relationship (${\tt RATE}={\rm LI}\times {\it crpoisim}$) 
is found in our data 
and therefore we have used it to correct empirically from all non-Poissonian effects introduced by {\tt emldetect}  
to the values of {\it crpoisim} (see Fig.~\ref{LI_plots}).
The corrections we have applied to the Poissonian count rates are listed in Table~\ref{table:2}.
We note that although the correction is small in all cases, it is nevertheless, very significant. 

   \begin{figure*}
   \centering
   \includegraphics[angle=90,width=0.5\textwidth]{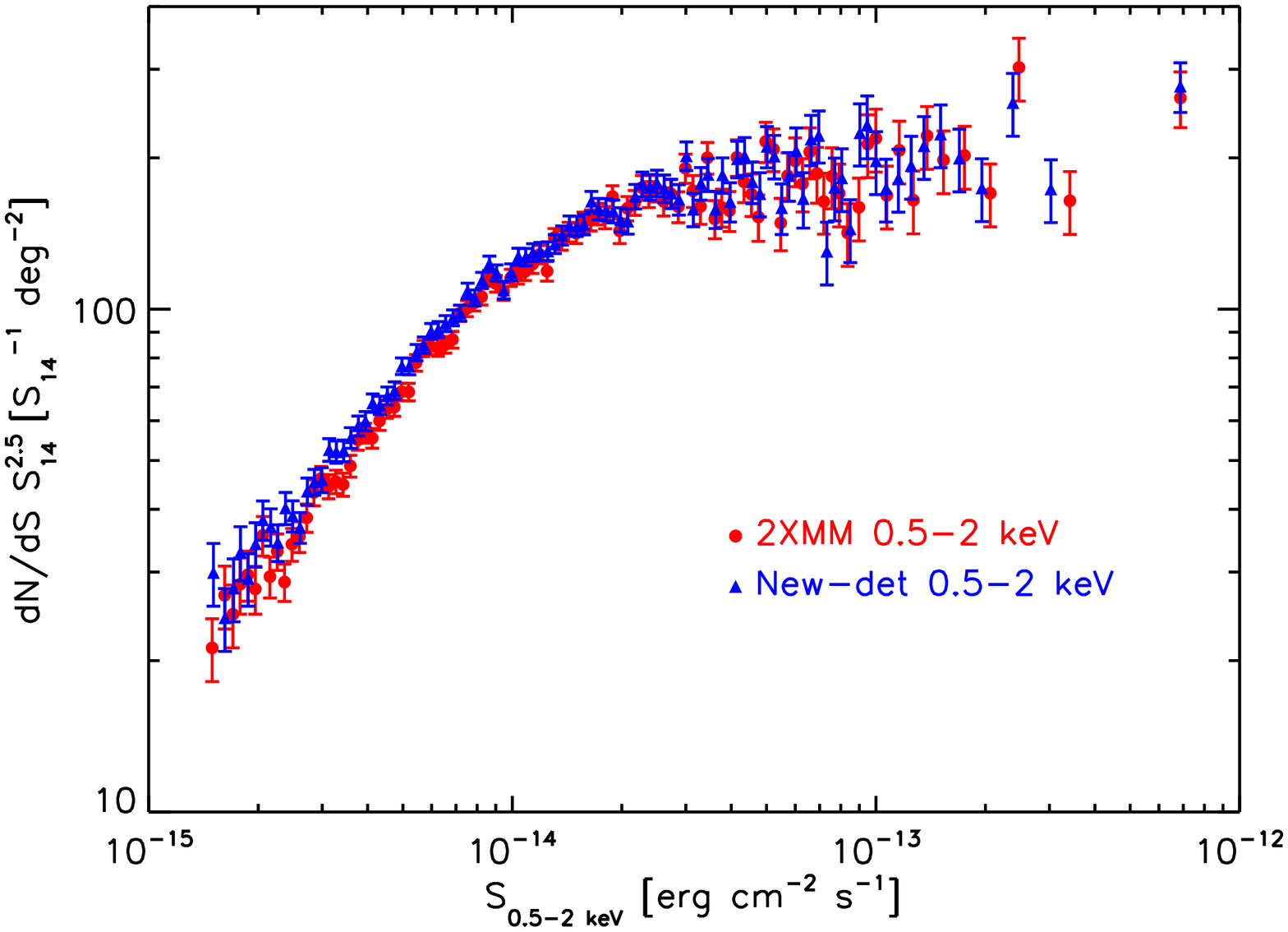}
   \includegraphics[angle=90,width=0.5\textwidth]{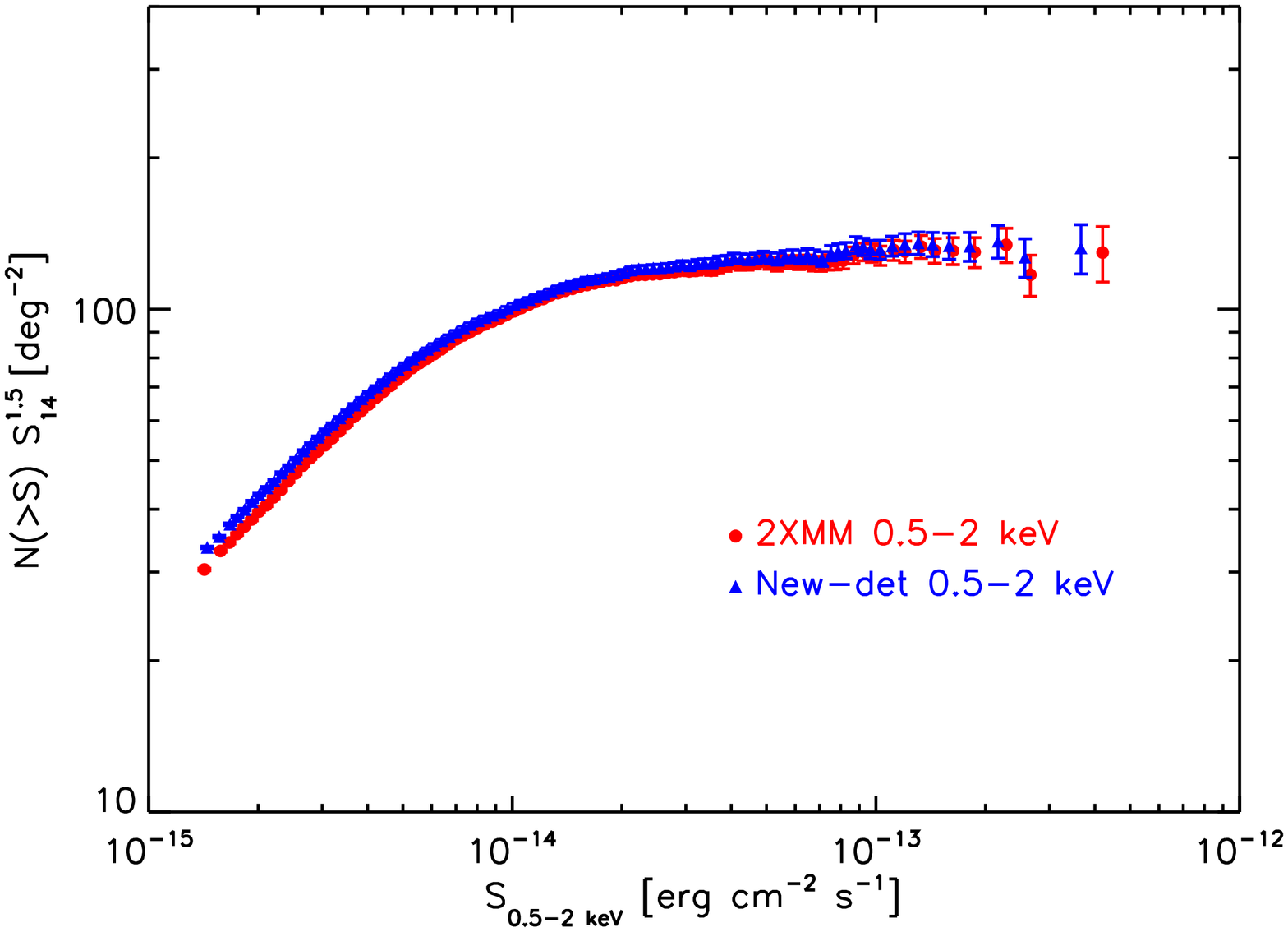}
   \caption{
     Comparison of the normalised differential (left) and integral (right) source count distributions for sources detected 
   in the 0.5-2 keV band from our source detection (triangles) and for sources in the {\tt 2XMM} catalogue (circles). The source 
parameters for {\tt 2XMM} sources were obtained from the combination of parameters from the {\tt 2XMM} energy bands 0.5-1 keV and 1-2 keV. Error bars correspond to 1$\sigma$ confidence.}
              \label{com_soft}%
    \end{figure*}

   \begin{figure*}
   \centering
   \includegraphics[angle=90,width=0.5\textwidth]{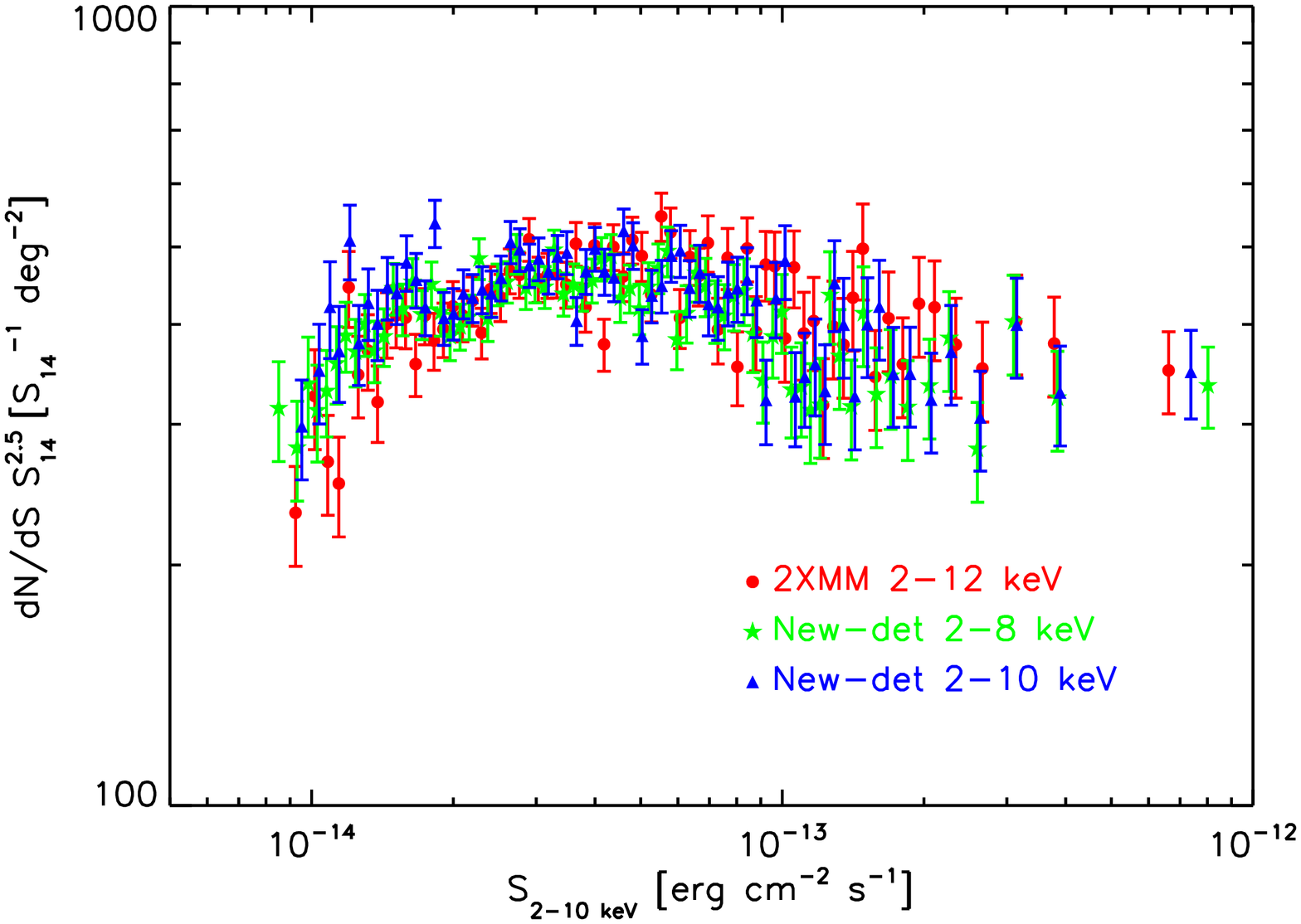}
   \includegraphics[angle=90,width=0.5\textwidth]{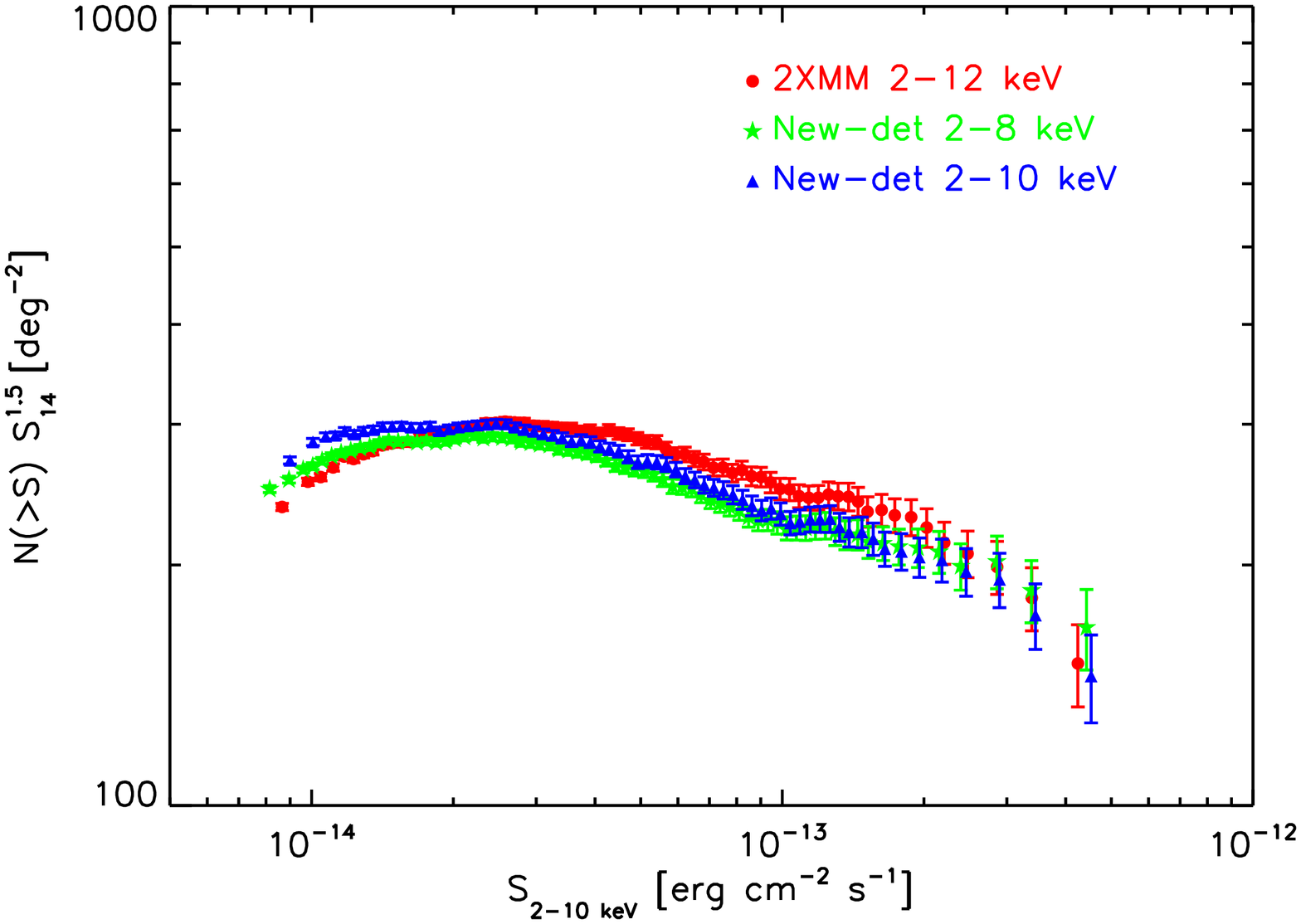}
   \caption{Comparison of the normalised differential (left) and integral (right) source count distributions for sources detected 
   in the 2-10 keV band from our source detection (triangles) and for sources from the {\tt 2XMM} catalogue (circles). Source parameters 
   for {\tt 2XMM} sources were obtained from the combination of parameters from the {\tt 2XMM} energy bands 2-4.5 keV and 
   4.5-12 keV. We also show for comparison the source count distribution in the 2-8 keV band obtained from our source detection (stars). 
   Error bars correspond to 1$\sigma$ confidence.}
              \label{com_hard}%
    \end{figure*}

\section{Soft and hard {\tt 2XMM} source counts}
\label{2xmm_vs_mydet}
The source detection pipeline used to make the {\tt 2XMM}
catalogue was run on data from the three EPIC cameras (MOS1, MOS2 and pn) and on five different energy bands 
simultaneously: 0.2-0.5 keV, 0.5-1 keV, 1-2 keV, 2-4.5 keV and 4.5-12 keV.

In principle it is possible to combine source parameters (fluxes and detection likelihoods) from  
different energy bands in order to obtain soft (combining {\tt 2XMM} energy bands 
0.5-1 keV and 1-2 keV) and hard (combining {\tt 2XMM} energy bands 2-4.5 keV and 4.5-12 keV) band parameters. 
Detection likelihoods in the 0.5-2 keV ($\mathcal{L}_{23}$) and 2-12 keV ($\mathcal{L}_{45}$) energy bands are obtained 
by combining the detection likelihoods in the individual energy bands via the recipe in the 
{\tt emldetect} documentation.
First {\tt emldetect} detection likelihoods ($\mathcal{L}_i$) have to be 
converted to probabilities ($\mathcal{L}_i^{'}$): $\mathcal{L}_{23}=-\log(1-P(\nu/2,\mathcal{L}_2^{'}+\mathcal{L}_3^{'}))$, 
where $P$ is the incomplete Gamma function 
and $\nu$ is the number of degrees of freedom: $\nu$=4 for point sources and $\nu$=5 for 
extended sources. The values $\mathcal{L}_i^{'}$ are obtained solving the equation $\mathcal{L}_i=-\log(1-P(\nu/2,\mathcal{L}_i^{'}))$. 
In this case $\nu$=3 for point sources and $\nu$=4 for extended sources. Fluxes are obtained adding the fluxes from the 
individual energy bands. 

Source count distributions derived from 
the combined {\tt 2XMM} bands can be compared with those obtained from running the detection algorithm 
directly on the 0.5-2 keV and 2-10 keV band data. 
Fig.~\ref{com_soft} shows the result for the soft energy band and Fig.~\ref{com_hard} for the hard energy band. 
Fluxes have been converted to the 2-10 keV band using the scaling factors in Table~\ref{table:4}.

We see that in the 0.5-2 keV band the distributions obtained 
following the two approaches look very similar, the only difference being that the distribution from the combined band 
has a marginally higher normalisation at bright fluxes and a flatter slope at fluxes below the break. 
The latter is maybe due to the fact that detection likelihoods obtained when combining the {\tt 2XMM} bands 
are systematically lower than those obtained from the single band analysis ($\approx$2 units lower) so the combination 
of the bands results in the loss of a small fraction 
of the sources detected at the very faint limit of the observations (when a fixed likelihood threshold is applied). 

However important differences are seen in both the shape and normalisation of the 2-12 keV source counts with respect 
to the distribution obtained for the 2-10 keV band\footnote{The 
effective area of the XMM-{\it Newton} detectors decreases rapidly above $\sim$5 keV and it is very low above $\sim$10 keV, hence we expect 
that the results in the 2-12 keV band will not differ significantly from those in the 2-10 keV band.} (see Fig.~\ref{com_hard}).
The origin of the different results is probably due to a combination of various effects, such as 
different background levels and different source properties. It is important to note that above 10 keV the effective area 
of the EPIC-pn detector is very low, but the background is high, so extending the energy band from 10 keV to 12 keV is 
expected to reduce the signal to noise of the data. 

Some previous results, mostly using data 
from {\it Chandra} are based on a selection of sources in the 2-8 keV band, so we also show the distribution 
obtained from our source detection in this band for comparison. The source count distribution in the 2-8 keV band 
seems to have a marginally flatter slope below the break than the distribution obtained for sources 
detected in the 2-10 keV energy band.

\end{appendix}

\end{document}